\documentclass[aps,amsmath,amssymb,superscriptaddress,notitlepage,nofootinbib,twocolumn]{revtex4-1}

\usepackage{graphicx}
\usepackage{bm}
\usepackage{xcolor}
\usepackage{hyperref}
\usepackage{marvosym}
\usepackage{soul}

\newcommand{\mnras}{MNRAS}
\renewcommand{\apj}{ApJ}
\newcommand{\aap}{A\&A}

\newcommand{\apjs}{ApJS}
\newcommand{\apjl}{ApJL}
\newcommand{\araa}{ARA\&A}
\renewcommand{\prd}{PRD}
\renewcommand{\prl}{PRL}
\newcommand{\jhep}{JHEP}

\renewcommand{\nat}{Nature}
\newcommand{\jcap}{JCAP}
\newcommand{\pasp}{PASP}

\newcommand{\epjc}{EPJC}

\newcommand{\dataset}{\vect{d}}
\newcommand{\etc}{{\em etc.}}
\newcommand{\eg}{{\em e.g.}}
\newcommand{\cf}{{\em cf.}}
\newcommand{\ie}{{\em i.e.}}
\newcommand{\msun}{M_\odot}

\newcommand{\vect}[1]{\ensuremath{{\bm{#1}}}}
\newcommand{\hubble}{\ensuremath{H_0}}
\newcommand{\hubbleest}{\ensuremath{\hat{H}_0}}
\newcommand{\hubblemap}{\hat{H}_{0,{\rm MAP}}}
\newcommand{\decel}{\ensuremath{q_0}}

\newcommand{\dl}{\ensuremath{D}}
\newcommand{\dprop}{\ensuremath{D_{\rm P}}}
\newcommand{\dlmax}{\ensuremath{D_*}}
\newcommand{\dlss}{\ensuremath{D_0}}
\newcommand{\zmax}{\ensuremath{z_*}}
\newcommand{\dlobs}{\ensuremath{\hat{D}}}
\newcommand{\sigmadmax}{\sigma_{*}}

\newcommand{\vobs}{\ensuremath{\hat{v}}}
\newcommand{\vobss}{\ensuremath{\hat{\vect{v}}}}
\newcommand{\zobs}{\ensuremath{\hat{z}}}
\newcommand{\zobss}{\ensuremath{\hat{\vect{z}}}}
\newcommand{\zspec}{\ensuremath{\tilde{z}}}
\newcommand{\prob}{\ensuremath{{\rm P}}}
\newcommand{\normal}{{\rm{N}}}
\newcommand{\diff}{{\rm{d}}}

\newcommand{\sels}{\vect{S}}
\newcommand{\dgws}{\hat{\vect{h}}}
\newcommand{\dgw}{\hat{h}}
\newcommand{\gwpar}{\theta}
\newcommand{\gwpars}{\vect{\gwpar}}

\newcommand{\inc}{\iota}

\newcommand{\incs}{\vect{\iota}}
\newcommand{\zs}{{\vect{z}}}
\newcommand{\vs}{{\vect{v}}}
\newcommand{\csmpar}{\ensuremath{\Omega}}
\newcommand{\nexp}{\bar{N}}

\newcommand{\mchirp}{{\cal{M}}}
\newcommand{\sigmamchirp}{\sigma_{\!{{\cal{M}}}}}
\newcommand{\mchirps}{\vect{{\cal{\vect{M}}}}}
\newcommand{\mchirpmean}{\bar{{\cal{M}}}}
\newcommand{\mz}{{\cal{M}}_z}
\newcommand{\mzobs}{\hat{{\cal{M}}}_z}
\newcommand{\mzobss}{\hat{\vect{{\cal{M}}}}_z}
\newcommand{\tc}{t_{\rm c}}
\newcommand{\tcobs}{\hat{t}_{\rm c}}
\newcommand{\phic}{\Phi_{\rm c}}
\newcommand{\phicobs}{\hat{\Phi}_{\rm c}}
\newcommand{\amp}{A}
\newcommand{\ampobs}{\hat{A}}

\newcommand{\ampobss}{\hat{\vect{A}}}

\newcommand{\plus}{+}
\newcommand{\pol}{p}
\newcommand{\strain}{h}

\newcommand{\sigmav}{\sigma_{||}}
\newcommand{\gwpoppar}{\beta}

\newcommand{\rate}{\Gamma}

\newcommand{\step}{\Theta}
\newcommand{\stonobs}{\hat{\rho}}
\newcommand{\ston}{\rho}
\newcommand{\stonmin}{\rho_*}
\newcommand{\eq}[1]{Eq.~\ref{equation:#1}}
\newcommand{\sect}[1]{Sec.~\ref{section:#1}}
\newcommand{\app}[1]{Appendix~\ref{section:#1}}
\newcommand{\fig}[1]{Fig.~\ref{figure:#1}}
\newcommand{\vol}{V}
\newcommand{\kmsmpc}{\ensuremath{{\rm km\,s^{-1}\,Mpc^{-1}}}}
\newcommand{\kms}{\ensuremath{{\rm km\,s^{-1}}}}
\newcommand{\mpc}{\ensuremath{{\rm Mpc}}}
\newcommand{\gpc}{\ensuremath{{\rm Gpc}}}
\newcommand{\yr}{\ensuremath{{\rm yr}}}
\newcommand{\planck}{{\it Planck}}

\newcommand{\lcdm}{$\Lambda$CDM}

\newcommand{\la}{\ {\raise-.5ex\hbox{$\buildrel<\over\sim$}}\ }
\newcommand{\ga}{\ {\raise-.5ex\hbox{$\buildrel>\over\sim$}}\ }

\begin{document}
\title{Unbiased Hubble constant estimation from binary neutron star mergers}
\author{Daniel J.\ Mortlock}
\email{mortlock@ic.ac.uk}
\affiliation{Astrophysics Group, Imperial College London, 
  Blackett Laboratory, Prince Consort Road, London SW7 2AZ, UK}
\affiliation{Department of Mathematics, Imperial College London, 
  London SW7 2AZ, UK}
\affiliation{Department of Astronomy, Stockholm University, 
  AlbaNova, SE-10691 Stockholm, Sweden}
\author{Stephen M.\ Feeney}
\affiliation{Center for Computational Astrophysics, Flatiron Institute, 162 5th Avenue, New York, NY 10010, USA}
\author{Hiranya V.\ Peiris}
\affiliation{Department of Physics \& Astronomy, University College London, Gower Street, London WC1E 6BT, UK}
\affiliation{Oskar Klein Centre for Cosmoparticle Physics, Department of Physics, Stockholm University, AlbaNova, Stockholm SE-106 91, Sweden}
\author{\\ Andrew R.\ Williamson}
\affiliation{GRAPPA, Anton Pannekoek Institute for Astronomy and Institute of High-Energy Physics, University of Amsterdam, Science Park 904, 1098 XH Amsterdam, The Netherlands}
\affiliation{Nikhef, Science Park 105, 1098 XG Amsterdam, The Netherlands}
\author{Samaya M.\ Nissanke}
\affiliation{GRAPPA, Anton Pannekoek Institute for Astronomy and Institute of High-Energy Physics, University of Amsterdam, Science Park 904, 1098 XH Amsterdam, The Netherlands}
\affiliation{Nikhef, Science Park 105, 1098 XG Amsterdam, The Netherlands}


\begin{abstract}
Gravitational wave (GW) observations of binary neutron star (BNS) 
mergers can be used to measure luminosity distances and hence, 
when coupled with estimates for the mergers' host redshifts, 
infer
the Hubble constant, \hubble. 
These observations are, however, affected by GW measurement noise, 
uncertainties in host redshifts and peculiar velocities,
and are potentially biased by selection effects and 
the mis-specification of the cosmological model or the BNS population.
The estimation of \hubble\ from samples of BNS mergers 
with optical counterparts
is tested here by using a phenomenological
model for the GW strains that captures both the data-driven
event selection and the distance-inclination degeneracy,
while being simple enough to facilitate large numbers of
simulations.
A rigorous Bayesian approach 
to analyzing the data from such simulated BNS merger samples 
is shown to yield
results that are unbiased, have the 
appropriate uncertainties, and are robust to model
mis-specification.
Applying such methods to a
sample of $N \simeq 50$ BNS merger events,
as LIGO+Virgo 
could
produce in the next $\sim \! 5$ years,
should yield robust and accurate
Hubble constant estimates that are  
precise to a level of 
$\la 2\ \kmsmpc$,
sufficient to reliably resolve the current tension between local
and cosmological measurements of \hubble.
\end{abstract}


\maketitle


\section{Introduction}
\label{section:intro}

The current expansion rate of the Universe is characterized by 
the Hubble constant, \hubble.  
While there is 
an empirical
consensus that
$\hubble \simeq 70\ \kmsmpc$,
there is a 
$4.4$-$\sigma$
tension between 
the most recent local and cosmological measurements:
the anchor-Cepheid-supernova distance ladder gives 
$\hubble = 74.2 \pm 1.8$ \kmsmpc\ \cite{Riess_etal:2019}; 
while
the \planck\ cosmic microwave background (CMB) data, 
combined with the assumption of a standard flat cold dark matter (\lcdm) 
cosmological model,
imply that $\hubble = 67.3 \pm 0.6$ \kmsmpc\ \cite{Planck_VI:2018}. 
It is tempting to use this tension as motivation for rejecting or extending \lcdm~\cite{Wyman_etal:2014,Pourtsidou_etal:2016,Di_Valentino_etal:2016,Huang_etal:2016,Bernal_etal:2016,Ko_etal:2016,Karwal_etal:2016,Kumar_etal:2016,Santos_etal:2017,Prilepina_etal:2017,Zhao_etal:2017,Zhao_M_etal:2017,Solar_etal:2017,Di_Valentino_etal:2017a,Di_Valentino_etal:2017b,Graef_etal:2018,Yang_etal:2018a,El-zant_etal:2018,Benevento_etal:2018,Yang_etal:2018b,Aylor_etal:2018,Chiang_Slosar:2018,Poulin_etal:2018}, 
but before settling on such an exciting possibility
it is necessary to ensure that the discrepancy is not due to 
limitations in the analysis of one or both of the datasets~\cite{Efstathiou:2014,Spergel_etal:2015,Rigault_etal:2015,Jones_etal:2015,Addison_etal:2016,Planck_Int_XLVI:2016,Cardona_etal:2016,Zhang_etal:2017,Wu_Huterer:2017,Feeney_etal:2017,Follin_Knox:2017,Dhawan_etal:2017,Rigault_etal:2018,Bengaly_etal:2018}.

The most direct way of resolving this `Hubble trouble' would be a local measurement of \hubble\ that is completely independent of the above distance ladder.  One of the most promising options is gravitational wave (GW) observations of mergers between compact/relativistic objects such as black holes (BHs) and neutron stars (NSs).  The GW waveform from any such merger provides information about the distance to the system which can then be combined with a redshift measurement/estimate to constrain \hubble\ \cite{Schutz:1986}.  The advent of the Advanced Laser Interferometer Gravitational-Wave Observatory and Advanced Virgo (LIGO+Virgo) has resulted in the detection of several such mergers \cite{Abbott_etal:2017d,Ligo:2018}, although the utility for measuring \hubble\ depends strongly on the type of system:

\begin{itemize}

\item
Binary neutron star (BNS) mergers are particularly promising, as a possible electromagnetic (EM) counterpart could be used to identify a host galaxy from which a spectroscopic redshift measurement could be made (\eg, \cite{Dalal:2006,Nissanke_etal:2010,Abbott_etal:2017a}). For some fraction of BNS mergers a counterpart will not be identified, in which case it is plausible to take a statistical approach, averaging over the host galaxies that are consistent with the GW localization \cite{del_Pozzo:2012,Oguri:2016,Fishbach_etal:2018}.  It is also possible that GW data alone could be used to obtain redshift constraints by exploiting either the narrowness of the NS mass distribution \cite{Taylor_Gair:2012b} or NS tidal deformability \cite{Messenger_Read:2012}.
The first BNS merger event detected in GWs, GW~170817/GRB~170817A
\cite{Abbott_etal:2017d},
provided the constraint that
$\hubble = 70.0_{-8.0}^{+12.0}\ \kmsmpc$ \cite{Abbott_etal:2017a}.
More precise, if necessarily model-dependent, constraints can be obtained by
using EM data to estimate the inclination of the system \cite{Mooley_etal:2018},
with different analyses of the data from GRB~170817A yielding 
$\hubble = 74.0_{-13.7}^{+5.3}\ \kmsmpc$ \cite{Guidorzi_etal:2017}
and 
$\hubble = 68.9_{-4.6}^{+4.7}\ \kmsmpc$ \cite{Hotokezaka_etal:2018}.
Conversely, if it had not been
possible to identify an EM counterpart to GW~170817,
the GW measurements alone would have yielded
$\hubble = 70.0_{-23.0}^{+48.0}\ \kmsmpc$~\cite{Fishbach_etal:2018}.

\item
Binary black hole (BBH) mergers are the strongest GW sources and have dominated current merger samples \cite{Abbott_etal:2018}.  But the expected lack of any EM emissions, along with their broad range of masses, makes it difficult to obtain useful redshift estimates.  The best hope is to take the same sort of statistical approach as considered above for the BNS mergers \cite{del_Pozzo:2012,Nair:2018,Gray_etal:2019}, although the uncertainties in \hubble\ are then dominated by this limitation (\eg, \cite{Soares-Santos_etal:2019,Ligo:2019}).

\item
NS-BH mergers might provide tighter distance constraints than BNS mergers due to both their higher system mass \cite{Nissanke_etal:2010,Nissanke_etal:2013} and the possibility that the BH spin is not aligned with the orbital angular momentum, which would induce precession of the orbital plane \cite{Vitale_Chen:2018}.  It is less clear, however, whether there would be detectable EM emission from such mergers; if there is not then the constraints on \hubble\ would be limited in the same way as BBH mergers and BNS mergers without counterparts.

\end{itemize}

\noindent
As compact mergers with EM counterparts provide the cleanest \hubble\ constraints and no NS-BH merger has yet been decisively confirmed in either GW or EM observations, the focus here is on BNS mergers with EM counterparts and confirmed host galaxies, although the overall approach taken here applies to all types of compact binary mergers.
The third LIGO+Virgo observing run (O3)
should detect several more BNS mergers which,
if EM counterparts could be identified in all cases,
would already give an uncertainty in \hubble\
smaller than the difference between the local and CMB values
\cite{Seto:2018,Chen_etal:2018}.
Looking further ahead, 
detector advances over the next five years 
could yield a sample of $\sim\!50$ BNS merger events,
sufficient to measure
$\hubble$ with a precision of 
$\la 2\ \kmsmpc$
\cite{Chen_etal:2018,Feeney_etal:2018}.

Such a sample of BNS mergers with EM counterparts 
would be sufficient to resolve the current 
$\hubble$
tension
\cite{Feeney_etal:2018}, 
but only if the data analysis produces 
\hubble\ estimates that have
correct uncertainties and that are 
free of systematic biases
(\ie, accurate, as well as precise).
One potential source of bias is selection effects,
as detection 
on the basis of
the observed GW signal-to-noise ratio (SNR)
will preferentially
include events for which the measurement noise has augmented the signal,
making such mergers appear closer than they are.
Unless accounted for, this selection effect 
would result in a systematic overestimate of \hubble\
(\eg, \cite{Taylor_etal:2012,Messenger:2013}).
Another potential source of bias is mis-specification
of the cosmological model or the BNS population,
as the data-generation process links \hubble\ to
these other global parameters 
(\eg, \cite{Taylor_etal:2012,Abbott_etal:2017a,Chen_etal:2018}).
Eventually, overall calibration uncertainties 
are likely to provide the absolute systematic floor to the 
precision in \hubble\ of a BNS merger sample;
these are currently at a level of 
few per cent, although there are clear prospects for 
further improvements
\cite{Karki_etal:2016,Cahillane_etal:2017}.

The main aim of this paper is to test whether
a Bayesian population analysis 
of the sort described by Refs
\cite{Taylor_etal:2012,Abbott_etal:2016,Abbott_etal:2017a,Chen_etal:2018,Fishbach_etal:2018,Feeney_etal:2018,Mandel_etal:2018}
gives unbiased \hubble\ estimates 
with valid uncertainties 
(\ie, is both accurate and precise)
when applied to realistic BNS merger samples.
A secondary aim is to provide a derivation from first 
principles
of the full posterior distribution appropriate 
to a sample of BNS mergers with 
EM
counterparts.
Simple predictions for the uncertainty 
and selection bias in \hubble\ are given in
Secs \ref{section:prediction} and \ref{section:biases}.
The sample simulations are described in \sect{model}
and the Bayesian analysis approach summarized in \sect{bayes}.
The large-sample properties of this approach are then explored 
in \sect{test},
with the conclusions and possibilities for future development summarized 
in \sect{conclusions}.
The general model and inference formalism is presented 
in \app{derivation_app}
and the simplified BNS inspiral model is described in \app{linear}.


\section{Predicted uncertainties}
\label{section:prediction}

\begin{figure}
\includegraphics[width=9cm]{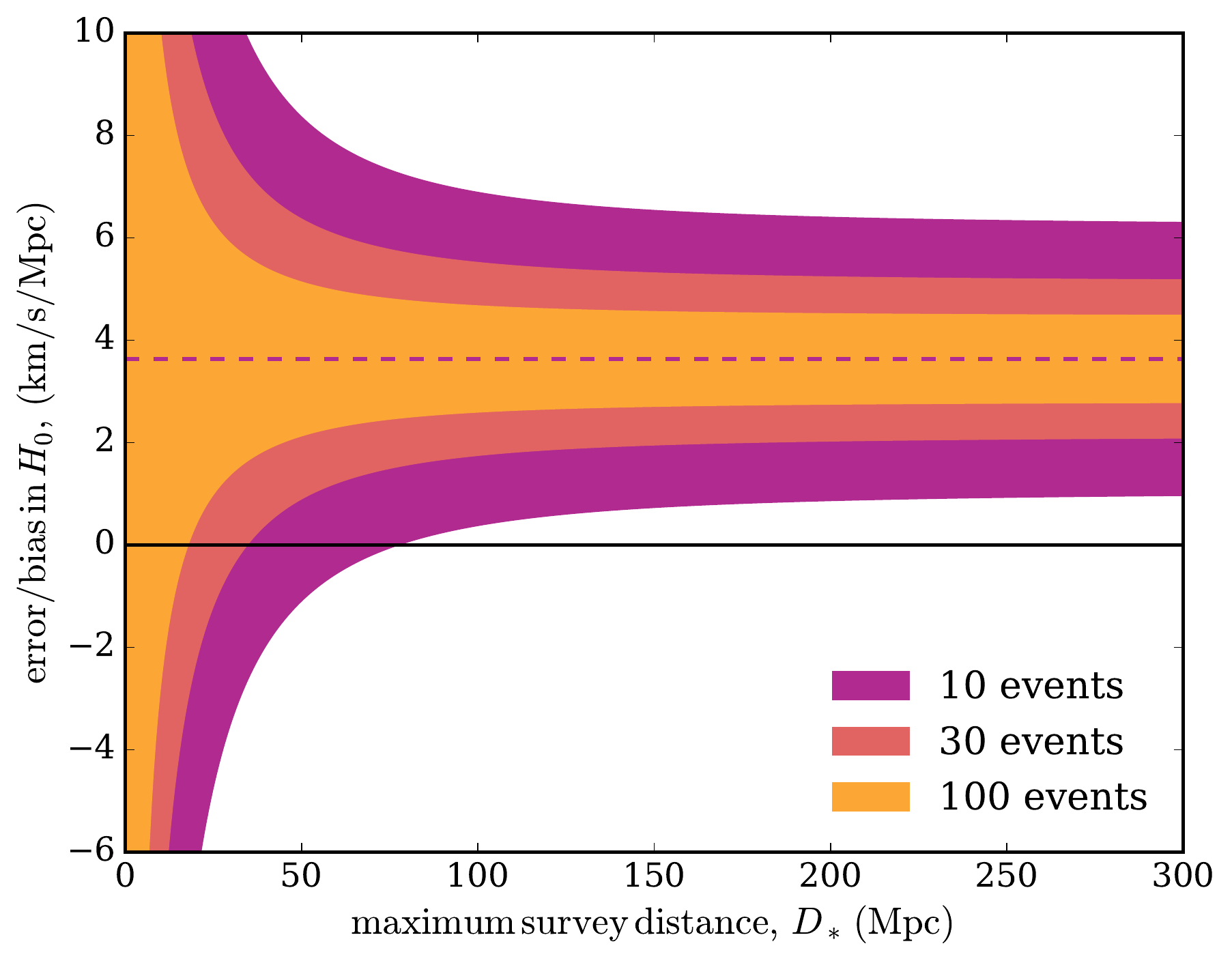}
\caption{The expected range of \hubble\
estimates from 
sample of $N = 10$, $N = 30$ and $N = 100$ BNS merger events,
as a function of the maximum distance, $\dlmax$, 
to which sources can be detected.
The offset from zero comes from the potential 
bias due to selection effects,
which would dominate over sample variance for 
$N \ga 10$ and $\dlmax \ga 70\ \mpc$.}
\label{figure:h0_errbias}
\end{figure}

In order to assess the performance of any data analysis 
method it is useful to have a prediction for the 
expected uncertainties in the idealized case
in which there are no complicating factors like selection
biases or systematic errors.
This gives a target for the 
uncertainties obtained 
from
real data
and also a guide for assessing whether any 
potential systematic effects are likely to be significant.

For a local sample of BNS mergers with counterparts,
the predicted uncertainty in \hubble\ can be estimated by 
considering 
an expanding
Euclidean geometry in which 
distance, $D$, is related to redshift, $z$, by $D = c \, z / \hubble$,
where $c$ is the speed of light.
Given a 
measured (spectroscopic) redshift, $\zobs$, 
an estimated peculiar velocity, $\vobs$
and a GW distance, $\dlobs$, 
for a BNS merger event 
(or any other low-redshift extragalactic object),
the natural estimator for the Hubble constant is
\begin{equation}
\label{equation:hest}
\hubbleest = \frac{c \, \zobs - \vobs}{\dlobs},
\end{equation}
where the peculiar velocity estimate effectively corrects 
the spectroscopic redshift to 
(hopefully) bring it closer to the cosmological value.
The resultant uncertainty 
(really the standard deviation of the estimator) 
is approximated to leading order as 
(\cf\ \cite{Chen_etal:2018,Seto:2018})
\begin{equation}
\label{equation:sigmah}
  \sigma_{\!H}
  \simeq 
  \frac{1}{\dl} 
  \left(
  c^2 \, \sigma_z^2 
  + 
  \sigma_v^2
  + 
  \hubble^2 \, \sigma_D^2
  \right)^{1/2}
  ,
\end{equation}
where
$\sigma_D$ is the uncertainty in the 
distance from the GW data,
$\sigma_z$ is the 
observational
redshift uncertainty,
and
$\sigma_v$ is the uncertainty in the peculiar velocity 
(which should be no larger than the 
observed peculiar velocity dispersion of $\sim 200\ \kms$
\cite{Burstein:1990},
but could be smaller if object-specific information is available).

Using representative numerical values for BNS mergers,
and exploiting the fact that 
it is already known that $\hubble$ is close to $70\ \kmsmpc$,
\eq{sigmah} can be re-written as
\begin{align}
\label{equation:sigmahnum}
\sigma_{\!H} 
&
\simeq 7.8\ \kmsmpc\
\left( \frac{\dl}{43.8\ \mpc} \right)^{-1}
\\
& \times
\left[
\left( \frac{0.21 \, \sigma_z}{0.00024} \right)^2 \!
\! + \!
\left( \frac{0.20 \, \sigma_v}{70.0\ \kms} \right)^2 \!
\! + \!
\left( \frac{\sigma_D}{4.9\ \mpc} \right)^2
\right]^{1/2}
\nonumber \\
&
\simeq 7.0\ \kmsmpc\
\left( \frac{\dl}{100\ \mpc} \right)^{-1}
\nonumber \\
& \times 
\left[
\left( \frac{0.43 \, \sigma_z}{0.001} \right)^2 \!
\! + \!
\left( \frac{0.29 \, \sigma_v}{200\ \kms} \right)^2 \!
\! + \!
\left( \frac{\sigma_D}{10\ \mpc} \right)^2
\right]^{1/2},
\nonumber
\end{align}
where the first case uses GW~170817 as a reference 
and the second uses more generic values appropriate to the 
simulations described below.
Both cases illustrate that the error in the reconstruction of the 
source distance from the GW data will be the dominant
uncertainty unless the SNR is sufficiently high that 
$\sigma_D$ can be reduced to $\la 3\ \mpc$,
which might be possible for very nearby sources
(\eg, the `golden' events invoked in Ref.~\cite{Chen_etal:2018}).
Even for such sources, however, there would be little value
in obtaining a redshift measurement more precise than the
fiducial value of $\sigma_z \simeq 0.001$ unless the 
peculiar velocity uncertainty of the host could be reduced
significantly below the cosmological prior value.

A single BNS merger clearly cannot place interesting constraints
on \hubble,
but if the uncertainty from a sample of $N$ events scales as $N^{-1/2}$
then \eq{sigmahnum}
broadly confirms the 
numerical predictions
\cite{Chen_etal:2018,Feeney_etal:2018}
that
$\sim 50$ events should yield
$\sigma_{\!H} \la 2\ \kmsmpc$ and hence be able to
resolve the current Hubble constant tension.
For the purposes of making quantitative predictions,
it is useful to take a more realistic approach
that incorporates both
the dependence of the uncertainty on source distance 
and the relative numbers of sources at different distances.

For a given source the GW strain signal
is inversely proportional to its distance,
while the strain noise is
additive and a characteristic of the detector
(\eg, \cite{Cutler_Flanagan:1994}).
Assuming the sources are all intrinsically 
identical
(\ie, truly `standard' sirens),
the SNR can then be 
written as $\ston \simeq \stonmin \, \dlmax / \dl$,
where $\stonmin$ is the minimum SNR required for selection and 
$\dlmax$ is the maximum distance out to which the 
survey can detect such sources.
Assuming further that the relative uncertainty in the
distance reconstruction is related to the 
SNR by $\sigma_D / \dl \simeq 1 / \ston$,
the absolute distance uncertainty is 
(\cf\ \cite{Markovic:1993,Cutler_Flanagan:1994,Seto:2018})
\begin{equation}
\label{equation:sigmad}
\sigma_D \simeq \frac{\dl^2}{\dlmax^2} \, \sigmadmax,
\end{equation}
where $\sigmadmax$ is the distance uncertainty for a source at $\dlmax$.
Combining this with \eq{sigmah} gives
\begin{equation}
\label{equation:sigmahd}
\sigma_{\!H} \simeq
  \hubble \frac{\sigmadmax}{\dlmax}
  \left(
  \frac{\dlss^2}{\dl^2}
  +
  \frac{\dl^2}{\dlmax^2}
  \right)^{1/2}
  ,
\end{equation}
where
$\dlss = (c^2 \, \sigma_z^2 + \sigma_v^2)^{1/2}
/ (\hubble \, \sigmadmax / \dlmax) \simeq 30\ \mpc$
is the distance beyond which redshift/velocity uncertainties 
become unimportant.
This relationship also identifies a distance
of $\sim (\dlss \, \dlmax)^{1/2} 
  \simeq 30 \, (\dlmax / \dlss)^{1/2} \, \mpc$ as that for
which a single source in such a 
survey would yield the tightest constraint on \hubble\
(\cf\ the `sweet spot' distance of
Ref.~\cite{Chen_etal:2018}). 

The number of detected sources will,
out to $\sim \dlmax$,
be proportional to the volume element, which 
for the purposes of this calculation can be 
approximated as 
$\diff V / \diff \dl \simeq 4 \pi \, \dl^2$.
Ignoring the effect of noise on completeness
(which is discussed below in \sect{biases} and \sect{selection}), 
the probability of selection, $S$, is
$\prob(S | \dl, \dlmax) = \step(\dlmax - \dl)$
and so the distance distribution of detected events is then 
\begin{equation}
\prob(\dl | S, \dlmax) \simeq \step(\dl) \, \step(\dlmax - \dl) \,
  \frac{3 \, \dl^2}{\dlmax^3},
\end{equation}
where $\step(\cdot)$ denotes the Heaviside step function.

The optimal estimate from a sample of BNS events 
would be to take an 
inverse variance-weighted average of \hubbleest\ from \eq{hest} 
for each source,
with the weighting proportional to 
$1 / \sigma_{\!H}^2$ from \eq{sigmahd}.
The resultant uncertainty in \hubble\ for a
given sample of $N$ sources 
with individual uncertainties
$\sigma_{\!H,1}, \sigma_{\!H,2}, \ldots, \sigma_{\!H,N}$
would then be 
$\sigma_{\!H} = [(\sum_{i = 1}^N 1 / \sigma_{H,i}^2)]^{-1/2}$.
However, because the number of sources increases
as $\dl^2$ and 
for $\dl \ga (\dl_0 \, \dlmax)^{1/2}$
the inverse variance weighting
decreases as 
$\dl^{-2}$, the uncertainty produced by weighting 
all sources equally is within $\sim 10\%$ of that given by
the optimal scheme, so uniform weighting 
is used here for simplicity.
The expected uncertainty from a sample of $N$
events is hence approximated as the average of the expected variance 
per event,
\begin{align}
\label{equation:sigmahsurvey}
\sigma_{\!H} 
&
\simeq \left[
  \frac{1}{N} 
  \int_0^\infty 
  \diff \dl \,
  \prob(\dl | S, \dlmax) 
  \, 
  \sigma_D^2
  \right]^{1/2}
\nonumber \\
&
  =
  \frac{1}{N^{1/2}}
  \left[
  \int_0^{\dlmax}
  \diff \dl \, 
  \frac{3 \, \dl^2}{\dlmax^3}
  \, 
  \left( \hubble \frac{\sigmadmax}{\dlmax}\right)^2
  \!\!
  \left(
  \frac{\dlss^2}{\dl^2}
  +
  \frac{\dl^2}{\dlmax^2}
  \right)
  \right]^{1/2}
\nonumber \\
  & 
  =
  \frac{1}{N^{1/2}}
  \left(\frac{3}{5}\right)^{1/2}
  \frac{\hubble \, \sigmadmax}{\dlmax}
  \left(
  5 \, \frac{\dlss^2}{\dlmax^2} + 1
  \right)^{1/2}
  ,
\end{align}
where Eqs.~\ref{equation:sigmad} and \ref{equation:sigmahd} have been used.
These uncertainties are shown for different sample sizes 
as the colored bands in \fig{h0_errbias},
from which it is clear that redshift and velocity uncertainties
become irrelevant for 
sufficiently deep surveys with $\dlmax \gg \dlss \simeq 30\ \mpc$.
In this regime \eq{sigmahsurvey}
simplifies to be
\begin{equation}
\label{equation:sigmahsurveynum}
\sigma_{\!H} \simeq 1.2\ \kmsmpc\ 
\left(\frac{N}{50}\right)^{-1/2} \, 
\left(\frac{\sigmadmax / \dlmax}{0.15}\right),
\end{equation}
where the fiducial value chosen for $\sigmadmax / \dlmax$
is, from simulations, found to be approximately $2/\stonmin$.

Equation~\ref{equation:sigmahsurvey} 
(or, if appropriate, \eq{sigmahsurveynum})
represents a target for any analysis of real or simulated 
BNS merger data,
and also provides a guide to the level at which potential
biases could have a significant impact.


\section{Selection bias}
\label{section:biases}

Given that it is likely that GW observations of BNS mergers
will soon produce uncertainties in \hubble\ sufficiently small
to resolve the current Hubble constant tension,
it is important to ensure that the resultant estimates are
not significantly affected by systematic biases.
There can be systematic errors from model mis-specification, 
but the most pernicious potential bias comes from the fact that any 
real BNS merger sample will consist of 
sources selected on the basis of the measured
SNR of the same GW data that is then
used to obtain distance constraints.
This selection cut will preferentially
include cases where the measurement noise has added to the signal,
and which hence are inferred to be closer than they in fact are;
applying any sort of simple
average (\eg, using \eq{hest}) 
to such a BNS merger sample would overestimate \hubble.

The magnitude of this potential bias is somewhat ambiguous, as it is 
only defined in the context of a specific method
for obtaining a distance estimate, $\dlobs$, from the GW data,
but a reasonable approach is to extend the simple survey 
model described in \sect{prediction} by including selection effects.
The error in \hubble\ from a single object at distance $\dl$
for which the data give a distance estimate that is off
by $\Delta\dl = \dlobs - \dl$ is
\begin{align}
\hubbleest - \hubble
  & \simeq - \frac{\Delta\dl}{\dl} \, \hubble
  \nonumber \\ 
  & \simeq - 10\ \kmsmpc \, \frac{\Delta\dl / \dl}{0.15},
\end{align}
where the (cosmological) redshift is assumed to be known perfectly
and the final expression is reasonable for an object that is 
close to the detection threshold.

The implied bias from a sample would then be given by averaging 
$\hubbleest - \hubble$ over 
all distances and possible noise realizations, 
which would require an explicit model for the measurement process.
A more generic approximation is possible by considering the 
distinct behavior in three different distance ranges:
objects with $\dl \la \dlmax - \sigmadmax$
are all well detected,
irrespective of the noise realization, and so dilute any overall bias;
objects 
close to the survey horizon 
with $|\dl - \dlmax| \la \sigmadmax$ are 
detected only if the noise augments the signal, in 
which case their distance is underestimated by 
$|\Delta\dl| \simeq \sigmadmax$;
and objects with $\dl \ga \dlmax + \sigmadmax$
are never detected and so
do not affect the Hubble constant estimate at all.
Taking the distance distribution of detected objects as
\begin{align}
& \prob(\dl | S)
  \propto
\\
&
\left\{
\!\!\!
\begin{array}{lll}
  \dl^2
  & \! {\rm if}\!  & 0 \leq \dl \leq \dlmax - \sigmadmax \\
\\
  (\! \dlmax \! + \! \sigmadmax \! - \! \dl) \, \! (\dlmax \! - \! \sigma)^2 
  \! / (2 \, \sigmadmax)
  & \! {\rm if}\!  & \dlmax \! - \! \sigmadmax \leq \dl \leq \dlmax \! + \! \sigmadmax \\
\\
  0  
  & \! {\rm if}\!  & \dl > \dlmax + \sigmadmax,
\end{array}
\right.
\nonumber
\end{align}
the bias from selection effects would be
\begin{align}
\Delta H_{\rm sel}
  & \simeq \frac{3}{1 + 2 \, \sigmadmax / \dlmax} 
  \left(\frac{\sigmadmax}{\dlmax} \right)^2 \, \hubble
\nonumber \\
  & \simeq 
  3.6\ \kmsmpc \left(\frac{\sigmadmax / \dlmax }{0.15}\right)^2.
\end{align}
The $(\sigmadmax / \dlmax)^2$ scaling
comes about as this ratio determines both the range of 
distances for which selection effects are important and the 
magnitude of the distance underestimate for the fraction of these objects
selected.
This bias is shown as the offset in \fig{h0_errbias}
and, aside from being comparable to the 
current difference between the local and cosmological
\hubble\ measurements (\sect{intro}),
is larger than the predicted uncertainties (\sect{prediction})
from a sample of even $N \simeq 10$ events
if $\dlmax \ga 70\ \mpc$.
While the exact value of the predicted \hubble\
bias is somewhat model-dependent, 
it is clear that it must be taken into account
if BNS mergers are to be useful in resolving the current Hubble tension.


\section{Simulated BNS merger samples}
\label{section:model}

In order to test both the precision and accuracy
of \hubble\ estimates from BNS mergers it 
is necessary to generate simulated samples of events
that, in particular, are subject to appropriate selection effects.
The population model and observations described here 
represent a specific version
of the general structure described in \app{derivation_app}.
The primary focus is on the self-consistency of the model,
with the sample selection being performed on the same measured
quantities which are subsequently used to drive the inference
calculation.
As such, the model is reasonably simple, 
and includes neither realistic detector and noise models 
(\cf\ \cite{Nissanke_etal:2010,Chen_etal:2018,Feeney_etal:2018})
nor full inference of the individual BNS merger parameters 
(\cf\ \cite{Abbott_etal:2017a,Veitch:2014wba,Feeney_etal:2018,Ashton_etal:2019,Biwer_etal:2019}).


\subsection{Cosmology}
\label{section:cosmol}

For the low redshifts of $z \la 0.2$ out to which BNS mergers 
(and NS-BH mergers) are likely to be detected in the next decade
it is sufficient to
adopt the standard Taylor series approximation to the
cosmological expansion history,
in which the dynamics are characterized to
leading order by the deceleration parameter
\decel\ (\eg, \cite{Weinberg:2008})
In a \lcdm\ model $\decel = \Omega_{\rm m} / 2 - \Omega_\Lambda$,
where
$\Omega_{\rm m}$ is the normalized matter density
and $\Omega_\Lambda$ is the normalized cosmological constant.
The {\em Planck} CMB data imply that
$\Omega_{\rm m} \simeq 0.31$ and $\Omega_\Lambda \simeq 0.69$
\cite{Planck_VI:2018}
and hence that $\decel \simeq -0.53$,
although these values would come into question
if the {\em Planck} value of \hubble\ were determined to be incorrect.

In this cosmological model the luminosity distance is
\begin{equation}
\label{equation:dl}
\dl
(z, \hubble, \decel)
  \simeq
  \frac{c \, z}{\hubble}
  \left[
  1 + \frac{1}{2} \, (1 - \decel) \, z
  \right]
\end{equation}
and the co-moving volume element is
\begin{equation}
\label{equation:dvdz}
\frac{\diff V}{\diff z}
(\hubble, \decel)
  \simeq 4 \pi \frac{c^3 \, z^2}{\hubble^3}
  \left[
  1 - 2 \, (1 + \decel) \, z
  \right].
\end{equation}


\subsection{BNS population}
\label{section:bnspop}

The local BNS population is taken to be defined 
by the rate of mergers
per unit proper time 
per unit co-moving volume,
$\rate$,
which has been measured as
$\rate = 1540_{-1220}^{+3200}\ \gpc^{-3}\ \yr^{-1}$ 
\cite{Abbott_etal:2017d}.

An individual BNS merger can,
as discussed further in \app{linear},
effectively be described by just two parameters:
its chirp mass, $\mchirp = (M_1 \, M_2)^{3/5} / (M_1 + M_2)^{1/5}$,
where $M_1$ and $M_2$ are the masses of the two NSs;
and the inclination of the system to the line-of-sight, $\inc$. 
The population prior distribution in these parameters 
is taken to have the redshift-independent form
\begin{equation}
\label{equation:pai}
\prob(\mchirp, \inc | \mchirpmean, \sigmamchirp)
  \propto
  \normal(\mchirp; \mchirpmean, \sigmamchirp^2)
  \, \step(\inc) \, \step(\pi - \inc) \, \frac{\sin(\inc)}{2},
\end{equation}
where\footnote{Here
$\normal(x; \mu, \sigma^2)
  = \exp[- (x - \mu)^2 / (2 \, \sigma^2)]/ [(2 \pi)^{1/2} \, \sigma]$
denotes a normal density of mean $\mu$ and variance $\sigma^2$.}
the sinusoidal distribution in $\inc$ encodes the
assumption that these systems have random (and independent) orientations.

The distribution of host (line-of-sight) peculiar motions 
is taken to be
\begin{equation}
\label{equation:vprior}
\prob(v | \sigmav) = \normal(v; 0, \sigmav^2 ),
\end{equation}
where $\sigmav \simeq 500\ \kms$ 
from observations of local galaxy motions \cite{Burstein:1990}.
The additional motion of the BNS system relative to the host 
is unimportant in this context (\app{linear}).


\subsection{Observations}
\label{section:data}

\begin{figure*}
\includegraphics[width=16cm]{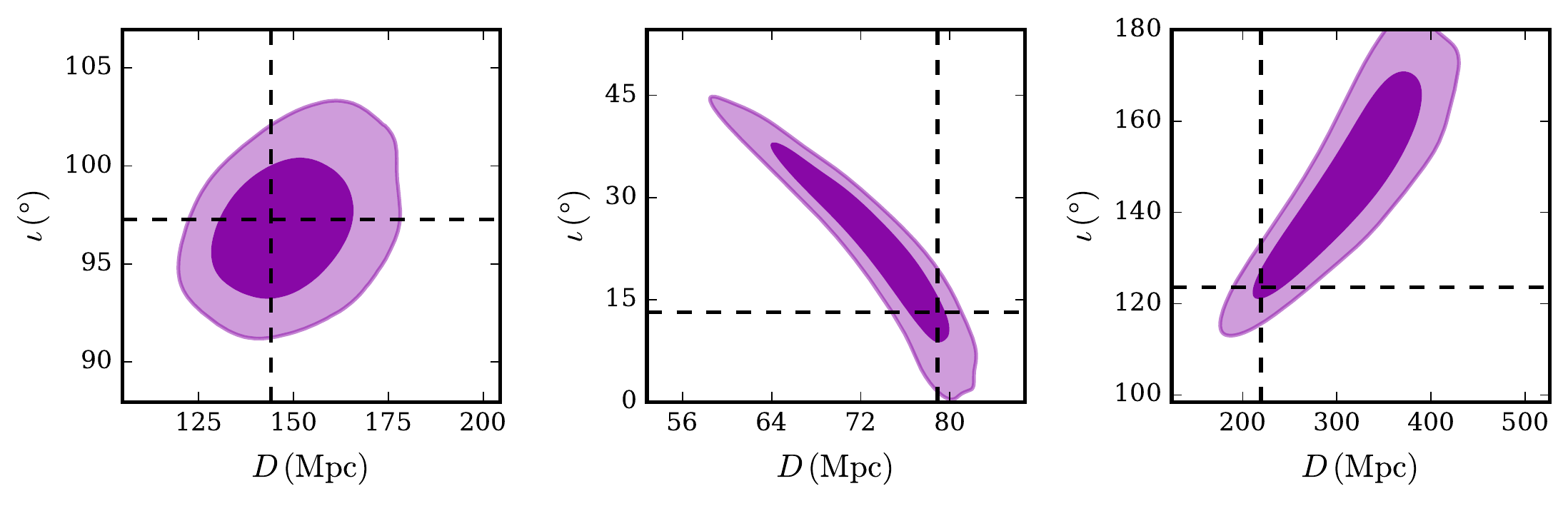}
\caption{Constraints on the distance, $\dl$, and inclination, $\inc$,
for three representative 
BNS mergers
from the GW data only, 
using the simplified model described in \sect{model}.
The purple and mauve 
indicate 68\% and 95\%
highest posterior density credible regions
and the dashed lines show the input parameters.}
\label{figure:dinc_post}
\end{figure*}

\begin{figure}
\includegraphics[width=8cm]{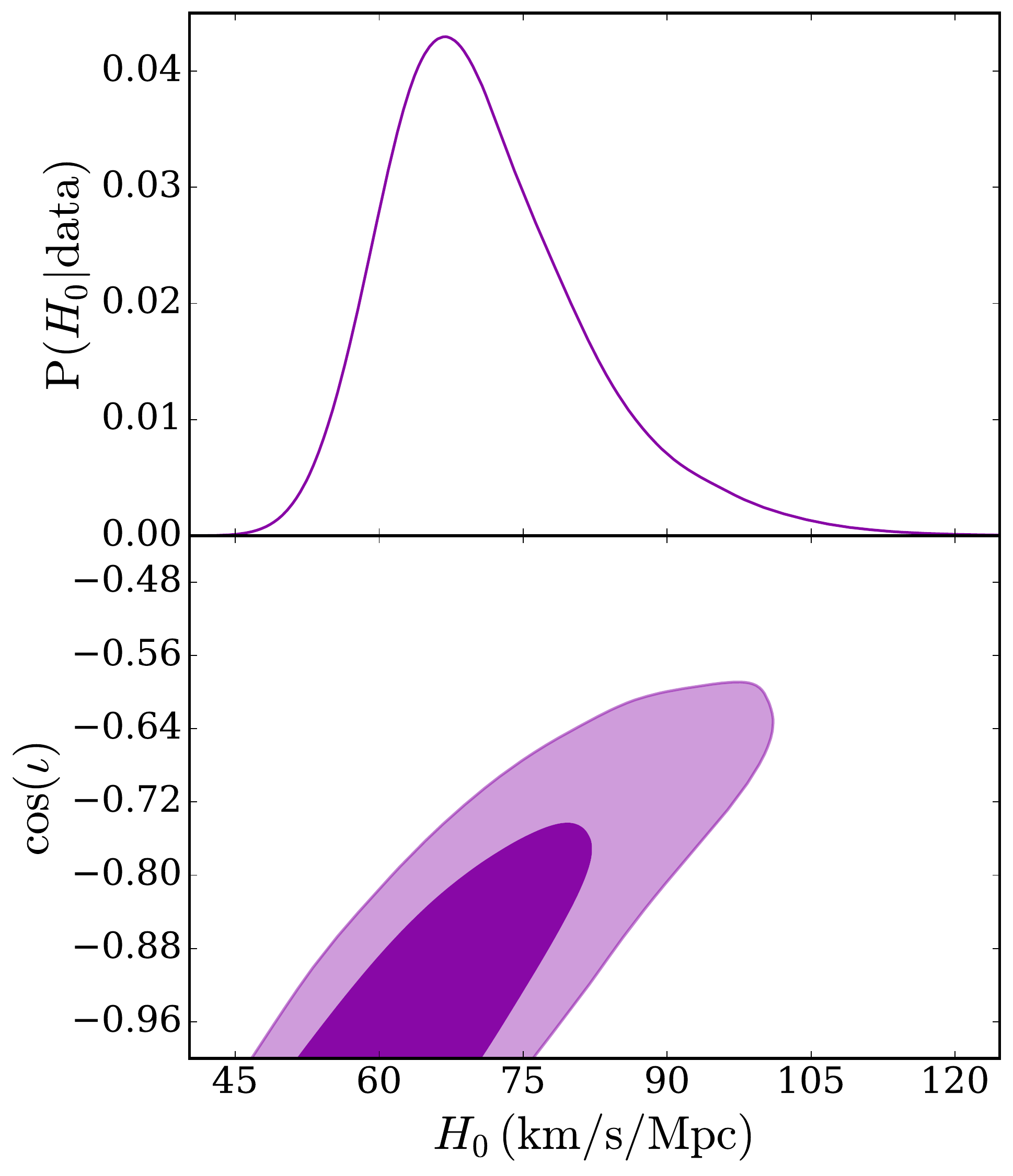}
\caption{
Constraints on the Hubble constant, $\hubble$, and the
cosine of the system inclination, $\cos(\inc)$,
from a BNS merger
designed to mimic GW~170817.
These distributions are hence directly comparable to Figures~1 and 2 of
Ref.~\cite{Abbott_etal:2017a}.
The purple and mauve
indicate 68\% and 95\%
highest posterior density credible regions.}
\label{figure:dinc_post_reconstructed}
\end{figure}

Any BNS merger that has been detected from its GW emission 
is subject to intensive follow-up observations,
the primary aim 
of which is to identify an EM counterpart and hence a host galaxy. 
While there is no guarantee of this process being successful
\cite{Schutz:1986,del_Pozzo:2012,Oguri:2016,Fishbach_etal:2018,Nair:2018,Soares-Santos_etal:2019},
it is reasonable to assume that hosts
will be identified for a significant fraction of 
BNS mergers
\cite{Nissanke_etal:2013b}.
And, as it is these systems which will provide
the best constraints on \hubble\ (\eg, \cite{Chen_etal:2018}),
they are the focus here.
Three distinct types of measurement can provide information about 
a BNS merger with a confirmed host:
the GW data from the BNS merger;
the host galaxy redshift measurement;
and, potentially, an estimate of the (line-of-sight) peculiar velocity
of the host galaxy.
Following the general formalism in \app{data_app},
the associated likelihoods depend on the 
details of the relevant observations and measurements,
which are denoted by $O$ for brevity. 


The GW data from a merger produced by a single detector consists of the 
discretized and noisy linear combination of the two GW
polarizations weighted by the detector response function. 
A full simulation of such data would require 
using a numerical general relativity code 
(\eg, LALSimulation \cite{LALSuite})
to generate such waveforms,
multiplying them by the response function,
and adding noise as appropriate for the instrument.
The resultant time stream(s) could then be analysed 
to obtain parameter constraints using software such as 
such as {\tt LALInference} \cite{Veitch:2014wba},
{\tt BILBY} \cite{Ashton_etal:2019}
or 
{\tt PyCBC} \cite{Biwer_etal:2019}.
Repeating this procedure for large numbers of 
mergers, 
as required here 
to examine the bias of population-level inference approaches (\sect{test}),
would be prohibitive.

Fortunately, it is possible to make use of a more 
streamlined approach
in which the time series data is never actually simulated 
(and hence, even more importantly, 
the full parameter estimation procedure is not required).
In order to assess the inference of the population-level 
parameters, and \hubble\ in particular, 
the only requirement on the model is that 
the dependence of the likelihood
on the model parameters 
is accurately represented.
For a BNS merger, 
almost all the constraining information about
$\mchirp$, $\inc$ and $\dl$
is encoded in the relatively simple inspiral phase,
the data from which can be summarized 
by three 
statistics that can be calculated from the full data
(as described in, \eg, 
Refs.~\cite{Cutler_Flanagan:1994,Nissanke_etal:2010} 
and defined in \app{linear}):
a measured value of the redshifted chirp mass, $\mzobs$,
which is obtained from the time-dependence of the chirp signal;
and the measured 
amplitudes\footnote{
The term `amplitude' is something of a 
misnomer, as these quantities can be positive or negative.}
of the two orthogonal strain components,
$\hat{\amp}_\plus$ and $\hat{\amp}_\times$.

The uncertainties in these three quantities,
$\sigma_{\mz}$, $\sigma_{{\amp}_\plus}$ and $\sigma_{{\amp}_\times}$,
are determined primarily by the properties 
of the observations
(noise level, correlations, time resolution, \etc),
and here are taken to effectively characterize the survey.
The utility of BNS mergers as a means of measuring distances
rests on the fact that the redshifted chirp mass can be 
measured to high precision 
(\eg, \cite{Schutz:1986,Abbott_etal:2017d}),
and the small error in this quantity does not contribute significantly to the
overall uncertainty on the merger distance.
As such, it is a reasonable approximation to ignore 
the uncertainty in $\mz$ completely,
treating the measured redshifted chirp mass as exact.
Both amplitude measurements, however, have 
appreciable uncertainties of up to $\sim\!10\%$ for a
realistic sample of detected mergers 
(\cf\ \sect{simulation}).
The errors in these measured amplitudes 
not only are the dominant contributions to the distance uncertainty,
but are also linked to the event selection (\sect{selection}),
so must be included self-consistently to assess any potential biases.
In general the uncertainties of the amplitudes of the two components 
will differ,
but for the purpose of the bias analysis undertaken here it is 
sufficient to assume they are the same, so 
that $\sigma_{\!\amp}
= \sigma_{\!{\amp}_\plus}$ and $\sigma_{\!{\amp}_\times}$
is the primary quantity used to characterises the GW observations here.

The main result of applying 
the above approximations and simplifying assumptions is,
as detailed in \app{linear}, 
that the 
likelihood for a single BNS merger can be written as 
\begin{align}
\label{equation:dnorm}
&
\prob(\mzobs, \hat{\amp}_\plus, \hat{\amp}_\times 
  | \mchirp, \inc, z, \hubble, \decel, \sigma_{\!\amp})
\nonumber \\
& 
= \delta\left[\mzobs - (1 + z) \, \mchirp \right]
\nonumber \\
& \times
  \normal\!\left[ \hat{\amp}_\plus;
\frac{G \, (1 + z) \, \mchirp / c^2}{\dl(z, \hubble, \decel)} \,
\frac{1 + \cos^2(\inc)}{2},
  \sigma_{\!\amp}^2
  \right]
\nonumber \\
& \times
  \,
  \normal\!\left[ \hat{\amp}_\times;
-\frac{G \, (1 + z) \, \mchirp / c^2}{\dl(z, \hubble, \decel)} \,
\cos(\inc),
  \sigma_{\!\amp}^2
  \right]
,
\end{align}
where\footnote{Here
$\delta(x)$ denotes a Dirac delta function.}
$\dl(z, \hubble, \decel)$ is the luminosity distance 
as defined in \sect{cosmol}.
It is likelihoods of this form that are incorporated into the
inference formalism described in \sect{bayes},
although in the numerical implementation a
small uncertainty in the redshifted chirp mass is included.

Despite the mathematical simplicity of the model
summarized in \eq{dnorm},
it captures all the behavior that is relevant to 
obtaining constraints on the source distance and inclination.
Examples of the resultant posterior distributions
are shown in \fig{dinc_post} 
for fiducial BNS mergers with
$\mchirp \simeq \msun$ and $\dl \simeq 100 \, \mpc$,
which hence have amplitudes 
(\ie, $|\amp_\plus|$ and $|\amp_\times|$) of
up to $\sim 5 \times 10^{-22}$.
The uncertainties of $\sigma_\amp \simeq 10^{-23}$
were chosen to give constraints
consistent with LIGO+Virgo in its current configuration.
The posterior distributions have a range of morphologies
and, most importantly, 
exhibit the strong distance-inclination degeneracy
that is the main source of distance uncertainty
(\cf\ \cite{Abbott_etal:2017a,Hotokezaka_etal:2018,Mooley_etal:2018}).

The reported spectroscopic redshift
of a BNS host, $\zobs$, is taken 
to have a measurement uncertainty $\sigma_z$, 
with the associated likelihood 
\begin{equation}
\label{equation:zlik}
\prob(\zobs | z, v, \sigma_z)
  = \normal\!\left[
  \zobs; 
  z + (1 + z) \frac{v}{c},
  \sigma_z^2
  \right].
\end{equation}
Any BNS merger event with a confirmed host galaxy is likely to be
of considerable scientific interest, implying 
that high-quality spectroscopic data will be obtained,
so a typical uncertainty would be $\sigma_{z} \la 0.001$
(\cf\ the measurement of $\sigma_z = 0.00024$ for 
the host of GW~170807 \cite{Crook_etal:2007}).

It is possible that the (line-of-sight) peculiar velocity
of a BNS host can be estimated from the positions and/or motions
of nearby galaxies, 
yielding an estimate $\vobs$,
with 
uncertainty $\sigma_v$.
The 
associated
likelihood is hence taken to be
\begin{equation}
\label{equation:vlik}
\prob(\vobs | v, \sigma_v) = \normal(\vobs ; v, \sigma_v^2).
\end{equation}
The case in which there is no useful peculiar velocity information
can be encoded by taking
$\sigma_v \rightarrow \infty$,
(and, optionally, $\vobs = 0$),
in which case the uncertainty on the peculiar velocity 
is given by the value of $\sigmav$ assumed from \sect{bnspop}.

Combining the simple GW liklihood (\eq{dnorm})
with the above models of the redshift measurement (\eq{zlik})
and peculiar velocity data (\eq{vlik}),
it is possible to produce constraints on \hubble\ from mock data.
This is illustrated in \fig{dinc_post_reconstructed},
which shows results for
$\hat{\amp}_\plus = 1.24 \times 10^{-21}$,
$\hat{\amp}_\times = 1.14 \times 10^{-21}$,
$\sigma_{\amp} = 4.8 \times 10^{-23}$,
$\zobs = 0.01$, 
$\sigma_z = 0$,
$\vobs = 0$,
and 
$\sigma_v = 200 \, \kms$.
These
values were chosen to mimic the joint $\hubble$ and $\cos(\inc)$ posterior
from GW170817 as presented in Ref.~\cite{Abbott_etal:2017a}.
The essential phenomenology is reproduced correctly, 
with the distance-inclination degeneracy still present but 
less distinct due to the extra uncorrelated uncertainty from 
the lack of knowledge about the (cosmological) redshift.


\subsection{Selection function}
\label{section:selection}

The selection of a BNS merger event into the survey is
assumed to be determined by the GW data alone,
and to take the form of a hard cut on the observed SNR,
$\stonmin$.
For the simple model described in \app{linear},
with measured strain amplitudes
of $\ampobs_\plus$ and $\ampobs_\times$
and uncertainty $\sigma_\amp$,
the observed SNR is
\begin{equation}
\label{equation:snobs}
  \stonobs
  =
  \frac{(\ampobs_{\plus}^2 + \ampobs_{\times}^2)^{1/2}}
  {\sigma_{\amp}}
.
\end{equation}
The selection probability is then
\begin{equation}
\label{equation:hx}
  \prob(S | \ampobs_{\plus}, \ampobs_{\times}, \sigma_A)
  =
  \step\!\left[
  \frac{(\ampobs_{\plus}^2 + \ampobs_{\times}^2)^{1/2}}
  {\sigma_{\amp}}
  - \stonmin \right].
\end{equation}

The implied selection function is 
given in terms of the chirp mass, $\mchirp$, 
inclination $\inc$ and source distance $\dl$ as 
\begin{align}
\label{equation:psel}
&
\prob(S | \mchirp, \inc, z, \hubble, \decel, \sigma_A, \stonmin)
\\
& = \!\!\!
  \int \!\! \diff \ampobs_{\!\plus} \!\!
  \int \!\! \diff  \ampobs_{\!\times}
  \prob(\ampobs_{\!\plus}, \! \ampobs_{\!\times} 
    | \mchirp, \! \inc, \! z, \! \hubble, \decel, \sigma_{\!A})
  \, 
  \prob(S | \ampobs_{\!\plus} \!, \ampobs_{\!\times} \!, \sigma_{\!A})
\nonumber \\
&
  = 
  1 - \frac{1}{\pi} \! \int_0^{\phi_{\rm max}}
  \!\!\!\!\! \diff \phi 
  \left\{
  \exp[- r_{\rm min}^2(\phi) / 2]
  -
  \exp[- r_{\rm max}^2(\phi) / 2]
  \right\}
  ,
\nonumber
\end{align}
where 
$\phi_{\rm max} = \min[\arcsin(\ston / \stonmin), \pi]$,
\begin{align}
r_{\rm min}(\phi) 
& = 
  \max\left\{
  0,
\ston \, 
\cos(\phi) -
\left[
\stonmin^2 - \ston^2 \,\sin^2(\phi)
\right]^{1/2}
  \right\},
\nonumber \\
\\
r_{\rm max}(\phi)
& = 
\ston \,
\cos(\phi) + 
\left[
\stonmin^2 - \ston^2\, \sin^2(\phi)
\right]^{1/2},
\nonumber
\end{align}
and 
\begin{equation}
\ston = 
\frac{G \, (1 + z) \, \mchirp / c^2}{\dl(z, \hubble, \decel)}
\frac{[\cos^4(\inc) + 6 \, \cos^2(\inc) + 1]^{1/2}}{2}
\frac{1}{\sigma_A}
\end{equation}
is, from \eq{dnorm}, the mean SNR
(\cf\ \cite{Cutler_Flanagan:1994}).
The combination of additive noise 
and the inverse distance dependence 
of the signal strength
effectively defines a maximum distance out 
to which a survey can detect sources.
For the simple model defined by \eq{dnorm},
and assuming redshifts of $z \ll 1$,
this is given by 
\begin{equation}
\label{equation:dmax}
\dlmax \simeq \frac{G \, \mchirp}{c^2\, \stonmin \, \sigma_A},
\end{equation}
but in many cases (\eg, \sect{prediction}) it is most useful
to think of \dlmax\ 
itself
as a characteristic of a survey.

In reality, the selection of a sample of BNS merger events with 
confirmed hosts also depends on whether it is possible
to confirm a host galaxy from EM follow-up observations
(\eg, \cite{Fishbach_etal:2018}).
For the moment all GW events are
going to be the subject of intense follow-up observing campaigns,
and so it is assumed here both that every BNS 
merger
that occurs 
in a galaxy will have its host identified and, further, that
a spectroscopic redshift measurement will subsequently be made.
While some fraction of events might be host-less 
(\eg\ \cite{Berger:2010,Fong_Berger:2013}) 
this will just reduce the useful sample size produced by any given
survey;
the inference of \hubble\ should not be affected.


\subsection{Expected number of events} 
\label{section:number}

Assuming BNS mergers are independent of each other,
and the parent population of BNS systems is large,
it is reasonable to model the sample size, $N$, as a 
draw from a Poisson distribution.
This is characterized purely by the expected number of events,
which
(as described further in \app{number_app})
is given by integrating over the BNS parameters and the
observing period $T$ to obtain
\begin{align}
\label{equation:nexp}
& \nexp(
\rate, \mchirpmean, \sigmamchirp,
\hubble, \decel, 
\sigma_A, \stonmin, T) 
\\
& 
  =
  T  
  \int_0^\infty \diff z \, 
  \frac{\rate}{1 + z}
  \frac{\diff \vol}{\diff z}(\hubble, \decel)
\nonumber \\
& 
  \int_0^\infty \!\! \diff \mchirp \!
  \int_0^\pi \!\! \diff \inc \,
  \prob(\mchirp, \inc | \mchirpmean, \sigmamchirp) \,
  \prob(S | \mchirp, \inc, z, \hubble, \decel, \sigma_A, \stonmin),
\nonumber
\end{align}
where 
$\prob(\mchirp, \inc | \mchirpmean, \sigmamchirp)$
is given in \eq{pai}
and 
$\prob(S | \mchirp, \inc, z, \hubble, \decel, \sigma_A, \stonmin)$
is given in \eq{psel}.
In general, these integrals must be evaluated numerically, 
\eg, using a Monte Carlo approach
such as that described below in \sect{simulation}.

Particular care needs to be taken here as
the Gaussian GW noise model defined in \sect{data}
results in a non-zero selection probability even for
sources at an infinite distance with a true SNR of $\ston = 0$.
Given that the volume element as specified in \sect{bnspop}
potentially increases to infinite redshifts,
the combination is an unnormalizable distribution 
with an infinite number of expected sources.
This is not a problem in practice as the integration
in \eq{nexp} can be truncated at 
a finite redshift
beyond the detection horizon;
this model is hence best understood as a numerical approximation.

Another important aspect of \eq{nexp} is 
that $\nexp$ is nearly independent of \hubble,
and can hence potentially be ignored in the data
analysis step.
For low redshifts
the comoving volume element scales as $\hubble^{-3}$,
but the effective maximum redshift of the survey is 
$\zmax \simeq \hubble \, \dlmax / c \propto \hubble$,
so the implied volume in redshift space scales as 
$\hubble^3$;
the two effects cancel out to leading order
(\cf\ \cite{Abbott_etal:2017a,Chen_etal:2018}).
This result can also be understood from a purely physical point of
view: the detectability of a GW event is not significantly affected
by its recession velocity; only its (luminosity) distance is 
important, and so changing the radial integration variable
to $\dl$ would (largely) remove \hubble\ from the calculation
of $\nexp$.


\subsection{Simulation algorithm}
\label{section:simulation}

\begin{figure*}
\includegraphics[width=15.5cm]{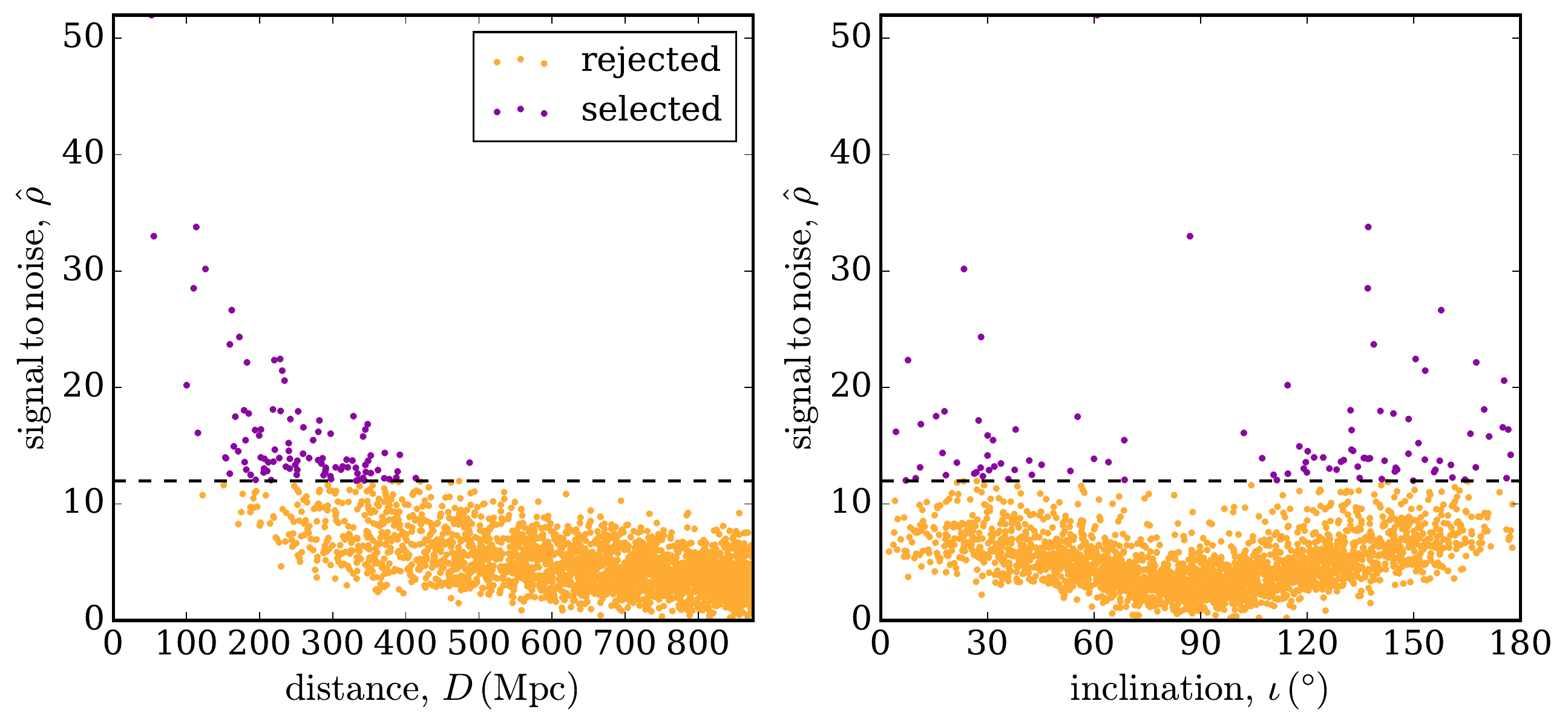}
\caption{Distributions of the measured SNR, $\stonobs$, and
distance (left) and inclination (right) for a simulated sample of 100 
selected BNS merger events (purple) with $\ston \geq \stonmin = 12$.
The much larger number of mergers which were not selected 
are also shown in orange.}
\label{figure:dl_inc_selection}
\end{figure*}

\begin{figure}[h]
\includegraphics[width=8.0cm]{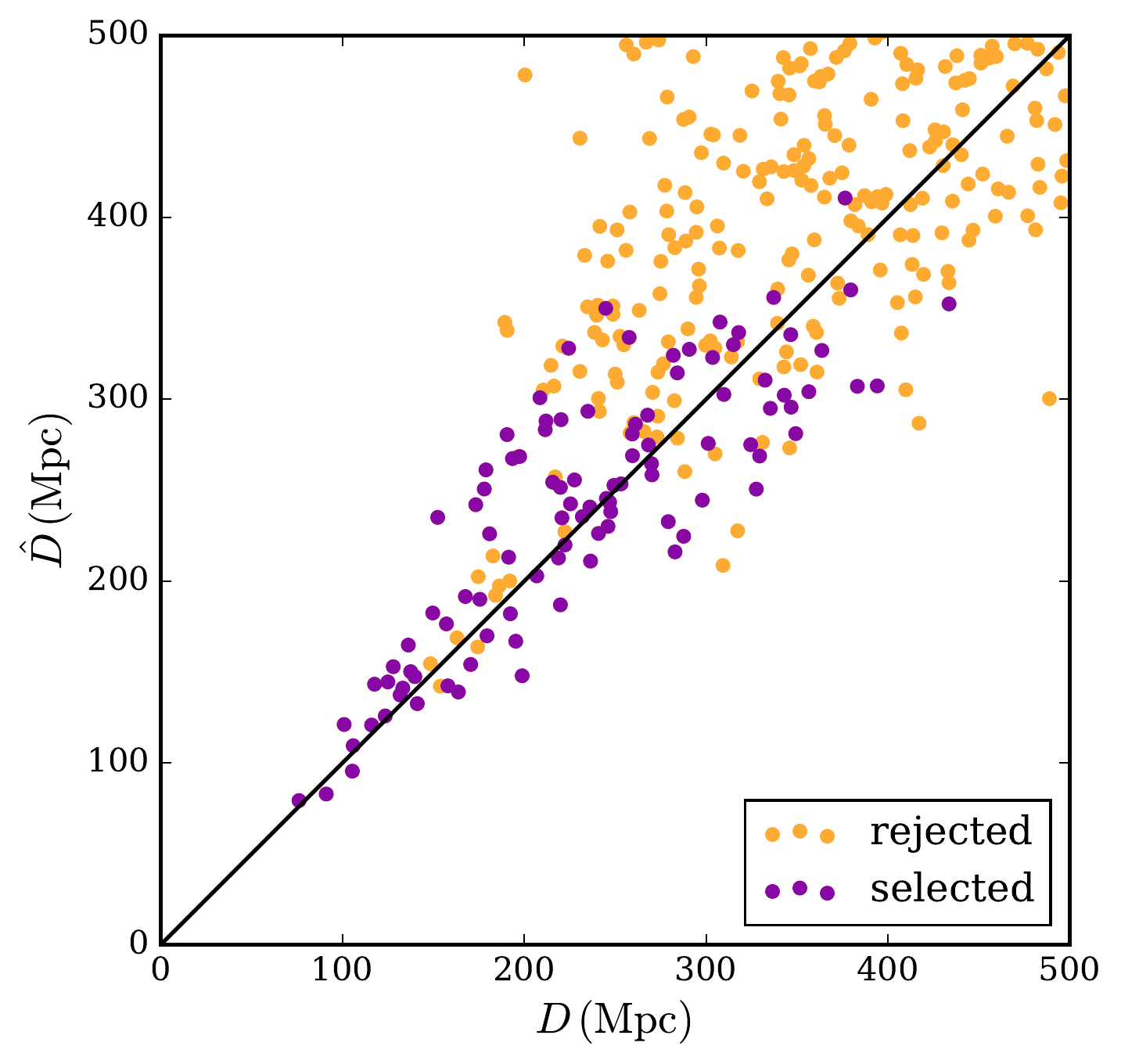}
\caption{Comparison of the fiducial best-fit distance, $\dlobs$,
and
the true distance, $\dl$, for both
selected (purple) and rejected (orange) sources.
The over-abundance of sources close to the maximum survey
distance with $\dlobs < \dl$ would potentially
give a biased estimate of \hubble, as described in \sect{biases}.}
\label{figure:dl_selection}
\end{figure}

The algorithm used here to generate a
self-consistently selected
sample of low-redshift BNS merger
events (as used in \sect{test}) is:

\begin{enumerate}

\item
Choose values of the cosmological model parameters,
\hubble\ and \decel\ (\sect{cosmol}),
and BNS population properties,
$\rate$, $\mchirpmean$ and $\sigmamchirp$ (\sect{bnspop}),
along with the observational characteristics of the 
GW and EM observations, $T$, $\sigma_A$, $\stonmin$,
$\sigma_z$ and $\sigma_v$ (\sect{data}).

\item
Identify a maximum 
redshift (or, equivalently, distance) 
such that there is a negligible 
probability of any merger beyond this
being detected in the survey.
If a fully realistic population model were used 
this redshift could be arbitrarily large, 
as evolution would ensure a finite sample size;
but, for the simple non-evolving model used here,
care needs to taken to avoid including the 
spurious secondary peak discussed in \sect{number}.
Less fundamentally, using a more carefully chosen maximum redshift 
is also important for computational efficiency, to avoid 
the need for simulating large numbers of undetectable sources.
The approach used for setting $z_{\rm max}$ here is 
to calculate the distance at which a merger
with a chirp mass 3 standard deviations above the population mean 
would require a $3 \, \sigma_A$ positive noise deviation to be
detected.  
Further adopting a linear redshift-distance relationship, this gives
\begin{equation}
z_{\rm max} 
  =
  \frac{G \, \hubble}{c^3}
  \frac{\mchirpmean + 3 \, \sigmamchirp} {(\stonmin - 3)\, \sigma_A} .
\end{equation}
In the 
simulations shown in \fig{dl_inc_selection} 
$z_{\rm max}$ corresponds to 
a maximum distance of $D_{\rm max} \simeq 900 \, \mpc$,
which is clearly more than sufficient to include all detectable sources.

\item
Calculate the expected number of 
BNS
mergers
with $z \leq z_{\rm max}$ during the observing period $T$
as
\begin{align}
\nexp_{\rm max} & = T \int_0^{z_{\rm max}}
  \diff z \, \frac{\rate}{1 + z} \, 
  \frac{\diff \vol}{\diff z}(\hubble, \decel)
  .
\end{align}

\item
Draw the actual number of mergers in this volume,
$N_{\rm max}$, from 
a Poisson distribution of mean $\nexp_{\rm max}$.

\item
For each of these $N_{\rm max}$ BNS mergers:

\begin{enumerate}

\item
Draw a redshift from the distribution
\begin{align}
\mbox{} \hspace*{1.2cm} 
& \prob(z | z_{\rm max}, \hubble, \decel) 
  \nonumber \\ 
&
  \propto
  \step(z) \, \step(z_{\rm max} - z) \, 
  \frac{1}{1 + z} \, \frac{\diff \vol}{\diff z}(\hubble, \decel),
\end{align}
and both a chirp mass and inclination from 
$\prob(\mchirp, \inc | \mchirpmean, \sigmamchirp)$ as given in \eq{pai}.

\item
Draw measured amplitudes 
$\hat{\amp}_\plus$
and 
$\hat{\amp}_\times$
(and, optionally, a measured chirp mass, $\mzobs$)
from \eq{dnorm},
and hence calculate the observed SNR, 
$\stonobs = \ston(\hat{\amp}_\plus, \hat{\amp}_\times, \sigma_A)$,
according to \eq{snobs}.

\item
Select the merger into the sample of detected events 
if $\stonobs \geq \stonmin$.

\item
If the merger is selected then 
draw a host peculiar velocity $v$ from \eq{vprior},
a measured redshift,
$\zobs$, from \eq{zlik}
and, optionally, a measured peculiar velocity,
$\vobs$, from \eq{vlik}.

\end{enumerate}

\end{enumerate}

The output of the above
algorithm is a sample of $N$ detected BNS mergers
along with their 
redshifted chirp masses,
$\mzobss = (\hat{\mchirp}_{z,1}, 
\hat{\mchirp}_{z,2}, \ldots,
\hat{\mchirp}_{z,N})$,
measured GW amplitudes,
$\ampobss = (\ampobs_{\plus,1}, \ampobs_{\times,1},
  \ampobs_{\plus,2}, \ampobs_{\times,2}, \ldots,
  \ampobs_{\plus,N}, \ampobs_{\times,N})$,
their hosts' spectroscopic redshifts,
$\zobss = (\zobs_1, \zobs_2, \ldots, \zobs_N)$,
and (possibly) their estimated peculiar velocities,
$\vobss = (\vobs_1, \vobs_2, \ldots, \vobs_N)$,
along with the associated measurement uncertainties.
The algorithm also produces the 
true redshifts, $\zs = (z_1, z_2, \ldots, z_N)$,
chirp masses, $\mchirps = (\mchirp_1, \mchirp_2, \ldots, \mchirp_N)$,
inclinations, $\incs = (\inc_1, \inc_2, \ldots, \incs_N)$,
and host peculiar velocities, $\vs = (v_1, v_2, \ldots, v_N)$,
for these selected mergers, although these quantities 
are not actually part of a simulated catalog,
and cannot be used in the parameter inference described in \sect{bayes}.

One important feature of the above approach
is that,
while $N$ is a draw from the appropriate
Poisson distribution of mean $\nexp$,
the integral in \eq{nexp} is never actually evaluated
explicitly.
This is not particularly important for the fairly simple 
model used here, but is potentially critical in the 
more general cosmological case described in \app{model_app}.
An example of a sample generated in this way is shown in 
\fig{dl_inc_selection}, in particular illustrating 
that the most distant detected sources
in the sample are those for which the noise has 
conspired to increase the SNR.
These are also typically 
face-on or face-off sources, despite these configurations 
having a low prior probability (\eq{pai}).
If such data were treated naively, 
it would produce systematic underestimates
for the distances to mergers close to the maximum survey distance.
For the sample shown in \fig{dl_selection}
this would induce a bias of $\sim 2\ \kmsmpc$
in the inferred value of \hubble\ (\cf\ \sect{biases}).


\section{Bayesian parameter inference}
\label{section:bayes}

\begin{figure*}
\includegraphics[width=16cm]{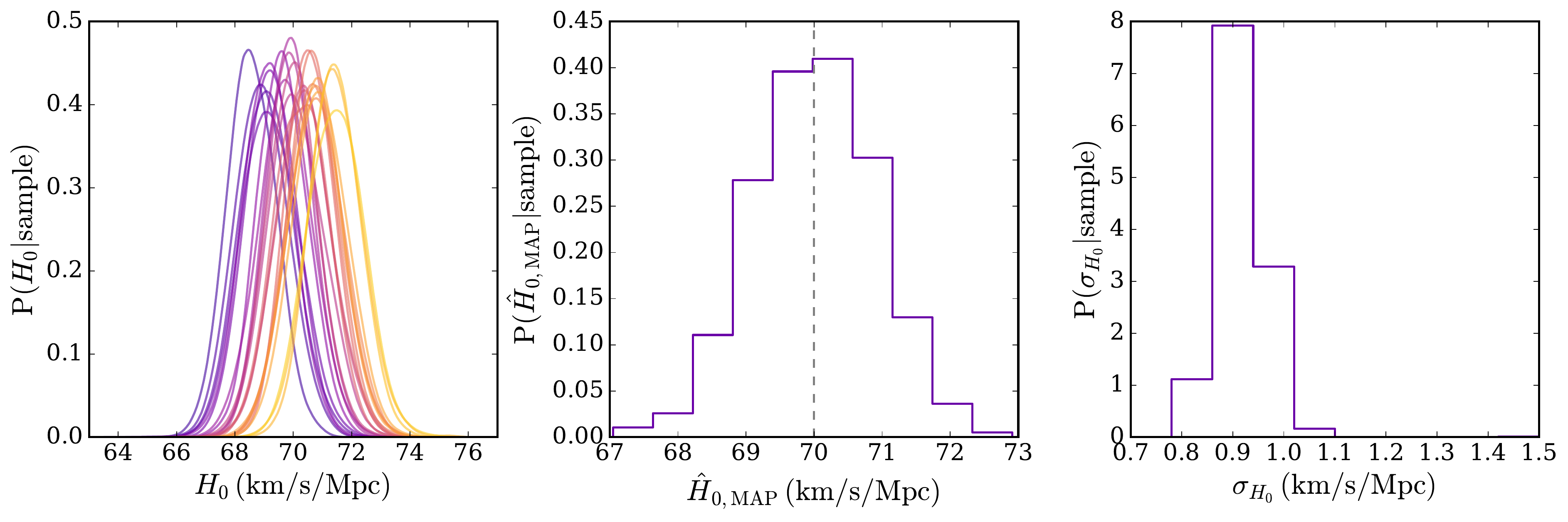}
\caption{Left: \hubble\ posteriors for 25 independent 100-BNS samples,
generated assuming a linear Hubble relation. The posteriors are colored dark purple to light orange by their MAP \hubble\ values. Center/right: distribution of MAP \hubble\ values and posterior standard deviations for the full set of 1000 100-BNS samples.}
\label{figure:lin_posts}
\end{figure*}

\begin{figure*}
\includegraphics[width=16cm]{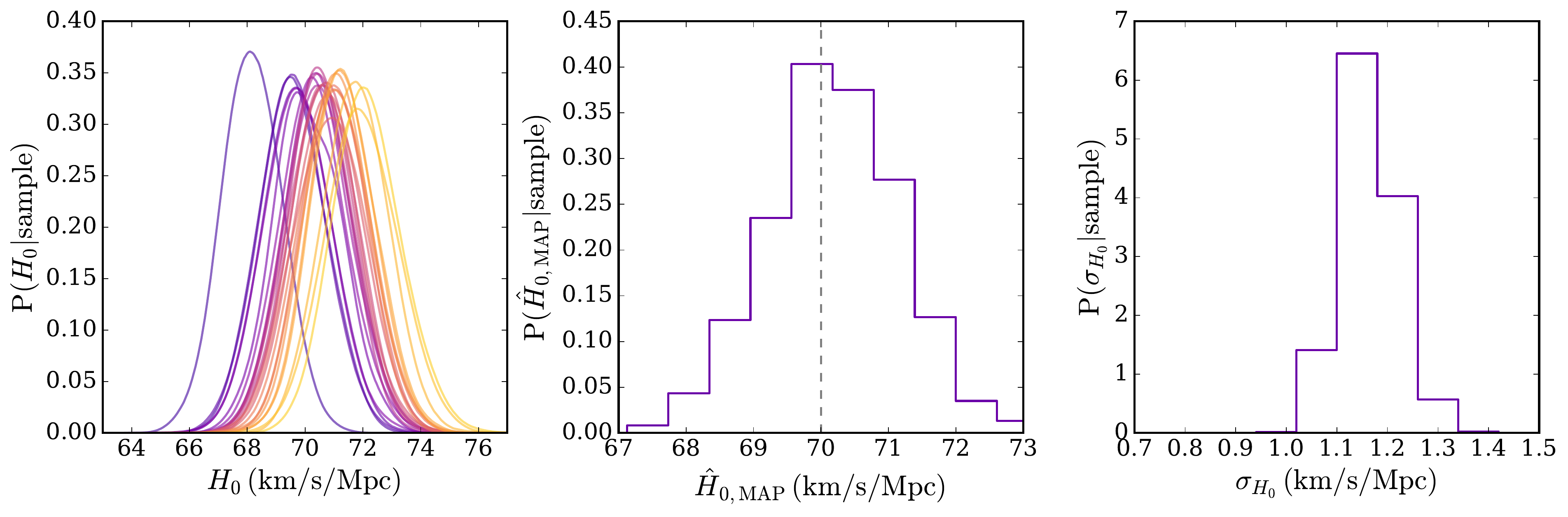}
\caption{As \fig{lin_posts}, but generated assuming a quadratic Hubble relation. From left to right: example \hubble\ posteriors, colored from low to high MAP value; distributions of MAP \hubble\ values and posterior standard deviations for 1000 100-BNS samples.
}
\label{figure:ntlo_posts}
\end{figure*}

\begin{figure}
\includegraphics[width=8.0cm]{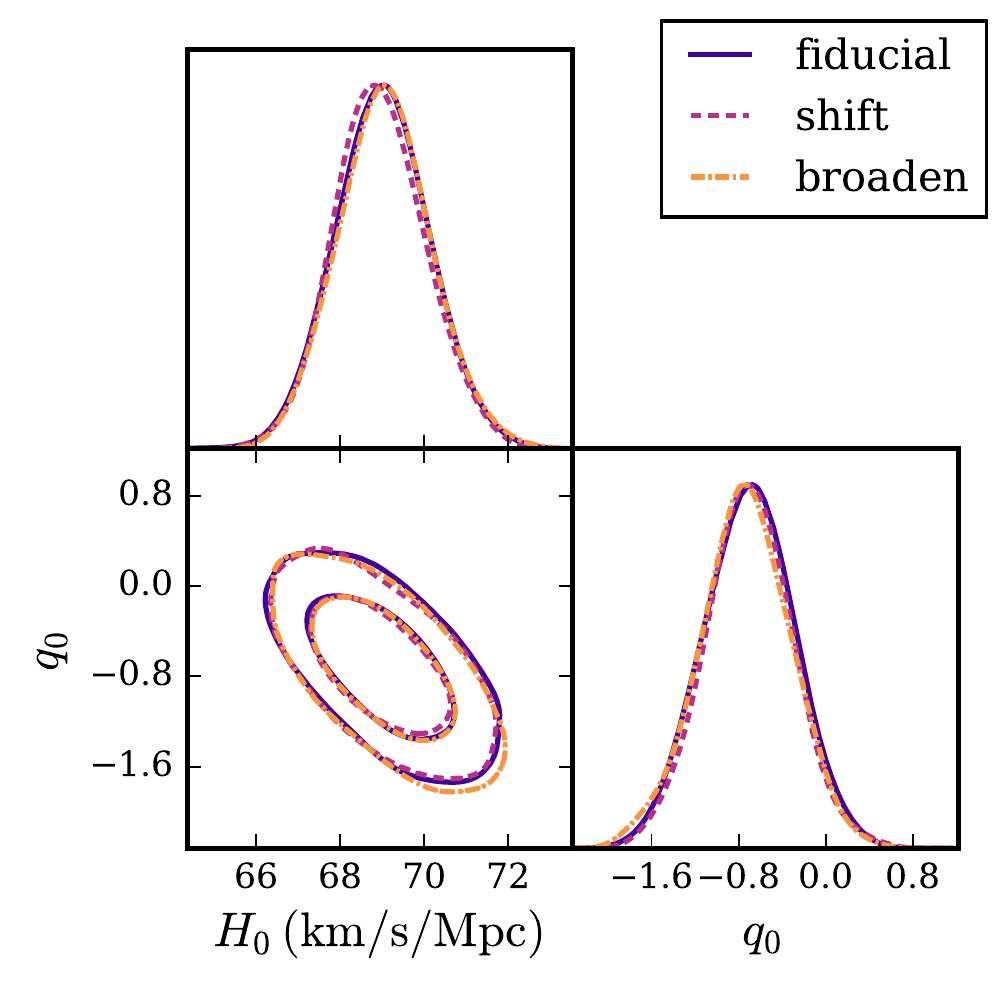}
\caption{Constraints on \hubble\ and $q_0$ for a single 100-event sample, inferred using our fiducial $\mchirp$ prior (purple solid); a prior shifted upwards by one sigma (pink dashed); and a prior with four times the variance (orange dot-dashed). The inference is robust to reasonable variations in the chirp mass prior.
}
\label{figure:m_c_prior_variation}
\end{figure}

A sample of 
BNS mergers
with measured GW data,
spectroscopic host redshifts
and (possibly) estimated peculiar velocities
places constraints 
on the cosmological model,
the BNS population and the properties 
of the mergers.
This is encoded in the 
joint posterior distribution of all the parameters.
For the population and data model described above in \sect{model}
this has the form
$\prob(
\mchirps, \!
\incs, \!
\zs, \!
\vs,\!
\rate\!, \! 
\mchirpmean, \! 
\sigmamchirp, \!
\hubble, \! \decel
|
N\!, \!
\mzobss, \!
\ampobss,  \!
\zobss, \!
\vobss, \!
\sigma_{\!\amp}, \!
\sigma_{\! z}, \!
\sigma_{\! v}, \!
\stonmin, \!
T\!, \!
I
)$,
where $I$ represents the prior information assumed about
the cosmological and BNS population parameters.
A derivation from first principles of
the general form of this posterior distribution,
valid for arbitrary cosmologies and BNS population models,
is given in \sect{bayes_app}.
This has the same structure and dependencies 
as the simpler model described in \sect{model},
so the above posterior can be
obtained from \eq{fullpost} 
by making the identifications
$\gwpars \rightarrow \mchirps$,
$\gwpoppar \rightarrow (\rate, \mchirpmean, \sigmamchirp)$,
$\csmpar \rightarrow (\hubble, \decel)$
and 
$O \rightarrow (\sigma_A, \sigma_z, \sigma_v, \stonmin, T)$.
This gives 
\begin{widetext}
\begin{align}
\label{equation:fullpostsimple}
& \prob(
\mchirps,
\incs,
\zs,
\vs,
\rate, \mchirpmean, \sigmamchirp,
\hubble, \decel
|
N,
\mzobss,
\ampobss,
\zobss,
\vobss,
\sigma_\amp,
\sigma_z,
\sigma_v,
\stonmin,
T,
I
)
\\ 
&
\propto
  \prob(\rate, \mchirpmean, \sigmamchirp | I) \, 
  \prob(\hubble, \decel | I) 
  \exp[- \nexp(\rate, \mchirpmean, \sigmamchirp,
  \hubble, \decel, \sigma_\amp, \stonmin, T)] 
  \nonumber \\
& \times
  \prod_{i = 1}^N
  \frac{\rate}{1 + z_i} \,
  \frac{\diff V}{\diff z}(\hubble, \decel) \,
  \prob(\mchirp_i, \inc_i ; \mchirpmean, \sigmamchirp) \,
  \prob(\hat{\mchirp}_{z,i}, \ampobs_{\plus,i}, \ampobs_{\times,i} 
    | \mchirp_i, \inc_i, z_i, \hubble, \decel, \sigma_\amp) \,
  \prob(v_i | \sigmav) \,
  \prob(\zobs_i | z_i, v_i, \sigma_z) \,
  \prob(\vobs_i | v_i, \sigma_v),
\nonumber
\end{align}
\end{widetext}
where the expected number of detected mergers
$\nexp(\rate, \mchirpmean, \sigmamchirp,
  \hubble, \decel, \sigma_\amp, \stonmin, T)$ is defined in \eq{nexp},
the volume element $\diff V/\diff z(\hubble, \decel)$ is given in \eq{dvdz},
the BNS demographic 
$\prob(\mchirp, \inc ; \mchirpmean, \sigmamchirp)$ is given in \eq{pai},
the GW likelihood 
$\prob(\mzobs, \ampobs_\plus, \ampobs_\times
  | \mchirp, \inc, z, \hubble, \decel, \sigma_\amp)$
is given in \eq{dnorm},
the peculiar velocity prior
$\prob(v | \sigmav)$ is given in \eq{vprior},
the redshift likelihood 
$\prob(\zobs | z, v, \sigma_z)$
is given in \eq{zlik},
and the 
peculiar velocity likelihood 
$\prob(\vobs | v, \sigma_v)$ is given in \eq{vlik}.

The prior in the cosmological parameters is taken to be
\begin{align}
& \prob(\hubble,\decel|I) \\
& \propto \normal(\hubble;\hat{H}_0, \sigma_{\!H}^2)
\, 
\step(\decel + 2)\, \step(1 - \decel) \, \normal(\decel;-0.5,0.5^2),
\nonumber
\end{align}
where the \hubble\ prior is centred on 
$\hat{H}_0 = 70\ \kmsmpc$
and has a width of
$\sigma_{\!H} = 20\ \kmsmpc$.
The truncation in \decel\ avoids the region of parameter 
space ($q_0 > 1/2\,z_{\rm max}-1 \simeq 1.1$) 
in which $\diff V/ \diff z$ drops to zero inside the simulation 
volume described in \sect{simulation}.
For samples of more than a few BNS mergers
these priors have a minimal effect on the inference of \hubble\
(\sect{test}).

The BNS population parameters
$\mchirpmean$, $\sigmamchirp$ and $\rate$
could be fit along with the cosmological parameters,
but doing so is prohibitive numerically (as explained below),
given the large number of simulations needed for the tests
described in \sect{test}.
The prior in the population parameters is hence taken to be
\begin{equation}
\prob(\rate, \mchirpmean, \sigmamchirp)
= \delta(\rate - \hat{\rate}) \,
  \delta(\mchirpmean - \hat{\bar{\mchirp}}) \,
  \delta(\sigmamchirp - \hat{\sigma}_\mchirp),
\end{equation}
where the assumed values are
$\hat{\rate}_0 = 1540\ \gpc^{-3}\ \yr^{-1}$,
$\hat{\bar{\mchirp}} = 1.2\ \msun$
and
$\hat{\sigma}_\mchirp = 0.12\ \msun$.

Evaluating the posterior (\eq{fullpostsimple}) 
requires calculating the expected number of detectable events, 
$\nexp$, for different values of the unspecified model parameters: 
in this case, \hubble\ and, if relevant, \decel. 
Rather than evaluate \eq{nexp} at every location in parameter space,
it is more efficient to estimate this integral using a Monte Carlo approach
for a grid of models and then fit a smooth function to capture its
parameter dependence.
A 25-point grid of \hubble\ (and \decel) values and 
a fourth-order polynomial were used, although the results
are insensitive to these specific choices.

With this fit in hand, 
the 402-parameter\footnote{The parameters are \hubble, \decel\ and 
a true redshift, peculiar velocity, chirp mass and inclination for
each of the 100 BNS mergers.} 
joint posterior for a given catalog was explored 
using Hamiltonian Monte Carlo~\cite{Neal:2012} 
as implemented in {\tt Stan}~\cite{Carpenter_etal:2017,pystan}.  
This yields $\sim700$ independent samples from each 
marginal \hubble\ posterior (from 2000 total samples) 
in 25 seconds on four Intel Xeon (2.4 GHz) CPUs,
sufficient to analyze large numbers of simulations. The Stan model code 
and Python driver used in this analysis are publicly available at 
\url{https://github.com/sfeeney/hh0}.


\section{Results}
\label{section:test}

\begin{table}
  \centering
  \begin{tabular}{ccc}\hline
       order & posterior bias & posterior width \\ \hline
       linear & $0.01 \pm 0.89$ & $0.92 \pm 0.30$ \\
       quadratic & $0.20 \pm 0.97$ & $1.17 \pm 0.36$ \\
  \hline
  \label{tbl:simple_sim_stats}
  \end{tabular}
  \caption{Average bias and width of \hubble\ posteriors calculated
    from our 1000 simulations of 100-event samples.}
\end{table}

The inference formalism described in \sect{bayes} can be tested on 
simulated BNS merger samples generated self-consistently
as described in \sect{simulation}.
The datasets were generated with these parameter values:
$\hubble = 70\ \kmsmpc$;
$\decel = -0.5$ or a linear Hubble relationship;
$\rate = 1540\ \gpc^{-3}\ \yr^{-1}$;
$\mchirpmean = 1.2\ \msun$;
and
$\sigmamchirp = 0.12\ \msun$.
The observations are defined by:
$\sigma_\amp = 2 \times 10^{-23}$;
$\stonmin = 12$;
$\sigmav = 500\ \kms$;
$\sigma_v = 200\ \kms$;
and 
$\sigma_z = 0.001$.
This set-up gives a maximum survey distance of
$\dlmax \simeq 250\ \mpc$
and typical
Hubble-flow velocity uncertainties of 
$\sim 360\ \kms$.

1000 independent merger catalogs of $N = 100$
selected events were generated for each of two cosmological scenarios:
a linear Hubble relationship described by \hubble\ alone;
and the quadratic relationship described by \hubble\ and \decel,
as set out in \sect{cosmol}.
The resultant posterior distributions in \hubble\ are shown for 
25 independent simulated samples of $N = 100$ mergers in the left 
panels of \fig{lin_posts} (for the linear Hubble relation) 
and \fig{ntlo_posts} (for the quadratic Hubble relation). 
Samples of this size are clearly approaching the asymptotic Gaussian 
regime,
with the quadratic Hubble relation posteriors slightly broader 
(and more skewed) than in the linear setting due to the degeneracy 
between \hubble\ and $q_0$. The right panels of 
Figs~\ref{figure:lin_posts} and~\ref{figure:ntlo_posts} show 
the distributions of posterior widths for the full sets of 1000 catalogs. 
These distributions, whose characteristics are summarized in Table~I, 
show that posteriors formed from samples of 100 BNS mergers 
have similar uncertainties (with $\sim33$\% variations in width), 
indicating that there is minimal inter-sample variance.
However, the peaks of the \hubble\ posterior from individual 
samples can be scattered significantly high or low
(as was the case for the single sample used in Ref.~\cite{Feeney_etal:2018}).

The distribution of maximum-posterior ($\hubblemap$) values 
from samples of $N = 100$ events are shown in the center panels of 
Figs~\ref{figure:lin_posts} and~\ref{figure:ntlo_posts}. 
For both Hubble relations, 
the distribution of $\hubblemap$ values is centered on the true 
underlying value, indicating that the method is unbiased in the 
presence of selection effects. 
The average bias for the two Hubble relations, 
defined as the difference between $\hubblemap$ and \hubble, 
is listed in Table~I
and is within a standard deviation of zero in both cases. 

The robustness of this approach,
and particularly the sensitivity of
the \hubble\ constraints to mis-specification
of $\mchirpmean$ and $\sigmamchirp$,
can be assessed by conditioning the inference on
values for these parameters that are different from
those used in the simulations. 
Figure~\ref{figure:m_c_prior_variation} shows posteriors on 
\hubble\ and \decel\ inferred using three different 
BNS population priors for 
a single 100-event sample with chirp masses generated from 
the fiducial model (\eq{pai}). 
The joint posteriors obtained 
for 
an offset
prior with 
$\hat{\bar{\mchirp}} = \mchirpmean + \sigmamchirp$ 
and 
$\hat{\sigma}_\mchirp = \sigmamchirp$
are shown by the pink dashed curves;
results for a
broadened
prior with 
$\hat{\bar{\mchirp}} = \mchirpmean$
and 
$\hat{\sigma}_\mchirp = 4 \, \sigmamchirp$ 
are shown by the orange dot-dashed curves.
The constraints on the cosmological parameters in both 
cases differ insignificantly 
from those obtained using the correct
model ($\hat{\bar{\mchirp}} = \mchirpmean$
and $\hat{\sigma}_\mchirp = \sigmamchirp$),
as shown by the solid purple curves.


\section{Conclusions}
\label{section:conclusions}

Using a rigorous Bayesian procedure 
to analyze realistic samples of BNS mergers 
gives estimates of the Hubble constant that are unbiased
in the presence of selection effects,
and robust to mis-specification of the other cosmological
parameters or the BNS population model.
The resultant uncertainties match expectations, 
confirming that this method is both accurate and precise. 
Applying such methods to the 
sample of $\sim50$ BNS mergers that could 
be produced by LIGO+Virgo in the next $\sim 5$ years 
should yield a robust \hubble\ estimate 
accurate to less than $2\ \kmsmpc$.
This would be sufficient to resolve the
current tension between local and cosmological measurements of 
\hubble. 

While the formalism derived and implemented here is very general
and applicable in an arbitrary cosmological setting,
the primary focus of the simulations was on self-consistency, 
in particular with the same selection rule(s) 
used both in the sample generation and the posterior calculations.
In order to take such an approach for sufficiently large numbers of 
simulations, it was necessary to
adopt a simplified model of the GW waveform and fix some 
population-level parameters.
One obvious extension of this work would 
be to use full numerical simulations of BNS and NS-BH 
inspirals and their resultant time series, using a simplified GW waveform
and likelihood analysis such as {\tt BAYESTAR}~\cite{Singer_Price:2016} or,
more ambitiously,
full inference using
{\tt LALInference} \cite{Veitch:2014wba},
{\tt BILBY} \cite{Ashton_etal:2019}
or
{\tt PyCBC} \cite{Biwer_etal:2019}.
A single simulated sample of 51 BNS mergers was analyzed 
using {\tt LALInference} in Ref.~\cite{Feeney_etal:2018},
suggesting that even with realistic individual object
constraints the regime of asymptotic normality has been
reached, but this remains to be demonstrated fully.

While the near-complete decoupling of the BNS mass distribution
from the cosmological parameters due to the tight chirp
mass constraints is central to this entire method of 
measuring the Hubble constant \cite{Schutz:1986}, 
it would also be preferable to extend the Bayesian inference
formalism to include the population parameters as well,
something which will be more important for 
NS-BH merger samples for which there 
is less prior knowledge.
A related extension would be to include a more realistic model of
the EM counterpart detection process,
in particular the correlation between GW selection
and counterpart identification due to viewing angle 
dependencies.
This will become particularly important in
the context of GW-based constraints on \hubble\
and other cosmological parameters from 
potential combined samples of hundreds to thousands of BNS, 
NS-BH and BBH mergers expected in the next decade with 
upgraded and new GW detectors 
(\eg, \cite{Sathyaprakash:2010,Nishizawa:2012}).


\section{Acknowledgments}

The authors acknowledge useful discussions with 
Will Farr, Jon Gair, Archisman Ghosh and Leo Singer,
along with the detailed comments made by the (anonymous) referee.
This work was performed in part at the Aspen Center for Physics, 
which is supported by National Science Foundation grant PHY-1607611. 
The Flatiron Institute is supported by the Simons Foundation. 
This work was also partially supported by a grant from
the Simons Foundation. 
HVP was partially supported by the European Research Council (ERC) 
under the European Community's Seventh Framework Programme 
(FP7/2007-2013)/ERC grant agreement number 306478-CosmicDawn, 
and the research environment grant 
`Gravitational Radiation and Electromagnetic Astrophysical Transients (GREAT)'
funded by the Swedish Research Council (VR) under Dnr 2016-06012. 
This work was partially enabled by funding from the UCL Cosmoparticle
Initiative. 
ARW and SMN acknowledge the generous financial support of
the Netherlands Organization for Scientific Research through the NWO
VIDI and NWO TOP Grants (PI: Nissanke).


\bibliographystyle{apsrev4-1}
\bibliography{references}

\begin{thebibliography}{95}%
\makeatletter
\providecommand \@ifxundefined [1]{%
 \@ifx{#1\undefined}
}%
\providecommand \@ifnum [1]{%
 \ifnum #1\expandafter \@firstoftwo
 \else \expandafter \@secondoftwo
 \fi
}%
\providecommand \@ifx [1]{%
 \ifx #1\expandafter \@firstoftwo
 \else \expandafter \@secondoftwo
 \fi
}%
\providecommand \natexlab [1]{#1}%
\providecommand \enquote  [1]{``#1''}%
\providecommand \bibnamefont  [1]{#1}%
\providecommand \bibfnamefont [1]{#1}%
\providecommand \citenamefont [1]{#1}%
\providecommand \href@noop [0]{\@secondoftwo}%
\providecommand \href [0]{\begingroup \@sanitize@url \@href}%
\providecommand \@href[1]{\@@startlink{#1}\@@href}%
\providecommand \@@href[1]{\endgroup#1\@@endlink}%
\providecommand \@sanitize@url [0]{\catcode `\\12\catcode `\$12\catcode
  `\&12\catcode `\#12\catcode `\^12\catcode `\_12\catcode `\%12\relax}%
\providecommand \@@startlink[1]{}%
\providecommand \@@endlink[0]{}%
\providecommand \url  [0]{\begingroup\@sanitize@url \@url }%
\providecommand \@url [1]{\endgroup\@href {#1}{\urlprefix }}%
\providecommand \urlprefix  [0]{URL }%
\providecommand \Eprint [0]{\href }%
\providecommand \doibase [0]{http://dx.doi.org/}%
\providecommand \selectlanguage [0]{\@gobble}%
\providecommand \bibinfo  [0]{\@secondoftwo}%
\providecommand \bibfield  [0]{\@secondoftwo}%
\providecommand \translation [1]{[#1]}%
\providecommand \BibitemOpen [0]{}%
\providecommand \bibitemStop [0]{}%
\providecommand \bibitemNoStop [0]{.\EOS\space}%
\providecommand \EOS [0]{\spacefactor3000\relax}%
\providecommand \BibitemShut  [1]{\csname bibitem#1\endcsname}%
\let\auto@bib@innerbib\@empty
\bibitem [{\citenamefont {{Riess}}\ \emph {et~al.}(2019)\citenamefont
  {{Riess}}, \citenamefont {{Casertano}}, \citenamefont {{Yuan}}, \citenamefont
  {{Macri}},\ and\ \citenamefont {{Scolnic}}}]{Riess_etal:2019}%
  \BibitemOpen
  \bibfield  {author} {\bibinfo {author} {\bibfnamefont {A.~G.}\ \bibnamefont
  {{Riess}}}, \bibinfo {author} {\bibfnamefont {S.}~\bibnamefont
  {{Casertano}}}, \bibinfo {author} {\bibfnamefont {W.}~\bibnamefont {{Yuan}}},
  \bibinfo {author} {\bibfnamefont {L.~M.}\ \bibnamefont {{Macri}}}, \ and\
  \bibinfo {author} {\bibfnamefont {D.}~\bibnamefont {{Scolnic}}},\ }\href
  {\doibase 10.3847/1538-4357/ab1422} {\bibfield  {journal} {\bibinfo
  {journal} {\apj}\ }\textbf {\bibinfo {volume} {876}},\ \bibinfo {eid} {85}
  (\bibinfo {year} {2019})}\BibitemShut {NoStop}%
\bibitem [{\citenamefont {{Planck Collaboration}}(2018)}]{Planck_VI:2018}%
  \BibitemOpen
  \bibfield  {author} {\bibinfo {author} {\bibnamefont {{Planck
  Collaboration}}},\ }\href@noop {} {\bibfield  {journal} {\bibinfo  {journal}
  {arXiv e-prints}\ } (\bibinfo {year} {2018})},\ \Eprint
  {http://arxiv.org/abs/1807.06209} {arXiv:1807.06209 [astro-ph.CO]}
  \BibitemShut {NoStop}%
\bibitem [{\citenamefont {{Wyman}}\ \emph {et~al.}(2014)\citenamefont
  {{Wyman}}, \citenamefont {{Rudd}}, \citenamefont {{Vanderveld}},\ and\
  \citenamefont {{Hu}}}]{Wyman_etal:2014}%
  \BibitemOpen
  \bibfield  {author} {\bibinfo {author} {\bibfnamefont {M.}~\bibnamefont
  {{Wyman}}}, \bibinfo {author} {\bibfnamefont {D.~H.}\ \bibnamefont {{Rudd}}},
  \bibinfo {author} {\bibfnamefont {R.~A.}\ \bibnamefont {{Vanderveld}}}, \
  and\ \bibinfo {author} {\bibfnamefont {W.}~\bibnamefont {{Hu}}},\ }\href
  {\doibase 10.1103/PhysRevLett.112.051302} {\bibfield  {journal} {\bibinfo
  {journal} {\prl}\ }\textbf {\bibinfo {volume} {112}},\ \bibinfo {eid}
  {051302} (\bibinfo {year} {2014})}\BibitemShut {NoStop}%
\bibitem [{\citenamefont {{Pourtsidou}}\ and\ \citenamefont
  {{Tram}}(2016)}]{Pourtsidou_etal:2016}%
  \BibitemOpen
  \bibfield  {author} {\bibinfo {author} {\bibfnamefont {A.}~\bibnamefont
  {{Pourtsidou}}}\ and\ \bibinfo {author} {\bibfnamefont {T.}~\bibnamefont
  {{Tram}}},\ }\href {\doibase 10.1103/PhysRevD.94.043518} {\bibfield
  {journal} {\bibinfo  {journal} {\prd}\ }\textbf {\bibinfo {volume} {94}},\
  \bibinfo {eid} {043518} (\bibinfo {year} {2016})}\BibitemShut {NoStop}%
\bibitem [{\citenamefont {{Di Valentino}}\ \emph {et~al.}(2016)\citenamefont
  {{Di Valentino}}, \citenamefont {{Melchiorri}},\ and\ \citenamefont
  {{Silk}}}]{Di_Valentino_etal:2016}%
  \BibitemOpen
  \bibfield  {author} {\bibinfo {author} {\bibfnamefont {E.}~\bibnamefont {{Di
  Valentino}}}, \bibinfo {author} {\bibfnamefont {A.}~\bibnamefont
  {{Melchiorri}}}, \ and\ \bibinfo {author} {\bibfnamefont {J.}~\bibnamefont
  {{Silk}}},\ }\href {\doibase 10.1016/j.physletb.2016.08.043} {\bibfield
  {journal} {\bibinfo  {journal} {Physics Letters B}\ }\textbf {\bibinfo
  {volume} {761}},\ \bibinfo {pages} {242} (\bibinfo {year}
  {2016})}\BibitemShut {NoStop}%
\bibitem [{\citenamefont {{Huang}}\ and\ \citenamefont
  {{Wang}}(2016)}]{Huang_etal:2016}%
  \BibitemOpen
  \bibfield  {author} {\bibinfo {author} {\bibfnamefont {Q.-G.}\ \bibnamefont
  {{Huang}}}\ and\ \bibinfo {author} {\bibfnamefont {K.}~\bibnamefont
  {{Wang}}},\ }\href {\doibase 10.1140/epjc/s10052-016-4352-x} {\bibfield
  {journal} {\bibinfo  {journal} {\epjc}\ }\textbf {\bibinfo {volume} {76}},\
  \bibinfo {eid} {506} (\bibinfo {year} {2016})}\BibitemShut {NoStop}%
\bibitem [{\citenamefont {{Bernal}}\ \emph {et~al.}(2016)\citenamefont
  {{Bernal}}, \citenamefont {{Verde}},\ and\ \citenamefont
  {{Riess}}}]{Bernal_etal:2016}%
  \BibitemOpen
  \bibfield  {author} {\bibinfo {author} {\bibfnamefont {J.~L.}\ \bibnamefont
  {{Bernal}}}, \bibinfo {author} {\bibfnamefont {L.}~\bibnamefont {{Verde}}}, \
  and\ \bibinfo {author} {\bibfnamefont {A.~G.}\ \bibnamefont {{Riess}}},\
  }\href {\doibase 10.1088/1475-7516/2016/10/019} {\bibfield  {journal}
  {\bibinfo  {journal} {\jcap}\ }\textbf {\bibinfo {volume} {10}},\ \bibinfo
  {eid} {019} (\bibinfo {year} {2016})}\BibitemShut {NoStop}%
\bibitem [{\citenamefont {{Ko}}\ and\ \citenamefont
  {{Tang}}(2016)}]{Ko_etal:2016}%
  \BibitemOpen
  \bibfield  {author} {\bibinfo {author} {\bibfnamefont {P.}~\bibnamefont
  {{Ko}}}\ and\ \bibinfo {author} {\bibfnamefont {Y.}~\bibnamefont {{Tang}}},\
  }\href {\doibase 10.1016/j.physletb.2016.10.001} {\bibfield  {journal}
  {\bibinfo  {journal} {Physics Letters B}\ }\textbf {\bibinfo {volume}
  {762}},\ \bibinfo {pages} {462} (\bibinfo {year} {2016})}\BibitemShut
  {NoStop}%
\bibitem [{\citenamefont {{Karwal}}\ and\ \citenamefont
  {{Kamionkowski}}(2016)}]{Karwal_etal:2016}%
  \BibitemOpen
  \bibfield  {author} {\bibinfo {author} {\bibfnamefont {T.}~\bibnamefont
  {{Karwal}}}\ and\ \bibinfo {author} {\bibfnamefont {M.}~\bibnamefont
  {{Kamionkowski}}},\ }\href {\doibase 10.1103/PhysRevD.94.103523} {\bibfield
  {journal} {\bibinfo  {journal} {\prd}\ }\textbf {\bibinfo {volume} {94}},\
  \bibinfo {eid} {103523} (\bibinfo {year} {2016})}\BibitemShut {NoStop}%
\bibitem [{\citenamefont {{Kumar}}\ and\ \citenamefont
  {{Nunes}}(2016)}]{Kumar_etal:2016}%
  \BibitemOpen
  \bibfield  {author} {\bibinfo {author} {\bibfnamefont {S.}~\bibnamefont
  {{Kumar}}}\ and\ \bibinfo {author} {\bibfnamefont {R.~C.}\ \bibnamefont
  {{Nunes}}},\ }\href {\doibase 10.1103/PhysRevD.94.123511} {\bibfield
  {journal} {\bibinfo  {journal} {\prd}\ }\textbf {\bibinfo {volume} {94}},\
  \bibinfo {eid} {123511} (\bibinfo {year} {2016})}\BibitemShut {NoStop}%
\bibitem [{\citenamefont {{Santos}}\ \emph {et~al.}(2017)\citenamefont
  {{Santos}}, \citenamefont {{Coley}}, \citenamefont {{Chandrachani Devi}},\
  and\ \citenamefont {{Alcaniz}}}]{Santos_etal:2017}%
  \BibitemOpen
  \bibfield  {author} {\bibinfo {author} {\bibfnamefont {B.}~\bibnamefont
  {{Santos}}}, \bibinfo {author} {\bibfnamefont {A.~A.}\ \bibnamefont
  {{Coley}}}, \bibinfo {author} {\bibfnamefont {N.}~\bibnamefont {{Chandrachani
  Devi}}}, \ and\ \bibinfo {author} {\bibfnamefont {J.~S.}\ \bibnamefont
  {{Alcaniz}}},\ }\href {\doibase 10.1088/1475-7516/2017/02/047} {\bibfield
  {journal} {\bibinfo  {journal} {\jcap}\ }\textbf {\bibinfo {volume} {2}},\
  \bibinfo {eid} {047} (\bibinfo {year} {2017})}\BibitemShut {NoStop}%
\bibitem [{\citenamefont {{Prilepina}}\ and\ \citenamefont
  {{Tsai}}(2017)}]{Prilepina_etal:2017}%
  \BibitemOpen
  \bibfield  {author} {\bibinfo {author} {\bibfnamefont {V.}~\bibnamefont
  {{Prilepina}}}\ and\ \bibinfo {author} {\bibfnamefont {Y.}~\bibnamefont
  {{Tsai}}},\ }\href {\doibase 10.1007/JHEP09(2017)033} {\bibfield  {journal}
  {\bibinfo  {journal} {\jhep}\ }\textbf {\bibinfo {volume} {9}},\ \bibinfo
  {eid} {33} (\bibinfo {year} {2017})}\BibitemShut {NoStop}%
\bibitem [{\citenamefont {{Zhao}}\ \emph
  {et~al.}(2017{\natexlab{a}})\citenamefont {{Zhao}} \emph
  {et~al.}}]{Zhao_etal:2017}%
  \BibitemOpen
  \bibfield  {author} {\bibinfo {author} {\bibfnamefont {G.-B.}\ \bibnamefont
  {{Zhao}}} \emph {et~al.},\ }\href {\doibase 10.1038/s41550-017-0216-z}
  {\bibfield  {journal} {\bibinfo  {journal} {Nature Astronomy}\ }\textbf
  {\bibinfo {volume} {1}},\ \bibinfo {pages} {627} (\bibinfo {year}
  {2017}{\natexlab{a}})}\BibitemShut {NoStop}%
\bibitem [{\citenamefont {{Zhao}}\ \emph
  {et~al.}(2017{\natexlab{b}})\citenamefont {{Zhao}}, \citenamefont {{He}},
  \citenamefont {{Zhang}},\ and\ \citenamefont {{Zhang}}}]{Zhao_M_etal:2017}%
  \BibitemOpen
  \bibfield  {author} {\bibinfo {author} {\bibfnamefont {M.-M.}\ \bibnamefont
  {{Zhao}}}, \bibinfo {author} {\bibfnamefont {D.-Z.}\ \bibnamefont {{He}}},
  \bibinfo {author} {\bibfnamefont {J.-F.}\ \bibnamefont {{Zhang}}}, \ and\
  \bibinfo {author} {\bibfnamefont {X.}~\bibnamefont {{Zhang}}},\ }\href
  {\doibase 10.1103/PhysRevD.96.043520} {\bibfield  {journal} {\bibinfo
  {journal} {\prd}\ }\textbf {\bibinfo {volume} {96}},\ \bibinfo {eid} {043520}
  (\bibinfo {year} {2017}{\natexlab{b}})}\BibitemShut {NoStop}%
\bibitem [{\citenamefont {{Sol{\`a}}}\ \emph {et~al.}(2017)\citenamefont
  {{Sol{\`a}}}, \citenamefont {{G{\'o}mez-Valent}},\ and\ \citenamefont {{de
  Cruz P{\'e}rez}}}]{Solar_etal:2017}%
  \BibitemOpen
  \bibfield  {author} {\bibinfo {author} {\bibfnamefont {J.}~\bibnamefont
  {{Sol{\`a}}}}, \bibinfo {author} {\bibfnamefont {A.}~\bibnamefont
  {{G{\'o}mez-Valent}}}, \ and\ \bibinfo {author} {\bibfnamefont
  {J.}~\bibnamefont {{de Cruz P{\'e}rez}}},\ }\href {\doibase
  10.1016/j.physletb.2017.09.073} {\bibfield  {journal} {\bibinfo  {journal}
  {Physics Letters B}\ }\textbf {\bibinfo {volume} {774}},\ \bibinfo {pages}
  {317} (\bibinfo {year} {2017})}\BibitemShut {NoStop}%
\bibitem [{\citenamefont {{Di Valentino}}\ \emph
  {et~al.}(2018{\natexlab{a}})\citenamefont {{Di Valentino}}, \citenamefont
  {{Linder}},\ and\ \citenamefont {{Melchiorri}}}]{Di_Valentino_etal:2017a}%
  \BibitemOpen
  \bibfield  {author} {\bibinfo {author} {\bibfnamefont {E.}~\bibnamefont {{Di
  Valentino}}}, \bibinfo {author} {\bibfnamefont {E.~V.}\ \bibnamefont
  {{Linder}}}, \ and\ \bibinfo {author} {\bibfnamefont {A.}~\bibnamefont
  {{Melchiorri}}},\ }\href {\doibase 10.1103/PhysRevD.97.043528} {\bibfield
  {journal} {\bibinfo  {journal} {\prd}\ }\textbf {\bibinfo {volume} {97}},\
  \bibinfo {eid} {043528} (\bibinfo {year} {2018}{\natexlab{a}})}\BibitemShut
  {NoStop}%
\bibitem [{\citenamefont {{Di Valentino}}\ \emph
  {et~al.}(2018{\natexlab{b}})\citenamefont {{Di Valentino}}, \citenamefont
  {{B{\oe}hm}}, \citenamefont {{Hivon}},\ and\ \citenamefont
  {{Bouchet}}}]{Di_Valentino_etal:2017b}%
  \BibitemOpen
  \bibfield  {author} {\bibinfo {author} {\bibfnamefont {E.}~\bibnamefont {{Di
  Valentino}}}, \bibinfo {author} {\bibfnamefont {C.}~\bibnamefont
  {{B{\oe}hm}}}, \bibinfo {author} {\bibfnamefont {E.}~\bibnamefont {{Hivon}}},
  \ and\ \bibinfo {author} {\bibfnamefont {F.~R.}\ \bibnamefont {{Bouchet}}},\
  }\href {\doibase 10.1103/PhysRevD.97.043513} {\bibfield  {journal} {\bibinfo
  {journal} {\prd}\ }\textbf {\bibinfo {volume} {97}},\ \bibinfo {eid} {043513}
  (\bibinfo {year} {2018}{\natexlab{b}})}\BibitemShut {NoStop}%
\bibitem [{\citenamefont {{Graef}}\ \emph {et~al.}(2019)\citenamefont
  {{Graef}}, \citenamefont {{Benetti}},\ and\ \citenamefont
  {{Alcaniz}}}]{Graef_etal:2018}%
  \BibitemOpen
  \bibfield  {author} {\bibinfo {author} {\bibfnamefont {L.~L.}\ \bibnamefont
  {{Graef}}}, \bibinfo {author} {\bibfnamefont {M.}~\bibnamefont {{Benetti}}},
  \ and\ \bibinfo {author} {\bibfnamefont {J.~S.}\ \bibnamefont {{Alcaniz}}},\
  }\href {\doibase 10.1103/PhysRevD.99.043519} {\bibfield  {journal} {\bibinfo
  {journal} {\prd}\ }\textbf {\bibinfo {volume} {99}},\ \bibinfo {eid} {043519}
  (\bibinfo {year} {2019})}\BibitemShut {NoStop}%
\bibitem [{\citenamefont {{Yang}}\ \emph {et~al.}(2018)\citenamefont {{Yang}},
  \citenamefont {{Mukherjee}}, \citenamefont {{Di Valentino}},\ and\
  \citenamefont {{Pan}}}]{Yang_etal:2018a}%
  \BibitemOpen
  \bibfield  {author} {\bibinfo {author} {\bibfnamefont {W.}~\bibnamefont
  {{Yang}}}, \bibinfo {author} {\bibfnamefont {A.}~\bibnamefont {{Mukherjee}}},
  \bibinfo {author} {\bibfnamefont {E.}~\bibnamefont {{Di Valentino}}}, \ and\
  \bibinfo {author} {\bibfnamefont {S.}~\bibnamefont {{Pan}}},\ }\href
  {\doibase 10.1103/PhysRevD.98.123527} {\bibfield  {journal} {\bibinfo
  {journal} {\prd}\ }\textbf {\bibinfo {volume} {98}},\ \bibinfo {eid} {123527}
  (\bibinfo {year} {2018})}\BibitemShut {NoStop}%
\bibitem [{\citenamefont {{El-Zant}}\ \emph {et~al.}(2019)\citenamefont
  {{El-Zant}}, \citenamefont {{El Hanafy}},\ and\ \citenamefont
  {{Elgammal}}}]{El-zant_etal:2018}%
  \BibitemOpen
  \bibfield  {author} {\bibinfo {author} {\bibfnamefont {A.}~\bibnamefont
  {{El-Zant}}}, \bibinfo {author} {\bibfnamefont {W.}~\bibnamefont {{El
  Hanafy}}}, \ and\ \bibinfo {author} {\bibfnamefont {S.}~\bibnamefont
  {{Elgammal}}},\ }\href {\doibase 10.3847/1538-4357/aafa12} {\bibfield
  {journal} {\bibinfo  {journal} {\apj}\ }\textbf {\bibinfo {volume} {871}},\
  \bibinfo {eid} {210} (\bibinfo {year} {2019})}\BibitemShut {NoStop}%
\bibitem [{\citenamefont {{Benevento}}\ \emph {et~al.}(2019)\citenamefont
  {{Benevento}}, \citenamefont {{Raveri}}, \citenamefont {{Lazanu}},
  \citenamefont {{Bartolo}}, \citenamefont {{Liguori}}, \citenamefont
  {{Brax}},\ and\ \citenamefont {{Valageas}}}]{Benevento_etal:2018}%
  \BibitemOpen
  \bibfield  {author} {\bibinfo {author} {\bibfnamefont {G.}~\bibnamefont
  {{Benevento}}}, \bibinfo {author} {\bibfnamefont {M.}~\bibnamefont
  {{Raveri}}}, \bibinfo {author} {\bibfnamefont {A.}~\bibnamefont {{Lazanu}}},
  \bibinfo {author} {\bibfnamefont {N.}~\bibnamefont {{Bartolo}}}, \bibinfo
  {author} {\bibfnamefont {M.}~\bibnamefont {{Liguori}}}, \bibinfo {author}
  {\bibfnamefont {P.}~\bibnamefont {{Brax}}}, \ and\ \bibinfo {author}
  {\bibfnamefont {P.}~\bibnamefont {{Valageas}}},\ }\href {\doibase
  10.1088/1475-7516/2019/05/027} {\bibfield  {journal} {\bibinfo  {journal}
  {\jcap}\ }\textbf {\bibinfo {volume} {2019}},\ \bibinfo {eid} {027} (\bibinfo
  {year} {2019})}\BibitemShut {NoStop}%
\bibitem [{\citenamefont {{Yang}}\ \emph {et~al.}(2019)\citenamefont {{Yang}},
  \citenamefont {{Pan}}, \citenamefont {{Di Valentino}}, \citenamefont
  {{Saridakis}},\ and\ \citenamefont {{Chakraborty}}}]{Yang_etal:2018b}%
  \BibitemOpen
  \bibfield  {author} {\bibinfo {author} {\bibfnamefont {W.}~\bibnamefont
  {{Yang}}}, \bibinfo {author} {\bibfnamefont {S.}~\bibnamefont {{Pan}}},
  \bibinfo {author} {\bibfnamefont {E.}~\bibnamefont {{Di Valentino}}},
  \bibinfo {author} {\bibfnamefont {E.~N.}\ \bibnamefont {{Saridakis}}}, \ and\
  \bibinfo {author} {\bibfnamefont {S.}~\bibnamefont {{Chakraborty}}},\ }\href
  {\doibase 10.1103/PhysRevD.99.043543} {\bibfield  {journal} {\bibinfo
  {journal} {\prd}\ }\textbf {\bibinfo {volume} {99}},\ \bibinfo {eid} {043543}
  (\bibinfo {year} {2019})}\BibitemShut {NoStop}%
\bibitem [{\citenamefont {{Aylor}}\ \emph {et~al.}(2019)\citenamefont
  {{Aylor}}, \citenamefont {{Joy}}, \citenamefont {{Knox}}, \citenamefont
  {{Millea}}, \citenamefont {{Raghunathan}},\ and\ \citenamefont {{Kimmy
  Wu}}}]{Aylor_etal:2018}%
  \BibitemOpen
  \bibfield  {author} {\bibinfo {author} {\bibfnamefont {K.}~\bibnamefont
  {{Aylor}}}, \bibinfo {author} {\bibfnamefont {M.}~\bibnamefont {{Joy}}},
  \bibinfo {author} {\bibfnamefont {L.}~\bibnamefont {{Knox}}}, \bibinfo
  {author} {\bibfnamefont {M.}~\bibnamefont {{Millea}}}, \bibinfo {author}
  {\bibfnamefont {S.}~\bibnamefont {{Raghunathan}}}, \ and\ \bibinfo {author}
  {\bibfnamefont {W.~L.}\ \bibnamefont {{Kimmy Wu}}},\ }\href {\doibase
  10.3847/1538-4357/ab0898} {\bibfield  {journal} {\bibinfo  {journal} {\apj}\
  }\textbf {\bibinfo {volume} {874}},\ \bibinfo {eid} {4} (\bibinfo {year}
  {2019})}\BibitemShut {NoStop}%
\bibitem [{\citenamefont {{Chiang}}\ and\ \citenamefont
  {{Slosar}}(2018)}]{Chiang_Slosar:2018}%
  \BibitemOpen
  \bibfield  {author} {\bibinfo {author} {\bibfnamefont {C.-T.}\ \bibnamefont
  {{Chiang}}}\ and\ \bibinfo {author} {\bibfnamefont {A.}~\bibnamefont
  {{Slosar}}},\ }\href@noop {} {\bibfield  {journal} {\bibinfo  {journal}
  {ArXiv e-prints}\ } (\bibinfo {year} {2018})},\ \Eprint
  {http://arxiv.org/abs/1811.03624} {arXiv:1811.03624} \BibitemShut {NoStop}%
\bibitem [{\citenamefont {{Poulin}}\ \emph {et~al.}(2019)\citenamefont
  {{Poulin}}, \citenamefont {{Smith}}, \citenamefont {{Karwal}},\ and\
  \citenamefont {{Kamionkowski}}}]{Poulin_etal:2018}%
  \BibitemOpen
  \bibfield  {author} {\bibinfo {author} {\bibfnamefont {V.}~\bibnamefont
  {{Poulin}}}, \bibinfo {author} {\bibfnamefont {T.~L.}\ \bibnamefont
  {{Smith}}}, \bibinfo {author} {\bibfnamefont {T.}~\bibnamefont {{Karwal}}}, \
  and\ \bibinfo {author} {\bibfnamefont {M.}~\bibnamefont {{Kamionkowski}}},\
  }\href {\doibase 10.1103/PhysRevLett.122.221301} {\bibfield  {journal}
  {\bibinfo  {journal} {\prl}\ }\textbf {\bibinfo {volume} {122}},\ \bibinfo
  {eid} {221301} (\bibinfo {year} {2019})}\BibitemShut {NoStop}%
\bibitem [{\citenamefont {{Efstathiou}}(2014)}]{Efstathiou:2014}%
  \BibitemOpen
  \bibfield  {author} {\bibinfo {author} {\bibfnamefont {G.}~\bibnamefont
  {{Efstathiou}}},\ }\href {\doibase 10.1093/mnras/stu278} {\bibfield
  {journal} {\bibinfo  {journal} {\mnras}\ }\textbf {\bibinfo {volume} {440}},\
  \bibinfo {pages} {1138} (\bibinfo {year} {2014})}\BibitemShut {NoStop}%
\bibitem [{\citenamefont {{Spergel}}\ \emph {et~al.}(2015)\citenamefont
  {{Spergel}}, \citenamefont {{Flauger}},\ and\ \citenamefont {{Hlo{\v
  z}ek}}}]{Spergel_etal:2015}%
  \BibitemOpen
  \bibfield  {author} {\bibinfo {author} {\bibfnamefont {D.~N.}\ \bibnamefont
  {{Spergel}}}, \bibinfo {author} {\bibfnamefont {R.}~\bibnamefont
  {{Flauger}}}, \ and\ \bibinfo {author} {\bibfnamefont {R.}~\bibnamefont
  {{Hlo{\v z}ek}}},\ }\href {\doibase 10.1103/PhysRevD.91.023518} {\bibfield
  {journal} {\bibinfo  {journal} {\prd}\ }\textbf {\bibinfo {volume} {91}},\
  \bibinfo {eid} {023518} (\bibinfo {year} {2015})}\BibitemShut {NoStop}%
\bibitem [{\citenamefont {{Rigault}}\ \emph {et~al.}(2015)\citenamefont
  {{Rigault}} \emph {et~al.}}]{Rigault_etal:2015}%
  \BibitemOpen
  \bibfield  {author} {\bibinfo {author} {\bibfnamefont {M.}~\bibnamefont
  {{Rigault}}} \emph {et~al.},\ }\href {\doibase 10.1088/0004-637X/802/1/20}
  {\bibfield  {journal} {\bibinfo  {journal} {\apj}\ }\textbf {\bibinfo
  {volume} {802}},\ \bibinfo {eid} {20} (\bibinfo {year} {2015})}\BibitemShut
  {NoStop}%
\bibitem [{\citenamefont {{Jones}}\ \emph {et~al.}(2015)\citenamefont
  {{Jones}}, \citenamefont {{Riess}},\ and\ \citenamefont
  {{Scolnic}}}]{Jones_etal:2015}%
  \BibitemOpen
  \bibfield  {author} {\bibinfo {author} {\bibfnamefont {D.~O.}\ \bibnamefont
  {{Jones}}}, \bibinfo {author} {\bibfnamefont {A.~G.}\ \bibnamefont
  {{Riess}}}, \ and\ \bibinfo {author} {\bibfnamefont {D.~M.}\ \bibnamefont
  {{Scolnic}}},\ }\href {\doibase 10.1088/0004-637X/812/1/31} {\bibfield
  {journal} {\bibinfo  {journal} {\apj}\ }\textbf {\bibinfo {volume} {812}},\
  \bibinfo {eid} {31} (\bibinfo {year} {2015})}\BibitemShut {NoStop}%
\bibitem [{\citenamefont {{Addison}}\ \emph {et~al.}(2016)\citenamefont
  {{Addison}}, \citenamefont {{Huang}}, \citenamefont {{Watts}}, \citenamefont
  {{Bennett}}, \citenamefont {{Halpern}}, \citenamefont {{Hinshaw}},\ and\
  \citenamefont {{Weiland}}}]{Addison_etal:2016}%
  \BibitemOpen
  \bibfield  {author} {\bibinfo {author} {\bibfnamefont {G.~E.}\ \bibnamefont
  {{Addison}}}, \bibinfo {author} {\bibfnamefont {Y.}~\bibnamefont {{Huang}}},
  \bibinfo {author} {\bibfnamefont {D.~J.}\ \bibnamefont {{Watts}}}, \bibinfo
  {author} {\bibfnamefont {C.~L.}\ \bibnamefont {{Bennett}}}, \bibinfo {author}
  {\bibfnamefont {M.}~\bibnamefont {{Halpern}}}, \bibinfo {author}
  {\bibfnamefont {G.}~\bibnamefont {{Hinshaw}}}, \ and\ \bibinfo {author}
  {\bibfnamefont {J.~L.}\ \bibnamefont {{Weiland}}},\ }\href {\doibase
  10.3847/0004-637X/818/2/132} {\bibfield  {journal} {\bibinfo  {journal}
  {\apj}\ }\textbf {\bibinfo {volume} {818}},\ \bibinfo {eid} {132} (\bibinfo
  {year} {2016})}\BibitemShut {NoStop}%
\bibitem [{\citenamefont {{Planck
  Collaboration}}(2016)}]{Planck_Int_XLVI:2016}%
  \BibitemOpen
  \bibfield  {author} {\bibinfo {author} {\bibnamefont {{Planck
  Collaboration}}},\ }\href {\doibase 10.1051/0004-6361/201628890} {\bibfield
  {journal} {\bibinfo  {journal} {\aap}\ }\textbf {\bibinfo {volume} {596}},\
  \bibinfo {eid} {A107} (\bibinfo {year} {2016})}\BibitemShut {NoStop}%
\bibitem [{\citenamefont {{Cardona}}\ \emph {et~al.}(2017)\citenamefont
  {{Cardona}}, \citenamefont {{Kunz}},\ and\ \citenamefont
  {{Pettorino}}}]{Cardona_etal:2016}%
  \BibitemOpen
  \bibfield  {author} {\bibinfo {author} {\bibfnamefont {W.}~\bibnamefont
  {{Cardona}}}, \bibinfo {author} {\bibfnamefont {M.}~\bibnamefont {{Kunz}}}, \
  and\ \bibinfo {author} {\bibfnamefont {V.}~\bibnamefont {{Pettorino}}},\
  }\href {\doibase 10.1088/1475-7516/2017/03/056} {\bibfield  {journal}
  {\bibinfo  {journal} {\jcap}\ }\textbf {\bibinfo {volume} {3}},\ \bibinfo
  {eid} {056} (\bibinfo {year} {2017})}\BibitemShut {NoStop}%
\bibitem [{\citenamefont {{Zhang}}\ \emph {et~al.}(2017)\citenamefont
  {{Zhang}}, \citenamefont {{Childress}}, \citenamefont {{Davis}},
  \citenamefont {{Karpenka}}, \citenamefont {{Lidman}}, \citenamefont
  {{Schmidt}},\ and\ \citenamefont {{Smith}}}]{Zhang_etal:2017}%
  \BibitemOpen
  \bibfield  {author} {\bibinfo {author} {\bibfnamefont {B.~R.}\ \bibnamefont
  {{Zhang}}}, \bibinfo {author} {\bibfnamefont {M.~J.}\ \bibnamefont
  {{Childress}}}, \bibinfo {author} {\bibfnamefont {T.~M.}\ \bibnamefont
  {{Davis}}}, \bibinfo {author} {\bibfnamefont {N.~V.}\ \bibnamefont
  {{Karpenka}}}, \bibinfo {author} {\bibfnamefont {C.}~\bibnamefont
  {{Lidman}}}, \bibinfo {author} {\bibfnamefont {B.~P.}\ \bibnamefont
  {{Schmidt}}}, \ and\ \bibinfo {author} {\bibfnamefont {M.}~\bibnamefont
  {{Smith}}},\ }\href {\doibase 10.1093/mnras/stx1600} {\bibfield  {journal}
  {\bibinfo  {journal} {\mnras}\ }\textbf {\bibinfo {volume} {471}},\ \bibinfo
  {pages} {2254} (\bibinfo {year} {2017})}\BibitemShut {NoStop}%
\bibitem [{\citenamefont {{Wu}}\ and\ \citenamefont
  {{Huterer}}(2017)}]{Wu_Huterer:2017}%
  \BibitemOpen
  \bibfield  {author} {\bibinfo {author} {\bibfnamefont {H.-Y.}\ \bibnamefont
  {{Wu}}}\ and\ \bibinfo {author} {\bibfnamefont {D.}~\bibnamefont
  {{Huterer}}},\ }\href {\doibase 10.1093/mnras/stx1967} {\bibfield  {journal}
  {\bibinfo  {journal} {\mnras}\ }\textbf {\bibinfo {volume} {471}},\ \bibinfo
  {pages} {4946} (\bibinfo {year} {2017})}\BibitemShut {NoStop}%
\bibitem [{\citenamefont {{Feeney}}\ \emph {et~al.}(2018)\citenamefont
  {{Feeney}}, \citenamefont {{Mortlock}},\ and\ \citenamefont
  {{Dalmasso}}}]{Feeney_etal:2017}%
  \BibitemOpen
  \bibfield  {author} {\bibinfo {author} {\bibfnamefont {S.~M.}\ \bibnamefont
  {{Feeney}}}, \bibinfo {author} {\bibfnamefont {D.~J.}\ \bibnamefont
  {{Mortlock}}}, \ and\ \bibinfo {author} {\bibfnamefont {N.}~\bibnamefont
  {{Dalmasso}}},\ }\href {\doibase 10.1093/mnras/sty418} {\bibfield  {journal}
  {\bibinfo  {journal} {\mnras}\ }\textbf {\bibinfo {volume} {476}},\ \bibinfo
  {pages} {3861} (\bibinfo {year} {2018})}\BibitemShut {NoStop}%
\bibitem [{\citenamefont {{Follin}}\ and\ \citenamefont
  {{Knox}}(2018)}]{Follin_Knox:2017}%
  \BibitemOpen
  \bibfield  {author} {\bibinfo {author} {\bibfnamefont {B.}~\bibnamefont
  {{Follin}}}\ and\ \bibinfo {author} {\bibfnamefont {L.}~\bibnamefont
  {{Knox}}},\ }\href {\doibase 10.1093/mnras/sty720} {\bibfield  {journal}
  {\bibinfo  {journal} {\mnras}\ }\textbf {\bibinfo {volume} {477}},\ \bibinfo
  {pages} {4534} (\bibinfo {year} {2018})}\BibitemShut {NoStop}%
\bibitem [{\citenamefont {{Dhawan}}\ \emph {et~al.}(2018)\citenamefont
  {{Dhawan}}, \citenamefont {{Jha}},\ and\ \citenamefont
  {{Leibundgut}}}]{Dhawan_etal:2017}%
  \BibitemOpen
  \bibfield  {author} {\bibinfo {author} {\bibfnamefont {S.}~\bibnamefont
  {{Dhawan}}}, \bibinfo {author} {\bibfnamefont {S.~W.}\ \bibnamefont {{Jha}}},
  \ and\ \bibinfo {author} {\bibfnamefont {B.}~\bibnamefont {{Leibundgut}}},\
  }\href {\doibase 10.1051/0004-6361/201731501} {\bibfield  {journal} {\bibinfo
   {journal} {\aap}\ }\textbf {\bibinfo {volume} {609}},\ \bibinfo {eid} {A72}
  (\bibinfo {year} {2018})}\BibitemShut {NoStop}%
\bibitem [{\citenamefont {{Rigault}}\ \emph {et~al.}(2018)\citenamefont
  {{Rigault}} \emph {et~al.}}]{Rigault_etal:2018}%
  \BibitemOpen
  \bibfield  {author} {\bibinfo {author} {\bibfnamefont {M.}~\bibnamefont
  {{Rigault}}} \emph {et~al.},\ }\href@noop {} {\bibfield  {journal} {\bibinfo
  {journal} {ArXiv e-prints}\ } (\bibinfo {year} {2018})},\ \Eprint
  {http://arxiv.org/abs/1806.03849} {arXiv:1806.03849} \BibitemShut {NoStop}%
\bibitem [{\citenamefont {{Bengaly}}\ \emph {et~al.}(2018)\citenamefont
  {{Bengaly}}, \citenamefont {{Andrade}},\ and\ \citenamefont
  {{Alcaniz}}}]{Bengaly_etal:2018}%
  \BibitemOpen
  \bibfield  {author} {\bibinfo {author} {\bibfnamefont {C.~A.~P.}\
  \bibnamefont {{Bengaly}}}, \bibinfo {author} {\bibfnamefont {U.}~\bibnamefont
  {{Andrade}}}, \ and\ \bibinfo {author} {\bibfnamefont {J.~S.}\ \bibnamefont
  {{Alcaniz}}},\ }\href@noop {} {\bibfield  {journal} {\bibinfo  {journal}
  {ArXiv e-prints}\ } (\bibinfo {year} {2018})},\ \Eprint
  {http://arxiv.org/abs/1810.04966} {arXiv:1810.04966} \BibitemShut {NoStop}%
\bibitem [{\citenamefont {{Schutz}}(1986)}]{Schutz:1986}%
  \BibitemOpen
  \bibfield  {author} {\bibinfo {author} {\bibfnamefont {B.~F.}\ \bibnamefont
  {{Schutz}}},\ }\href {\doibase 10.1038/323310a0} {\bibfield  {journal}
  {\bibinfo  {journal} {\nat}\ }\textbf {\bibinfo {volume} {323}},\ \bibinfo
  {pages} {310} (\bibinfo {year} {1986})}\BibitemShut {NoStop}%
\bibitem [{\citenamefont {{Abbott}}\ \emph
  {et~al.}(2017{\natexlab{a}})\citenamefont {{Abbott}}, \citenamefont
  {{Abbott}}, \citenamefont {{Abbott}}, \citenamefont {{Acernese}},
  \citenamefont {{Ackley}}, \citenamefont {{Adams}}, \citenamefont {{Adams}},
  \citenamefont {{Addesso}}, \citenamefont {{Adhikari}}, \citenamefont
  {{Adya}},\ and\ \citenamefont {et~al.}}]{Abbott_etal:2017d}%
  \BibitemOpen
  \bibfield  {author} {\bibinfo {author} {\bibfnamefont {B.~P.}\ \bibnamefont
  {{Abbott}}}, \bibinfo {author} {\bibfnamefont {R.}~\bibnamefont {{Abbott}}},
  \bibinfo {author} {\bibfnamefont {T.~D.}\ \bibnamefont {{Abbott}}}, \bibinfo
  {author} {\bibfnamefont {F.}~\bibnamefont {{Acernese}}}, \bibinfo {author}
  {\bibfnamefont {K.}~\bibnamefont {{Ackley}}}, \bibinfo {author}
  {\bibfnamefont {C.}~\bibnamefont {{Adams}}}, \bibinfo {author} {\bibfnamefont
  {T.}~\bibnamefont {{Adams}}}, \bibinfo {author} {\bibfnamefont
  {P.}~\bibnamefont {{Addesso}}}, \bibinfo {author} {\bibfnamefont {R.~X.}\
  \bibnamefont {{Adhikari}}}, \bibinfo {author} {\bibfnamefont {V.~B.}\
  \bibnamefont {{Adya}}}, \ and\ \bibinfo {author} {\bibnamefont {et~al.}},\
  }\href {\doibase 10.1103/PhysRevLett.119.161101} {\bibfield  {journal}
  {\bibinfo  {journal} {\prl}\ }\textbf {\bibinfo {volume} {119}},\ \bibinfo
  {eid} {161101} (\bibinfo {year} {2017}{\natexlab{a}})}\BibitemShut {NoStop}%
\bibitem [{\citenamefont {{The LIGO Scientific Collaboration}}\ \emph
  {et~al.}(2018{\natexlab{a}})\citenamefont {{The LIGO Scientific
  Collaboration}}, \citenamefont {{the Virgo Collaboration}} \emph
  {et~al.}}]{Ligo:2018}%
  \BibitemOpen
  \bibfield  {author} {\bibinfo {author} {\bibnamefont {{The LIGO Scientific
  Collaboration}}}, \bibinfo {author} {\bibnamefont {{the Virgo
  Collaboration}}},  \emph {et~al.},\ }\href@noop {} {\bibfield  {journal}
  {\bibinfo  {journal} {arXiv e-prints}\ } (\bibinfo {year}
  {2018}{\natexlab{a}})},\ \Eprint {http://arxiv.org/abs/1811.12907}
  {arXiv:1811.12907 [astro-ph.HE]} \BibitemShut {NoStop}%
\bibitem [{\citenamefont {{Dalal}}\ \emph {et~al.}(2006)\citenamefont
  {{Dalal}}, \citenamefont {{Holz}}, \citenamefont {{Hughes}},\ and\
  \citenamefont {{Jain}}}]{Dalal:2006}%
  \BibitemOpen
  \bibfield  {author} {\bibinfo {author} {\bibfnamefont {N.}~\bibnamefont
  {{Dalal}}}, \bibinfo {author} {\bibfnamefont {D.~E.}\ \bibnamefont {{Holz}}},
  \bibinfo {author} {\bibfnamefont {S.~A.}\ \bibnamefont {{Hughes}}}, \ and\
  \bibinfo {author} {\bibfnamefont {B.}~\bibnamefont {{Jain}}},\ }\href
  {\doibase 10.1103/PhysRevD.74.063006} {\bibfield  {journal} {\bibinfo
  {journal} {\prd}\ }\textbf {\bibinfo {volume} {74}},\ \bibinfo {eid} {063006}
  (\bibinfo {year} {2006})}\BibitemShut {NoStop}%
\bibitem [{\citenamefont {{Nissanke}}\ \emph {et~al.}(2010)\citenamefont
  {{Nissanke}}, \citenamefont {{Holz}}, \citenamefont {{Hughes}}, \citenamefont
  {{Dalal}},\ and\ \citenamefont {{Sievers}}}]{Nissanke_etal:2010}%
  \BibitemOpen
  \bibfield  {author} {\bibinfo {author} {\bibfnamefont {S.}~\bibnamefont
  {{Nissanke}}}, \bibinfo {author} {\bibfnamefont {D.~E.}\ \bibnamefont
  {{Holz}}}, \bibinfo {author} {\bibfnamefont {S.~A.}\ \bibnamefont
  {{Hughes}}}, \bibinfo {author} {\bibfnamefont {N.}~\bibnamefont {{Dalal}}}, \
  and\ \bibinfo {author} {\bibfnamefont {J.~L.}\ \bibnamefont {{Sievers}}},\
  }\href {\doibase 10.1088/0004-637X/725/1/496} {\bibfield  {journal} {\bibinfo
   {journal} {\apj}\ }\textbf {\bibinfo {volume} {725}},\ \bibinfo {pages}
  {496} (\bibinfo {year} {2010})}\BibitemShut {NoStop}%
\bibitem [{\citenamefont {{Abbott}}\ \emph
  {et~al.}(2017{\natexlab{b}})\citenamefont {{Abbott}} \emph
  {et~al.}}]{Abbott_etal:2017a}%
  \BibitemOpen
  \bibfield  {author} {\bibinfo {author} {\bibfnamefont {B.~P.}\ \bibnamefont
  {{Abbott}}} \emph {et~al.},\ }\href {\doibase 10.1038/nature24471} {\bibfield
   {journal} {\bibinfo  {journal} {\nat}\ }\textbf {\bibinfo {volume} {551}},\
  \bibinfo {pages} {85} (\bibinfo {year} {2017}{\natexlab{b}})}\BibitemShut
  {NoStop}%
\bibitem [{\citenamefont {{Del Pozzo}}(2012)}]{del_Pozzo:2012}%
  \BibitemOpen
  \bibfield  {author} {\bibinfo {author} {\bibfnamefont {W.}~\bibnamefont {{Del
  Pozzo}}},\ }\href {\doibase 10.1103/PhysRevD.86.043011} {\bibfield  {journal}
  {\bibinfo  {journal} {\prd}\ }\textbf {\bibinfo {volume} {86}},\ \bibinfo
  {eid} {043011} (\bibinfo {year} {2012})}\BibitemShut {NoStop}%
\bibitem [{\citenamefont {{Oguri}}(2016)}]{Oguri:2016}%
  \BibitemOpen
  \bibfield  {author} {\bibinfo {author} {\bibfnamefont {M.}~\bibnamefont
  {{Oguri}}},\ }\href {\doibase 10.1103/PhysRevD.93.083511} {\bibfield
  {journal} {\bibinfo  {journal} {\prd}\ }\textbf {\bibinfo {volume} {93}},\
  \bibinfo {eid} {083511} (\bibinfo {year} {2016})}\BibitemShut {NoStop}%
\bibitem [{\citenamefont {{Fishbach}}\ \emph {et~al.}(2019)\citenamefont
  {{Fishbach}}, \citenamefont {{Gray}}, \citenamefont {{Maga{\~n}a Hernandez}},
  \citenamefont {{Qi}}, \citenamefont {{Sur}}, \citenamefont {{Acernese}},
  \citenamefont {{Aiello}}, \citenamefont {{Allocca}}, \citenamefont {{Aloy}},\
  and\ \citenamefont {{Amato}}}]{Fishbach_etal:2018}%
  \BibitemOpen
  \bibfield  {author} {\bibinfo {author} {\bibfnamefont {M.}~\bibnamefont
  {{Fishbach}}}, \bibinfo {author} {\bibfnamefont {R.}~\bibnamefont {{Gray}}},
  \bibinfo {author} {\bibfnamefont {I.}~\bibnamefont {{Maga{\~n}a Hernandez}}},
  \bibinfo {author} {\bibfnamefont {H.}~\bibnamefont {{Qi}}}, \bibinfo {author}
  {\bibfnamefont {A.}~\bibnamefont {{Sur}}}, \bibinfo {author} {\bibfnamefont
  {F.}~\bibnamefont {{Acernese}}}, \bibinfo {author} {\bibfnamefont
  {L.}~\bibnamefont {{Aiello}}}, \bibinfo {author} {\bibfnamefont
  {A.}~\bibnamefont {{Allocca}}}, \bibinfo {author} {\bibfnamefont {M.~A.}\
  \bibnamefont {{Aloy}}}, \ and\ \bibinfo {author} {\bibfnamefont
  {A.}~\bibnamefont {{Amato}}},\ }\href {\doibase 10.3847/2041-8213/aaf96e}
  {\bibfield  {journal} {\bibinfo  {journal} {\apjl}\ }\textbf {\bibinfo
  {volume} {871}},\ \bibinfo {eid} {L13} (\bibinfo {year} {2019})}\BibitemShut
  {NoStop}%
\bibitem [{\citenamefont {{Taylor}}\ and\ \citenamefont
  {{Gair}}(2012)}]{Taylor_Gair:2012b}%
  \BibitemOpen
  \bibfield  {author} {\bibinfo {author} {\bibfnamefont {S.~R.}\ \bibnamefont
  {{Taylor}}}\ and\ \bibinfo {author} {\bibfnamefont {J.~R.}\ \bibnamefont
  {{Gair}}},\ }\href {\doibase 10.1103/PhysRevD.86.023502} {\bibfield
  {journal} {\bibinfo  {journal} {\prd}\ }\textbf {\bibinfo {volume} {86}},\
  \bibinfo {eid} {023502} (\bibinfo {year} {2012})}\BibitemShut {NoStop}%
\bibitem [{\citenamefont {{Messenger}}\ and\ \citenamefont
  {{Read}}(2012)}]{Messenger_Read:2012}%
  \BibitemOpen
  \bibfield  {author} {\bibinfo {author} {\bibfnamefont {C.}~\bibnamefont
  {{Messenger}}}\ and\ \bibinfo {author} {\bibfnamefont {J.}~\bibnamefont
  {{Read}}},\ }\href {\doibase 10.1103/PhysRevLett.108.091101} {\bibfield
  {journal} {\bibinfo  {journal} {\prl}\ }\textbf {\bibinfo {volume} {108}},\
  \bibinfo {eid} {091101} (\bibinfo {year} {2012})}\BibitemShut {NoStop}%
\bibitem [{\citenamefont {{Mooley}}\ \emph {et~al.}(2018)\citenamefont
  {{Mooley}}, \citenamefont {{Deller}}, \citenamefont {{Gottlieb}},
  \citenamefont {{Nakar}}, \citenamefont {{Hallinan}}, \citenamefont
  {{Bourke}}, \citenamefont {{Frail}}, \citenamefont {{Horesh}}, \citenamefont
  {{Corsi}},\ and\ \citenamefont {{Hotokezaka}}}]{Mooley_etal:2018}%
  \BibitemOpen
  \bibfield  {author} {\bibinfo {author} {\bibfnamefont {K.~P.}\ \bibnamefont
  {{Mooley}}}, \bibinfo {author} {\bibfnamefont {A.~T.}\ \bibnamefont
  {{Deller}}}, \bibinfo {author} {\bibfnamefont {O.}~\bibnamefont
  {{Gottlieb}}}, \bibinfo {author} {\bibfnamefont {E.}~\bibnamefont {{Nakar}}},
  \bibinfo {author} {\bibfnamefont {G.}~\bibnamefont {{Hallinan}}}, \bibinfo
  {author} {\bibfnamefont {S.}~\bibnamefont {{Bourke}}}, \bibinfo {author}
  {\bibfnamefont {D.~A.}\ \bibnamefont {{Frail}}}, \bibinfo {author}
  {\bibfnamefont {A.}~\bibnamefont {{Horesh}}}, \bibinfo {author}
  {\bibfnamefont {A.}~\bibnamefont {{Corsi}}}, \ and\ \bibinfo {author}
  {\bibfnamefont {K.}~\bibnamefont {{Hotokezaka}}},\ }\href {\doibase
  10.1038/s41586-018-0486-3} {\bibfield  {journal} {\bibinfo  {journal} {\nat}\
  }\textbf {\bibinfo {volume} {561}},\ \bibinfo {pages} {355} (\bibinfo {year}
  {2018})}\BibitemShut {NoStop}%
\bibitem [{\citenamefont {{Guidorzi}}\ \emph {et~al.}(2017)\citenamefont
  {{Guidorzi}}, \citenamefont {{Margutti}}, \citenamefont {{Brout}},
  \citenamefont {{Scolnic}}, \citenamefont {{Fong}}, \citenamefont
  {{Alexander}}, \citenamefont {{Cowperthwaite}}, \citenamefont {{Annis}},
  \citenamefont {{Berger}}, \citenamefont {{Blanchard}}, \citenamefont
  {{Chornock}}, \citenamefont {{Coppejans}}, \citenamefont {{Eftekhari}},
  \citenamefont {{Frieman}}, \citenamefont {{Huterer}}, \citenamefont
  {{Nicholl}}, \citenamefont {{Soares-Santos}}, \citenamefont {{Terreran}},
  \citenamefont {{Villar}},\ and\ \citenamefont
  {{Williams}}}]{Guidorzi_etal:2017}%
  \BibitemOpen
  \bibfield  {author} {\bibinfo {author} {\bibfnamefont {C.}~\bibnamefont
  {{Guidorzi}}}, \bibinfo {author} {\bibfnamefont {R.}~\bibnamefont
  {{Margutti}}}, \bibinfo {author} {\bibfnamefont {D.}~\bibnamefont {{Brout}}},
  \bibinfo {author} {\bibfnamefont {D.}~\bibnamefont {{Scolnic}}}, \bibinfo
  {author} {\bibfnamefont {W.}~\bibnamefont {{Fong}}}, \bibinfo {author}
  {\bibfnamefont {K.~D.}\ \bibnamefont {{Alexander}}}, \bibinfo {author}
  {\bibfnamefont {P.~S.}\ \bibnamefont {{Cowperthwaite}}}, \bibinfo {author}
  {\bibfnamefont {J.}~\bibnamefont {{Annis}}}, \bibinfo {author} {\bibfnamefont
  {E.}~\bibnamefont {{Berger}}}, \bibinfo {author} {\bibfnamefont {P.~K.}\
  \bibnamefont {{Blanchard}}}, \bibinfo {author} {\bibfnamefont
  {R.}~\bibnamefont {{Chornock}}}, \bibinfo {author} {\bibfnamefont {D.~L.}\
  \bibnamefont {{Coppejans}}}, \bibinfo {author} {\bibfnamefont
  {T.}~\bibnamefont {{Eftekhari}}}, \bibinfo {author} {\bibfnamefont {J.~A.}\
  \bibnamefont {{Frieman}}}, \bibinfo {author} {\bibfnamefont {D.}~\bibnamefont
  {{Huterer}}}, \bibinfo {author} {\bibfnamefont {M.}~\bibnamefont
  {{Nicholl}}}, \bibinfo {author} {\bibfnamefont {M.}~\bibnamefont
  {{Soares-Santos}}}, \bibinfo {author} {\bibfnamefont {G.}~\bibnamefont
  {{Terreran}}}, \bibinfo {author} {\bibfnamefont {V.~A.}\ \bibnamefont
  {{Villar}}}, \ and\ \bibinfo {author} {\bibfnamefont {P.~K.~G.}\ \bibnamefont
  {{Williams}}},\ }\href {\doibase 10.3847/2041-8213/aaa009} {\bibfield
  {journal} {\bibinfo  {journal} {\apjl}\ }\textbf {\bibinfo {volume} {851}},\
  \bibinfo {eid} {L36} (\bibinfo {year} {2017})}\BibitemShut {NoStop}%
\bibitem [{\citenamefont {{Hotokezaka}}\ \emph {et~al.}(2019)\citenamefont
  {{Hotokezaka}}, \citenamefont {{Nakar}}, \citenamefont {{Gottlieb}},
  \citenamefont {{Nissanke}}, \citenamefont {{Masuda}}, \citenamefont
  {{Hallinan}}, \citenamefont {{Mooley}},\ and\ \citenamefont
  {{Deller}}}]{Hotokezaka_etal:2018}%
  \BibitemOpen
  \bibfield  {author} {\bibinfo {author} {\bibfnamefont {K.}~\bibnamefont
  {{Hotokezaka}}}, \bibinfo {author} {\bibfnamefont {E.}~\bibnamefont
  {{Nakar}}}, \bibinfo {author} {\bibfnamefont {O.}~\bibnamefont {{Gottlieb}}},
  \bibinfo {author} {\bibfnamefont {S.}~\bibnamefont {{Nissanke}}}, \bibinfo
  {author} {\bibfnamefont {K.}~\bibnamefont {{Masuda}}}, \bibinfo {author}
  {\bibfnamefont {G.}~\bibnamefont {{Hallinan}}}, \bibinfo {author}
  {\bibfnamefont {K.~P.}\ \bibnamefont {{Mooley}}}, \ and\ \bibinfo {author}
  {\bibfnamefont {A.~T.}\ \bibnamefont {{Deller}}},\ }\href {\doibase
  10.1038/s41550-019-0820-1} {\bibfield  {journal} {\bibinfo  {journal} {Nature
  Astronomy}\ ,\ \bibinfo {pages} {385}} (\bibinfo {year} {2019})}\BibitemShut
  {NoStop}%
\bibitem [{\citenamefont {{The LIGO Scientific Collaboration}}\ \emph
  {et~al.}(2018{\natexlab{b}})\citenamefont {{The LIGO Scientific
  Collaboration}}, \citenamefont {{the Virgo Collaboration}} \emph
  {et~al.}}]{Abbott_etal:2018}%
  \BibitemOpen
  \bibfield  {author} {\bibinfo {author} {\bibnamefont {{The LIGO Scientific
  Collaboration}}}, \bibinfo {author} {\bibnamefont {{the Virgo
  Collaboration}}},  \emph {et~al.},\ }\href@noop {} {\bibfield  {journal}
  {\bibinfo  {journal} {arXiv e-prints}\ } (\bibinfo {year}
  {2018}{\natexlab{b}})},\ \Eprint {http://arxiv.org/abs/1811.12940}
  {arXiv:1811.12940 [astro-ph.HE]} \BibitemShut {NoStop}%
\bibitem [{\citenamefont {{Nair}}\ \emph {et~al.}(2018)\citenamefont {{Nair}},
  \citenamefont {{Bose}},\ and\ \citenamefont {{Saini}}}]{Nair:2018}%
  \BibitemOpen
  \bibfield  {author} {\bibinfo {author} {\bibfnamefont {R.}~\bibnamefont
  {{Nair}}}, \bibinfo {author} {\bibfnamefont {S.}~\bibnamefont {{Bose}}}, \
  and\ \bibinfo {author} {\bibfnamefont {T.~D.}\ \bibnamefont {{Saini}}},\
  }\href {\doibase 10.1103/PhysRevD.98.023502} {\bibfield  {journal} {\bibinfo
  {journal} {\prd}\ }\textbf {\bibinfo {volume} {98}},\ \bibinfo {eid} {023502}
  (\bibinfo {year} {2018})}\BibitemShut {NoStop}%
\bibitem [{\citenamefont {{Gray}}\ \emph {et~al.}(2019)\citenamefont {{Gray}},
  \citenamefont {{Maga{\~n}a Hernandez}}, \citenamefont {{Qi}}, \citenamefont
  {{Sur}}, \citenamefont {{Brady}}, \citenamefont {{Chen}}, \citenamefont
  {{Farr}}, \citenamefont {{Fishbach}}, \citenamefont {{Gair}}, \citenamefont
  {{Ghosh}}, \citenamefont {{Holz}}, \citenamefont {{Mastrogiovanni}},
  \citenamefont {{Messenger}}, \citenamefont {{Steer}},\ and\ \citenamefont
  {{Veitch}}}]{Gray_etal:2019}%
  \BibitemOpen
  \bibfield  {author} {\bibinfo {author} {\bibfnamefont {R.}~\bibnamefont
  {{Gray}}}, \bibinfo {author} {\bibfnamefont {I.}~\bibnamefont {{Maga{\~n}a
  Hernandez}}}, \bibinfo {author} {\bibfnamefont {H.}~\bibnamefont {{Qi}}},
  \bibinfo {author} {\bibfnamefont {A.}~\bibnamefont {{Sur}}}, \bibinfo
  {author} {\bibfnamefont {P.~R.}\ \bibnamefont {{Brady}}}, \bibinfo {author}
  {\bibfnamefont {H.-Y.}\ \bibnamefont {{Chen}}}, \bibinfo {author}
  {\bibfnamefont {W.~M.}\ \bibnamefont {{Farr}}}, \bibinfo {author}
  {\bibfnamefont {M.}~\bibnamefont {{Fishbach}}}, \bibinfo {author}
  {\bibfnamefont {J.~R.}\ \bibnamefont {{Gair}}}, \bibinfo {author}
  {\bibfnamefont {A.}~\bibnamefont {{Ghosh}}}, \bibinfo {author} {\bibfnamefont
  {D.~E.}\ \bibnamefont {{Holz}}}, \bibinfo {author} {\bibfnamefont
  {S.}~\bibnamefont {{Mastrogiovanni}}}, \bibinfo {author} {\bibfnamefont
  {C.}~\bibnamefont {{Messenger}}}, \bibinfo {author} {\bibfnamefont {D.~A.}\
  \bibnamefont {{Steer}}}, \ and\ \bibinfo {author} {\bibfnamefont
  {J.}~\bibnamefont {{Veitch}}},\ }\href@noop {} {\bibfield  {journal}
  {\bibinfo  {journal} {arXiv e-prints}\ } (\bibinfo {year} {2019})},\ \Eprint
  {http://arxiv.org/abs/1908.06050} {arXiv:1908.06050 [gr-qc]} \BibitemShut
  {NoStop}%
\bibitem [{\citenamefont {{Soares-Santos}}\ \emph {et~al.}(2019)\citenamefont
  {{Soares-Santos}}, \citenamefont {{Palmese}}, \citenamefont {{Hartley}},
  \citenamefont {{Annis}}, \citenamefont {{Garcia-Bellido}}, \citenamefont
  {{Lahav}}, \citenamefont {{Doctor}}, \citenamefont {{Fishbach}},
  \citenamefont {{Holz}},\ and\ \citenamefont
  {{Lin}}}]{Soares-Santos_etal:2019}%
  \BibitemOpen
  \bibfield  {author} {\bibinfo {author} {\bibfnamefont {M.}~\bibnamefont
  {{Soares-Santos}}}, \bibinfo {author} {\bibfnamefont {A.}~\bibnamefont
  {{Palmese}}}, \bibinfo {author} {\bibfnamefont {W.}~\bibnamefont
  {{Hartley}}}, \bibinfo {author} {\bibfnamefont {J.}~\bibnamefont {{Annis}}},
  \bibinfo {author} {\bibfnamefont {J.}~\bibnamefont {{Garcia-Bellido}}},
  \bibinfo {author} {\bibfnamefont {O.}~\bibnamefont {{Lahav}}}, \bibinfo
  {author} {\bibfnamefont {Z.}~\bibnamefont {{Doctor}}}, \bibinfo {author}
  {\bibfnamefont {M.}~\bibnamefont {{Fishbach}}}, \bibinfo {author}
  {\bibfnamefont {D.~E.}\ \bibnamefont {{Holz}}}, \ and\ \bibinfo {author}
  {\bibfnamefont {H.}~\bibnamefont {{Lin}}},\ }\href {\doibase
  10.3847/2041-8213/ab14f1} {\bibfield  {journal} {\bibinfo  {journal} {\apjl}\
  }\textbf {\bibinfo {volume} {876}},\ \bibinfo {eid} {L7} (\bibinfo {year}
  {2019})}\BibitemShut {NoStop}%
\bibitem [{\citenamefont {{The LIGO Scientific Collaboration}}\ \emph
  {et~al.}(2019)\citenamefont {{The LIGO Scientific Collaboration}},
  \citenamefont {{the Virgo Collaboration}} \emph {et~al.}}]{Ligo:2019}%
  \BibitemOpen
  \bibfield  {author} {\bibinfo {author} {\bibnamefont {{The LIGO Scientific
  Collaboration}}}, \bibinfo {author} {\bibnamefont {{the Virgo
  Collaboration}}},  \emph {et~al.},\ }\href@noop {} {\bibfield  {journal}
  {\bibinfo  {journal} {arXiv e-prints}\ } (\bibinfo {year} {2019})},\ \Eprint
  {http://arxiv.org/abs/1908.06060} {arXiv:1908.06060 [astro-ph.CO]}
  \BibitemShut {NoStop}%
\bibitem [{\citenamefont {{Nissanke}}\ \emph
  {et~al.}(2013{\natexlab{a}})\citenamefont {{Nissanke}}, \citenamefont
  {{Holz}}, \citenamefont {{Dalal}}, \citenamefont {{Hughes}}, \citenamefont
  {{Sievers}},\ and\ \citenamefont {{Hirata}}}]{Nissanke_etal:2013}%
  \BibitemOpen
  \bibfield  {author} {\bibinfo {author} {\bibfnamefont {S.}~\bibnamefont
  {{Nissanke}}}, \bibinfo {author} {\bibfnamefont {D.~E.}\ \bibnamefont
  {{Holz}}}, \bibinfo {author} {\bibfnamefont {N.}~\bibnamefont {{Dalal}}},
  \bibinfo {author} {\bibfnamefont {S.~A.}\ \bibnamefont {{Hughes}}}, \bibinfo
  {author} {\bibfnamefont {J.~L.}\ \bibnamefont {{Sievers}}}, \ and\ \bibinfo
  {author} {\bibfnamefont {C.~M.}\ \bibnamefont {{Hirata}}},\ }\href@noop {}
  {\bibfield  {journal} {\bibinfo  {journal} {ArXiv e-prints}\ } (\bibinfo
  {year} {2013}{\natexlab{a}})},\ \Eprint {http://arxiv.org/abs/1307.2638}
  {arXiv:1307.2638} \BibitemShut {NoStop}%
\bibitem [{\citenamefont {{Vitale}}\ and\ \citenamefont
  {{Chen}}(2018)}]{Vitale_Chen:2018}%
  \BibitemOpen
  \bibfield  {author} {\bibinfo {author} {\bibfnamefont {S.}~\bibnamefont
  {{Vitale}}}\ and\ \bibinfo {author} {\bibfnamefont {H.-Y.}\ \bibnamefont
  {{Chen}}},\ }\href {\doibase 10.1103/PhysRevLett.121.021303} {\bibfield
  {journal} {\bibinfo  {journal} {\prl}\ }\textbf {\bibinfo {volume} {121}},\
  \bibinfo {eid} {021303} (\bibinfo {year} {2018})}\BibitemShut {NoStop}%
\bibitem [{\citenamefont {{Seto}}\ and\ \citenamefont
  {{Kyutoku}}(2018)}]{Seto:2018}%
  \BibitemOpen
  \bibfield  {author} {\bibinfo {author} {\bibfnamefont {N.}~\bibnamefont
  {{Seto}}}\ and\ \bibinfo {author} {\bibfnamefont {K.}~\bibnamefont
  {{Kyutoku}}},\ }\href {\doibase 10.1093/mnras/sty090} {\bibfield  {journal}
  {\bibinfo  {journal} {\mnras}\ }\textbf {\bibinfo {volume} {475}},\ \bibinfo
  {pages} {4133} (\bibinfo {year} {2018})}\BibitemShut {NoStop}%
\bibitem [{\citenamefont {{Chen}}\ \emph {et~al.}(2018)\citenamefont {{Chen}},
  \citenamefont {{Fishbach}},\ and\ \citenamefont {{Holz}}}]{Chen_etal:2018}%
  \BibitemOpen
  \bibfield  {author} {\bibinfo {author} {\bibfnamefont {H.-Y.}\ \bibnamefont
  {{Chen}}}, \bibinfo {author} {\bibfnamefont {M.}~\bibnamefont {{Fishbach}}},
  \ and\ \bibinfo {author} {\bibfnamefont {D.}~\bibnamefont {{Holz}}},\ }\href
  {\doibase something} {\bibfield  {journal} {\bibinfo  {journal} {\nat}\
  }\textbf {\bibinfo {volume} {562}},\ \bibinfo {pages} {545} (\bibinfo {year}
  {2018})}\BibitemShut {NoStop}%
\bibitem [{\citenamefont {{Feeney}}\ \emph {et~al.}(2019)\citenamefont
  {{Feeney}}, \citenamefont {{Peiris}}, \citenamefont {{Williamson}},
  \citenamefont {{Nissanke}}, \citenamefont {{Mortlock}}, \citenamefont
  {{Alsing}},\ and\ \citenamefont {{Scolnic}}}]{Feeney_etal:2018}%
  \BibitemOpen
  \bibfield  {author} {\bibinfo {author} {\bibfnamefont {S.~M.}\ \bibnamefont
  {{Feeney}}}, \bibinfo {author} {\bibfnamefont {H.~V.}\ \bibnamefont
  {{Peiris}}}, \bibinfo {author} {\bibfnamefont {A.~R.}\ \bibnamefont
  {{Williamson}}}, \bibinfo {author} {\bibfnamefont {S.~M.}\ \bibnamefont
  {{Nissanke}}}, \bibinfo {author} {\bibfnamefont {D.~J.}\ \bibnamefont
  {{Mortlock}}}, \bibinfo {author} {\bibfnamefont {J.}~\bibnamefont
  {{Alsing}}}, \ and\ \bibinfo {author} {\bibfnamefont {D.}~\bibnamefont
  {{Scolnic}}},\ }\href {\doibase 10.1103/PhysRevLett.122.061105} {\bibfield
  {journal} {\bibinfo  {journal} {\prl}\ }\textbf {\bibinfo {volume} {122}},\
  \bibinfo {eid} {061105} (\bibinfo {year} {2019})}\BibitemShut {NoStop}%
\bibitem [{\citenamefont {{Taylor}}\ \emph {et~al.}(2012)\citenamefont
  {{Taylor}}, \citenamefont {{Gair}},\ and\ \citenamefont
  {{Mandel}}}]{Taylor_etal:2012}%
  \BibitemOpen
  \bibfield  {author} {\bibinfo {author} {\bibfnamefont {S.~R.}\ \bibnamefont
  {{Taylor}}}, \bibinfo {author} {\bibfnamefont {J.~R.}\ \bibnamefont
  {{Gair}}}, \ and\ \bibinfo {author} {\bibfnamefont {I.}~\bibnamefont
  {{Mandel}}},\ }\href {\doibase 10.1103/PhysRevD.85.023535} {\bibfield
  {journal} {\bibinfo  {journal} {\prd}\ }\textbf {\bibinfo {volume} {85}},\
  \bibinfo {eid} {023535} (\bibinfo {year} {2012})}\BibitemShut {NoStop}%
\bibitem [{\citenamefont {{Messenger}}\ and\ \citenamefont
  {{Veitch}}(2013)}]{Messenger:2013}%
  \BibitemOpen
  \bibfield  {author} {\bibinfo {author} {\bibfnamefont {C.}~\bibnamefont
  {{Messenger}}}\ and\ \bibinfo {author} {\bibfnamefont {J.}~\bibnamefont
  {{Veitch}}},\ }\href {\doibase 10.1088/1367-2630/15/5/053027} {\bibfield
  {journal} {\bibinfo  {journal} {New Journal of Physics}\ }\textbf {\bibinfo
  {volume} {15}},\ \bibinfo {eid} {053027} (\bibinfo {year}
  {2013})}\BibitemShut {NoStop}%
\bibitem [{\citenamefont {{Karki}}\ \emph {et~al.}(2016)\citenamefont
  {{Karki}}, \citenamefont {{Tuyenbayev}}, \citenamefont {{Kandhasamy}},
  \citenamefont {{Abbott}}, \citenamefont {{Abbott}}, \citenamefont {{Anders}},
  \citenamefont {{Berliner}}, \citenamefont {{Betzwieser}}, \citenamefont
  {{Cahillane}}, \citenamefont {{Canete}}, \citenamefont {{Conley}},
  \citenamefont {{Daveloza}}, \citenamefont {{De Lillo}}, \citenamefont
  {{Gleason}}, \citenamefont {{Goetz}}, \citenamefont {{Izumi}}, \citenamefont
  {{Kissel}}, \citenamefont {{Mendell}}, \citenamefont {{Quetschke}},
  \citenamefont {{Rodruck}}, \citenamefont {{Sachdev}}, \citenamefont
  {{Sadecki}}, \citenamefont {{Schwinberg}}, \citenamefont {{Sottile}},
  \citenamefont {{Wade}}, \citenamefont {{Weinstein}}, \citenamefont {{West}},\
  and\ \citenamefont {{Savage}}}]{Karki_etal:2016}%
  \BibitemOpen
  \bibfield  {author} {\bibinfo {author} {\bibfnamefont {S.}~\bibnamefont
  {{Karki}}}, \bibinfo {author} {\bibfnamefont {D.}~\bibnamefont
  {{Tuyenbayev}}}, \bibinfo {author} {\bibfnamefont {S.}~\bibnamefont
  {{Kandhasamy}}}, \bibinfo {author} {\bibfnamefont {B.~P.}\ \bibnamefont
  {{Abbott}}}, \bibinfo {author} {\bibfnamefont {T.~D.}\ \bibnamefont
  {{Abbott}}}, \bibinfo {author} {\bibfnamefont {E.~H.}\ \bibnamefont
  {{Anders}}}, \bibinfo {author} {\bibfnamefont {J.}~\bibnamefont
  {{Berliner}}}, \bibinfo {author} {\bibfnamefont {J.}~\bibnamefont
  {{Betzwieser}}}, \bibinfo {author} {\bibfnamefont {C.}~\bibnamefont
  {{Cahillane}}}, \bibinfo {author} {\bibfnamefont {L.}~\bibnamefont
  {{Canete}}}, \bibinfo {author} {\bibfnamefont {C.}~\bibnamefont {{Conley}}},
  \bibinfo {author} {\bibfnamefont {H.~P.}\ \bibnamefont {{Daveloza}}},
  \bibinfo {author} {\bibfnamefont {N.}~\bibnamefont {{De Lillo}}}, \bibinfo
  {author} {\bibfnamefont {J.~R.}\ \bibnamefont {{Gleason}}}, \bibinfo {author}
  {\bibfnamefont {E.}~\bibnamefont {{Goetz}}}, \bibinfo {author} {\bibfnamefont
  {K.}~\bibnamefont {{Izumi}}}, \bibinfo {author} {\bibfnamefont {J.~S.}\
  \bibnamefont {{Kissel}}}, \bibinfo {author} {\bibfnamefont {G.}~\bibnamefont
  {{Mendell}}}, \bibinfo {author} {\bibfnamefont {V.}~\bibnamefont
  {{Quetschke}}}, \bibinfo {author} {\bibfnamefont {M.}~\bibnamefont
  {{Rodruck}}}, \bibinfo {author} {\bibfnamefont {S.}~\bibnamefont
  {{Sachdev}}}, \bibinfo {author} {\bibfnamefont {T.}~\bibnamefont
  {{Sadecki}}}, \bibinfo {author} {\bibfnamefont {P.~B.}\ \bibnamefont
  {{Schwinberg}}}, \bibinfo {author} {\bibfnamefont {A.}~\bibnamefont
  {{Sottile}}}, \bibinfo {author} {\bibfnamefont {M.}~\bibnamefont {{Wade}}},
  \bibinfo {author} {\bibfnamefont {A.~J.}\ \bibnamefont {{Weinstein}}},
  \bibinfo {author} {\bibfnamefont {M.}~\bibnamefont {{West}}}, \ and\ \bibinfo
  {author} {\bibfnamefont {R.~L.}\ \bibnamefont {{Savage}}},\ }\href {\doibase
  10.1063/1.4967303} {\bibfield  {journal} {\bibinfo  {journal} {Review of
  Scientific Instruments}\ }\textbf {\bibinfo {volume} {87}},\ \bibinfo {eid}
  {114503} (\bibinfo {year} {2016})}\BibitemShut {NoStop}%
\bibitem [{\citenamefont {{Cahillane}}\ \emph {et~al.}(2017)\citenamefont
  {{Cahillane}}, \citenamefont {{Betzwieser}}, \citenamefont {{Brown}},
  \citenamefont {{Goetz}}, \citenamefont {{Hall}}, \citenamefont {{Izumi}},
  \citenamefont {{Kand hasamy}}, \citenamefont {{Karki}}, \citenamefont
  {{Kissel}}, \citenamefont {{Mendell}}, \citenamefont {{Savage}},
  \citenamefont {{Tuyenbayev}}, \citenamefont {{Urban}}, \citenamefont
  {{Viets}}, \citenamefont {{Wade}},\ and\ \citenamefont
  {{Weinstein}}}]{Cahillane_etal:2017}%
  \BibitemOpen
  \bibfield  {author} {\bibinfo {author} {\bibfnamefont {C.}~\bibnamefont
  {{Cahillane}}}, \bibinfo {author} {\bibfnamefont {J.}~\bibnamefont
  {{Betzwieser}}}, \bibinfo {author} {\bibfnamefont {D.~A.}\ \bibnamefont
  {{Brown}}}, \bibinfo {author} {\bibfnamefont {E.}~\bibnamefont {{Goetz}}},
  \bibinfo {author} {\bibfnamefont {E.~D.}\ \bibnamefont {{Hall}}}, \bibinfo
  {author} {\bibfnamefont {K.}~\bibnamefont {{Izumi}}}, \bibinfo {author}
  {\bibfnamefont {S.}~\bibnamefont {{Kand hasamy}}}, \bibinfo {author}
  {\bibfnamefont {S.}~\bibnamefont {{Karki}}}, \bibinfo {author} {\bibfnamefont
  {J.~S.}\ \bibnamefont {{Kissel}}}, \bibinfo {author} {\bibfnamefont
  {G.}~\bibnamefont {{Mendell}}}, \bibinfo {author} {\bibfnamefont {R.~L.}\
  \bibnamefont {{Savage}}}, \bibinfo {author} {\bibfnamefont {D.}~\bibnamefont
  {{Tuyenbayev}}}, \bibinfo {author} {\bibfnamefont {A.}~\bibnamefont
  {{Urban}}}, \bibinfo {author} {\bibfnamefont {A.}~\bibnamefont {{Viets}}},
  \bibinfo {author} {\bibfnamefont {M.}~\bibnamefont {{Wade}}}, \ and\ \bibinfo
  {author} {\bibfnamefont {A.~J.}\ \bibnamefont {{Weinstein}}},\ }\href
  {\doibase 10.1103/PhysRevD.96.102001} {\bibfield  {journal} {\bibinfo
  {journal} {\prd}\ }\textbf {\bibinfo {volume} {96}},\ \bibinfo {eid} {102001}
  (\bibinfo {year} {2017})}\BibitemShut {NoStop}%
\bibitem [{\citenamefont {{Abbott}}\ \emph {et~al.}(2016)\citenamefont
  {{Abbott}}, \citenamefont {{Abbott}}, \citenamefont {{Abbott}}, \citenamefont
  {{Abernathy}}, \citenamefont {{Acernese}}, \citenamefont {{Ackley}},
  \citenamefont {{Adams}}, \citenamefont {{Adams}}, \citenamefont {{Addesso}},
  \citenamefont {{Adhikari}},\ and\ \citenamefont {et~al.}}]{Abbott_etal:2016}%
  \BibitemOpen
  \bibfield  {author} {\bibinfo {author} {\bibfnamefont {B.~P.}\ \bibnamefont
  {{Abbott}}}, \bibinfo {author} {\bibfnamefont {R.}~\bibnamefont {{Abbott}}},
  \bibinfo {author} {\bibfnamefont {T.~D.}\ \bibnamefont {{Abbott}}}, \bibinfo
  {author} {\bibfnamefont {M.~R.}\ \bibnamefont {{Abernathy}}}, \bibinfo
  {author} {\bibfnamefont {F.}~\bibnamefont {{Acernese}}}, \bibinfo {author}
  {\bibfnamefont {K.}~\bibnamefont {{Ackley}}}, \bibinfo {author}
  {\bibfnamefont {C.}~\bibnamefont {{Adams}}}, \bibinfo {author} {\bibfnamefont
  {T.}~\bibnamefont {{Adams}}}, \bibinfo {author} {\bibfnamefont
  {P.}~\bibnamefont {{Addesso}}}, \bibinfo {author} {\bibfnamefont {R.~X.}\
  \bibnamefont {{Adhikari}}}, \ and\ \bibinfo {author} {\bibnamefont
  {et~al.}},\ }\href {\doibase 10.1103/PhysRevX.6.041015} {\bibfield  {journal}
  {\bibinfo  {journal} {Physical Review X}\ }\textbf {\bibinfo {volume} {6}},\
  \bibinfo {eid} {041015} (\bibinfo {year} {2016})}\BibitemShut {NoStop}%
\bibitem [{\citenamefont {{Mandel}}\ \emph {et~al.}(2019)\citenamefont
  {{Mandel}}, \citenamefont {{Farr}},\ and\ \citenamefont
  {{Gair}}}]{Mandel_etal:2018}%
  \BibitemOpen
  \bibfield  {author} {\bibinfo {author} {\bibfnamefont {I.}~\bibnamefont
  {{Mandel}}}, \bibinfo {author} {\bibfnamefont {W.~M.}\ \bibnamefont
  {{Farr}}}, \ and\ \bibinfo {author} {\bibfnamefont {J.~R.}\ \bibnamefont
  {{Gair}}},\ }\href {\doibase 10.1093/mnras/stz896} {\bibfield  {journal}
  {\bibinfo  {journal} {\mnras}\ }\textbf {\bibinfo {volume} {486}},\ \bibinfo
  {pages} {1086} (\bibinfo {year} {2019})}\BibitemShut {NoStop}%
\bibitem [{\citenamefont {{Burstein}}(1990)}]{Burstein:1990}%
  \BibitemOpen
  \bibfield  {author} {\bibinfo {author} {\bibfnamefont {D.}~\bibnamefont
  {{Burstein}}},\ }\href {\doibase 10.1088/0034-4885/53/4/002} {\bibfield
  {journal} {\bibinfo  {journal} {Reports on Progress in Physics}\ }\textbf
  {\bibinfo {volume} {53}},\ \bibinfo {pages} {421} (\bibinfo {year}
  {1990})}\BibitemShut {NoStop}%
\bibitem [{\citenamefont {{Cutler}}\ and\ \citenamefont
  {{Flanagan}}(1994)}]{Cutler_Flanagan:1994}%
  \BibitemOpen
  \bibfield  {author} {\bibinfo {author} {\bibfnamefont {C.}~\bibnamefont
  {{Cutler}}}\ and\ \bibinfo {author} {\bibfnamefont {{\'E}.~E.}\ \bibnamefont
  {{Flanagan}}},\ }\href {\doibase 10.1103/PhysRevD.49.2658} {\bibfield
  {journal} {\bibinfo  {journal} {\prd}\ }\textbf {\bibinfo {volume} {49}},\
  \bibinfo {pages} {2658} (\bibinfo {year} {1994})}\BibitemShut {NoStop}%
\bibitem [{\citenamefont {{Markovi{\'c}}}(1993)}]{Markovic:1993}%
  \BibitemOpen
  \bibfield  {author} {\bibinfo {author} {\bibfnamefont {D.}~\bibnamefont
  {{Markovi{\'c}}}},\ }\href {\doibase 10.1103/PhysRevD.48.4738} {\bibfield
  {journal} {\bibinfo  {journal} {\prd}\ }\textbf {\bibinfo {volume} {48}},\
  \bibinfo {pages} {4738} (\bibinfo {year} {1993})}\BibitemShut {NoStop}%
\bibitem [{\citenamefont {{Veitch}}\ \emph {et~al.}(2015)\citenamefont
  {{Veitch}}, \citenamefont {{Raymond}}, \citenamefont {{Farr}}, \citenamefont
  {{Farr}}, \citenamefont {{Graff}}, \citenamefont {{Vitale}}, \citenamefont
  {{Aylott}}, \citenamefont {{Blackburn}}, \citenamefont {{Christensen}},
  \citenamefont {{Coughlin}}, \citenamefont {{Del Pozzo}}, \citenamefont
  {{Feroz}}, \citenamefont {{Gair}}, \citenamefont {{Haster}}, \citenamefont
  {{Kalogera}}, \citenamefont {{Littenberg}}, \citenamefont {{Mandel}},
  \citenamefont {{O'Shaughnessy}}, \citenamefont {{Pitkin}}, \citenamefont
  {{Rodriguez}}, \citenamefont {{R{\"o}ver}}, \citenamefont {{Sidery}},
  \citenamefont {{Smith}}, \citenamefont {{van Der Sluys}}, \citenamefont
  {{Vecchio}}, \citenamefont {{Vousden}},\ and\ \citenamefont
  {{Wade}}}]{Veitch:2014wba}%
  \BibitemOpen
  \bibfield  {author} {\bibinfo {author} {\bibfnamefont {J.}~\bibnamefont
  {{Veitch}}}, \bibinfo {author} {\bibfnamefont {V.}~\bibnamefont {{Raymond}}},
  \bibinfo {author} {\bibfnamefont {B.}~\bibnamefont {{Farr}}}, \bibinfo
  {author} {\bibfnamefont {W.}~\bibnamefont {{Farr}}}, \bibinfo {author}
  {\bibfnamefont {P.}~\bibnamefont {{Graff}}}, \bibinfo {author} {\bibfnamefont
  {S.}~\bibnamefont {{Vitale}}}, \bibinfo {author} {\bibfnamefont
  {B.}~\bibnamefont {{Aylott}}}, \bibinfo {author} {\bibfnamefont
  {K.}~\bibnamefont {{Blackburn}}}, \bibinfo {author} {\bibfnamefont
  {N.}~\bibnamefont {{Christensen}}}, \bibinfo {author} {\bibfnamefont
  {M.}~\bibnamefont {{Coughlin}}}, \bibinfo {author} {\bibfnamefont
  {W.}~\bibnamefont {{Del Pozzo}}}, \bibinfo {author} {\bibfnamefont
  {F.}~\bibnamefont {{Feroz}}}, \bibinfo {author} {\bibfnamefont
  {J.}~\bibnamefont {{Gair}}}, \bibinfo {author} {\bibfnamefont {C.~J.}\
  \bibnamefont {{Haster}}}, \bibinfo {author} {\bibfnamefont {V.}~\bibnamefont
  {{Kalogera}}}, \bibinfo {author} {\bibfnamefont {T.}~\bibnamefont
  {{Littenberg}}}, \bibinfo {author} {\bibfnamefont {I.}~\bibnamefont
  {{Mandel}}}, \bibinfo {author} {\bibfnamefont {R.}~\bibnamefont
  {{O'Shaughnessy}}}, \bibinfo {author} {\bibfnamefont {M.}~\bibnamefont
  {{Pitkin}}}, \bibinfo {author} {\bibfnamefont {C.}~\bibnamefont
  {{Rodriguez}}}, \bibinfo {author} {\bibfnamefont {C.}~\bibnamefont
  {{R{\"o}ver}}}, \bibinfo {author} {\bibfnamefont {T.}~\bibnamefont
  {{Sidery}}}, \bibinfo {author} {\bibfnamefont {R.}~\bibnamefont {{Smith}}},
  \bibinfo {author} {\bibfnamefont {M.}~\bibnamefont {{van Der Sluys}}},
  \bibinfo {author} {\bibfnamefont {A.}~\bibnamefont {{Vecchio}}}, \bibinfo
  {author} {\bibfnamefont {W.}~\bibnamefont {{Vousden}}}, \ and\ \bibinfo
  {author} {\bibfnamefont {L.}~\bibnamefont {{Wade}}},\ }\href {\doibase
  10.1103/PhysRevD.91.042003} {\bibfield  {journal} {\bibinfo  {journal}
  {\prd}\ }\textbf {\bibinfo {volume} {91}},\ \bibinfo {eid} {042003} (\bibinfo
  {year} {2015})}\BibitemShut {NoStop}%
\bibitem [{\citenamefont {{Ashton}}\ \emph {et~al.}(2019)\citenamefont
  {{Ashton}}, \citenamefont {{H{\"u}bner}}, \citenamefont {{Lasky}},
  \citenamefont {{Talbot}}, \citenamefont {{Ackley}}, \citenamefont
  {{Biscoveanu}}, \citenamefont {{Chu}}, \citenamefont {{Divakarla}},
  \citenamefont {{Easter}}, \citenamefont {{Goncharov}}, \citenamefont
  {{Hernandez Vivanco}}, \citenamefont {{Harms}}, \citenamefont {{Lower}},
  \citenamefont {{Meadors}}, \citenamefont {{Melchor}}, \citenamefont
  {{Payne}}, \citenamefont {{Pitkin}}, \citenamefont {{Powell}}, \citenamefont
  {{Sarin}}, \citenamefont {{Smith}},\ and\ \citenamefont
  {{Thrane}}}]{Ashton_etal:2019}%
  \BibitemOpen
  \bibfield  {author} {\bibinfo {author} {\bibfnamefont {G.}~\bibnamefont
  {{Ashton}}}, \bibinfo {author} {\bibfnamefont {M.}~\bibnamefont
  {{H{\"u}bner}}}, \bibinfo {author} {\bibfnamefont {P.~D.}\ \bibnamefont
  {{Lasky}}}, \bibinfo {author} {\bibfnamefont {C.}~\bibnamefont {{Talbot}}},
  \bibinfo {author} {\bibfnamefont {K.}~\bibnamefont {{Ackley}}}, \bibinfo
  {author} {\bibfnamefont {S.}~\bibnamefont {{Biscoveanu}}}, \bibinfo {author}
  {\bibfnamefont {Q.}~\bibnamefont {{Chu}}}, \bibinfo {author} {\bibfnamefont
  {A.}~\bibnamefont {{Divakarla}}}, \bibinfo {author} {\bibfnamefont {P.~J.}\
  \bibnamefont {{Easter}}}, \bibinfo {author} {\bibfnamefont {B.}~\bibnamefont
  {{Goncharov}}}, \bibinfo {author} {\bibfnamefont {F.}~\bibnamefont
  {{Hernandez Vivanco}}}, \bibinfo {author} {\bibfnamefont {J.}~\bibnamefont
  {{Harms}}}, \bibinfo {author} {\bibfnamefont {M.~E.}\ \bibnamefont
  {{Lower}}}, \bibinfo {author} {\bibfnamefont {G.~D.}\ \bibnamefont
  {{Meadors}}}, \bibinfo {author} {\bibfnamefont {D.}~\bibnamefont
  {{Melchor}}}, \bibinfo {author} {\bibfnamefont {E.}~\bibnamefont {{Payne}}},
  \bibinfo {author} {\bibfnamefont {M.~D.}\ \bibnamefont {{Pitkin}}}, \bibinfo
  {author} {\bibfnamefont {J.}~\bibnamefont {{Powell}}}, \bibinfo {author}
  {\bibfnamefont {N.}~\bibnamefont {{Sarin}}}, \bibinfo {author} {\bibfnamefont
  {R.~J.~E.}\ \bibnamefont {{Smith}}}, \ and\ \bibinfo {author} {\bibfnamefont
  {E.}~\bibnamefont {{Thrane}}},\ }\href {\doibase 10.3847/1538-4365/ab06fc}
  {\bibfield  {journal} {\bibinfo  {journal} {\apjs}\ }\textbf {\bibinfo
  {volume} {241}},\ \bibinfo {eid} {27} (\bibinfo {year} {2019})}\BibitemShut
  {NoStop}%
\bibitem [{\citenamefont {{Biwer}}\ \emph {et~al.}(2019)\citenamefont
  {{Biwer}}, \citenamefont {{Capano}}, \citenamefont {{De}}, \citenamefont
  {{Cabero}}, \citenamefont {{Brown}}, \citenamefont {{Nitz}},\ and\
  \citenamefont {{Raymond}}}]{Biwer_etal:2019}%
  \BibitemOpen
  \bibfield  {author} {\bibinfo {author} {\bibfnamefont {C.~M.}\ \bibnamefont
  {{Biwer}}}, \bibinfo {author} {\bibfnamefont {C.~D.}\ \bibnamefont
  {{Capano}}}, \bibinfo {author} {\bibfnamefont {S.}~\bibnamefont {{De}}},
  \bibinfo {author} {\bibfnamefont {M.}~\bibnamefont {{Cabero}}}, \bibinfo
  {author} {\bibfnamefont {D.~A.}\ \bibnamefont {{Brown}}}, \bibinfo {author}
  {\bibfnamefont {A.~H.}\ \bibnamefont {{Nitz}}}, \ and\ \bibinfo {author}
  {\bibfnamefont {V.}~\bibnamefont {{Raymond}}},\ }\href {\doibase
  10.1088/1538-3873/aaef0b} {\bibfield  {journal} {\bibinfo  {journal} {\pasp}\
  }\textbf {\bibinfo {volume} {131}},\ \bibinfo {pages} {024503} (\bibinfo
  {year} {2019})}\BibitemShut {NoStop}%
\bibitem [{\citenamefont {{Weinberg}}(2008)}]{Weinberg:2008}%
  \BibitemOpen
  \bibfield  {author} {\bibinfo {author} {\bibfnamefont {S.}~\bibnamefont
  {{Weinberg}}},\ }\href@noop {} {\emph {\bibinfo {title} {Cosmology}}}\
  (\bibinfo  {publisher} {Oxford University Press},\ \bibinfo {year}
  {2008})\BibitemShut {NoStop}%
\bibitem [{\citenamefont {{Nissanke}}\ \emph
  {et~al.}(2013{\natexlab{b}})\citenamefont {{Nissanke}}, \citenamefont
  {{Kasliwal}},\ and\ \citenamefont {{Georgieva}}}]{Nissanke_etal:2013b}%
  \BibitemOpen
  \bibfield  {author} {\bibinfo {author} {\bibfnamefont {S.}~\bibnamefont
  {{Nissanke}}}, \bibinfo {author} {\bibfnamefont {M.}~\bibnamefont
  {{Kasliwal}}}, \ and\ \bibinfo {author} {\bibfnamefont {A.}~\bibnamefont
  {{Georgieva}}},\ }\href {\doibase 10.1088/0004-637X/767/2/124} {\bibfield
  {journal} {\bibinfo  {journal} {\apj}\ }\textbf {\bibinfo {volume} {767}},\
  \bibinfo {eid} {124} (\bibinfo {year} {2013}{\natexlab{b}})},\ \Eprint
  {http://arxiv.org/abs/1210.6362} {arXiv:1210.6362 [astro-ph.HE]} \BibitemShut
  {NoStop}%
\bibitem [{\citenamefont {{LIGO Scientific Collaboration}}(2018)}]{LALSuite}%
  \BibitemOpen
  \bibfield  {author} {\bibinfo {author} {\bibnamefont {{LIGO Scientific
  Collaboration}}},\ }\href {\doibase 10.7935/GT1W-FZ16} {\enquote {\bibinfo
  {title} {{LIGO} {A}lgorithm {L}ibrary - {LALS}uite},}\ }\bibinfo
  {howpublished} {free software (GPL)} (\bibinfo {year} {2018})\BibitemShut
  {NoStop}%
\bibitem [{\citenamefont {{Crook}}\ \emph {et~al.}(2007)\citenamefont
  {{Crook}}, \citenamefont {{Huchra}}, \citenamefont {{Martimbeau}},
  \citenamefont {{Masters}}, \citenamefont {{Jarrett}},\ and\ \citenamefont
  {{Macri}}}]{Crook_etal:2007}%
  \BibitemOpen
  \bibfield  {author} {\bibinfo {author} {\bibfnamefont {A.~C.}\ \bibnamefont
  {{Crook}}}, \bibinfo {author} {\bibfnamefont {J.~P.}\ \bibnamefont
  {{Huchra}}}, \bibinfo {author} {\bibfnamefont {N.}~\bibnamefont
  {{Martimbeau}}}, \bibinfo {author} {\bibfnamefont {K.~L.}\ \bibnamefont
  {{Masters}}}, \bibinfo {author} {\bibfnamefont {T.}~\bibnamefont
  {{Jarrett}}}, \ and\ \bibinfo {author} {\bibfnamefont {L.~M.}\ \bibnamefont
  {{Macri}}},\ }\href {\doibase 10.1086/510201} {\bibfield  {journal} {\bibinfo
   {journal} {\apj}\ }\textbf {\bibinfo {volume} {655}},\ \bibinfo {pages}
  {790} (\bibinfo {year} {2007})}\BibitemShut {NoStop}%
\bibitem [{\citenamefont {{Berger}}(2010)}]{Berger:2010}%
  \BibitemOpen
  \bibfield  {author} {\bibinfo {author} {\bibfnamefont {E.}~\bibnamefont
  {{Berger}}},\ }\href {\doibase 10.1088/0004-637X/722/2/1946} {\bibfield
  {journal} {\bibinfo  {journal} {\apj}\ }\textbf {\bibinfo {volume} {722}},\
  \bibinfo {pages} {1946} (\bibinfo {year} {2010})}\BibitemShut {NoStop}%
\bibitem [{\citenamefont {{Fong}}\ and\ \citenamefont
  {{Berger}}(2013)}]{Fong_Berger:2013}%
  \BibitemOpen
  \bibfield  {author} {\bibinfo {author} {\bibfnamefont {W.}~\bibnamefont
  {{Fong}}}\ and\ \bibinfo {author} {\bibfnamefont {E.}~\bibnamefont
  {{Berger}}},\ }\href {\doibase 10.1088/0004-637X/776/1/18} {\bibfield
  {journal} {\bibinfo  {journal} {\apj}\ }\textbf {\bibinfo {volume} {776}},\
  \bibinfo {eid} {18} (\bibinfo {year} {2013})}\BibitemShut {NoStop}%
\bibitem [{\citenamefont {{Neal}}(2012)}]{Neal:2012}%
  \BibitemOpen
  \bibfield  {author} {\bibinfo {author} {\bibfnamefont {R.~M.}\ \bibnamefont
  {{Neal}}},\ }\href@noop {} {\bibfield  {journal} {\bibinfo  {journal} {ArXiv
  e-prints}\ } (\bibinfo {year} {2012})},\ \Eprint
  {http://arxiv.org/abs/1206.1901} {arXiv:1206.1901 [stat.CO]} \BibitemShut
  {NoStop}%
\bibitem [{\citenamefont {Carpenter}\ \emph {et~al.}(2017)\citenamefont
  {Carpenter}, \citenamefont {Gelman}, \citenamefont {Hoffman}, \citenamefont
  {Lee}, \citenamefont {Goodrich}, \citenamefont {Betancourt}, \citenamefont
  {Brubaker}, \citenamefont {Guo}, \citenamefont {Li},\ and\ \citenamefont
  {Riddell}}]{Carpenter_etal:2017}%
  \BibitemOpen
  \bibfield  {author} {\bibinfo {author} {\bibfnamefont {B.}~\bibnamefont
  {Carpenter}}, \bibinfo {author} {\bibfnamefont {A.}~\bibnamefont {Gelman}},
  \bibinfo {author} {\bibfnamefont {M.}~\bibnamefont {Hoffman}}, \bibinfo
  {author} {\bibfnamefont {D.}~\bibnamefont {Lee}}, \bibinfo {author}
  {\bibfnamefont {B.}~\bibnamefont {Goodrich}}, \bibinfo {author}
  {\bibfnamefont {M.}~\bibnamefont {Betancourt}}, \bibinfo {author}
  {\bibfnamefont {M.}~\bibnamefont {Brubaker}}, \bibinfo {author}
  {\bibfnamefont {J.}~\bibnamefont {Guo}}, \bibinfo {author} {\bibfnamefont
  {P.}~\bibnamefont {Li}}, \ and\ \bibinfo {author} {\bibfnamefont
  {A.}~\bibnamefont {Riddell}},\ }\href {\doibase 10.18637/jss.v076.i01}
  {\bibfield  {journal} {\bibinfo  {journal} {Journal of Statistical Software,
  Articles}\ }\textbf {\bibinfo {volume} {76}},\ \bibinfo {pages} {1} (\bibinfo
  {year} {2017})}\BibitemShut {NoStop}%
\bibitem [{\citenamefont {{Stan Development Team}}(2016)}]{pystan}%
  \BibitemOpen
  \bibfield  {author} {\bibinfo {author} {\bibnamefont {{Stan Development
  Team}}},\ }\href {http://mc-stan.org} {\enquote {\bibinfo {title} {{PyStan}:
  the python interface to stan, {Version} 2.14.0.0},}\ } (\bibinfo {year}
  {2016})\BibitemShut {NoStop}%
\bibitem [{\citenamefont {{Singer}}\ and\ \citenamefont
  {{Price}}(2016)}]{Singer_Price:2016}%
  \BibitemOpen
  \bibfield  {author} {\bibinfo {author} {\bibfnamefont {L.~P.}\ \bibnamefont
  {{Singer}}}\ and\ \bibinfo {author} {\bibfnamefont {L.~R.}\ \bibnamefont
  {{Price}}},\ }\href {\doibase 10.1103/PhysRevD.93.024013} {\bibfield
  {journal} {\bibinfo  {journal} {\prd}\ }\textbf {\bibinfo {volume} {93}},\
  \bibinfo {eid} {024013} (\bibinfo {year} {2016})}\BibitemShut {NoStop}%
\bibitem [{\citenamefont {{Sathyaprakash}}\ \emph {et~al.}(2010)\citenamefont
  {{Sathyaprakash}}, \citenamefont {{Schutz}},\ and\ \citenamefont {{Van Den
  Broeck}}}]{Sathyaprakash:2010}%
  \BibitemOpen
  \bibfield  {author} {\bibinfo {author} {\bibfnamefont {B.~S.}\ \bibnamefont
  {{Sathyaprakash}}}, \bibinfo {author} {\bibfnamefont {B.~F.}\ \bibnamefont
  {{Schutz}}}, \ and\ \bibinfo {author} {\bibfnamefont {C.}~\bibnamefont {{Van
  Den Broeck}}},\ }\href {\doibase 10.1088/0264-9381/27/21/215006} {\bibfield
  {journal} {\bibinfo  {journal} {Classical and Quantum Gravity}\ }\textbf
  {\bibinfo {volume} {27}},\ \bibinfo {eid} {215006} (\bibinfo {year}
  {2010})}\BibitemShut {NoStop}%
\bibitem [{\citenamefont {{Nishizawa}}\ \emph {et~al.}(2012)\citenamefont
  {{Nishizawa}}, \citenamefont {{Yagi}}, \citenamefont {{Taruya}},\ and\
  \citenamefont {{Tanaka}}}]{Nishizawa:2012}%
  \BibitemOpen
  \bibfield  {author} {\bibinfo {author} {\bibfnamefont {A.}~\bibnamefont
  {{Nishizawa}}}, \bibinfo {author} {\bibfnamefont {K.}~\bibnamefont {{Yagi}}},
  \bibinfo {author} {\bibfnamefont {A.}~\bibnamefont {{Taruya}}}, \ and\
  \bibinfo {author} {\bibfnamefont {T.}~\bibnamefont {{Tanaka}}},\ }in\ \href
  {\doibase 10.1088/1742-6596/363/1/012052} {\emph {\bibinfo {booktitle}
  {Journal of Physics Conference Series}}},\ \bibinfo {series} {Journal of
  Physics Conference Series}, Vol.\ \bibinfo {volume} {363}\ (\bibinfo {year}
  {2012})\ p.\ \bibinfo {pages} {012052}\BibitemShut {NoStop}%
\bibitem [{\citenamefont {{Del Pozzo}}\ \emph {et~al.}(2017)\citenamefont {{Del
  Pozzo}}, \citenamefont {{Li}},\ and\ \citenamefont
  {{Messenger}}}]{delPozzo:2017}%
  \BibitemOpen
  \bibfield  {author} {\bibinfo {author} {\bibfnamefont {W.}~\bibnamefont {{Del
  Pozzo}}}, \bibinfo {author} {\bibfnamefont {T.~G.~F.}\ \bibnamefont {{Li}}},
  \ and\ \bibinfo {author} {\bibfnamefont {C.}~\bibnamefont {{Messenger}}},\
  }\href {\doibase 10.1103/PhysRevD.95.043502} {\bibfield  {journal} {\bibinfo
  {journal} {\prd}\ }\textbf {\bibinfo {volume} {95}},\ \bibinfo {eid} {043502}
  (\bibinfo {year} {2017})}\BibitemShut {NoStop}%
\bibitem [{\citenamefont {{Belgacem}}\ \emph {et~al.}(2018)\citenamefont
  {{Belgacem}}, \citenamefont {{Dirian}}, \citenamefont {{Foffa}},\ and\
  \citenamefont {{Maggiore}}}]{Belgacem:2018}%
  \BibitemOpen
  \bibfield  {author} {\bibinfo {author} {\bibfnamefont {E.}~\bibnamefont
  {{Belgacem}}}, \bibinfo {author} {\bibfnamefont {Y.}~\bibnamefont
  {{Dirian}}}, \bibinfo {author} {\bibfnamefont {S.}~\bibnamefont {{Foffa}}}, \
  and\ \bibinfo {author} {\bibfnamefont {M.}~\bibnamefont {{Maggiore}}},\
  }\href {\doibase 10.1103/PhysRevD.98.023510} {\bibfield  {journal} {\bibinfo
  {journal} {\prd}\ }\textbf {\bibinfo {volume} {98}},\ \bibinfo {eid} {023510}
  (\bibinfo {year} {2018})}\BibitemShut {NoStop}%
\bibitem [{\citenamefont {{Carroll}}\ \emph {et~al.}(1992)\citenamefont
  {{Carroll}}, \citenamefont {{Press}},\ and\ \citenamefont
  {{Turner}}}]{Carroll_etal:1992}%
  \BibitemOpen
  \bibfield  {author} {\bibinfo {author} {\bibfnamefont {S.~M.}\ \bibnamefont
  {{Carroll}}}, \bibinfo {author} {\bibfnamefont {W.~H.}\ \bibnamefont
  {{Press}}}, \ and\ \bibinfo {author} {\bibfnamefont {E.~L.}\ \bibnamefont
  {{Turner}}},\ }\href {\doibase 10.1146/annurev.aa.30.090192.002435}
  {\bibfield  {journal} {\bibinfo  {journal} {\araa}\ }\textbf {\bibinfo
  {volume} {30}},\ \bibinfo {pages} {499} (\bibinfo {year} {1992})}\BibitemShut
  {NoStop}%
\bibitem [{\citenamefont {{Wright}}(2006)}]{Wright:2006}%
  \BibitemOpen
  \bibfield  {author} {\bibinfo {author} {\bibfnamefont {E.~L.}\ \bibnamefont
  {{Wright}}},\ }\href {\doibase 10.1086/510102} {\bibfield  {journal}
  {\bibinfo  {journal} {\pasp}\ }\textbf {\bibinfo {volume} {118}},\ \bibinfo
  {pages} {1711} (\bibinfo {year} {2006})}\BibitemShut {NoStop}%
\bibitem [{\citenamefont {{Loredo}}(2004)}]{Loredo:2004}%
  \BibitemOpen
  \bibfield  {author} {\bibinfo {author} {\bibfnamefont {T.~J.}\ \bibnamefont
  {{Loredo}}},\ }in\ \href {\doibase 10.1063/1.1835214} {\emph {\bibinfo
  {booktitle} {American Institute of Physics Conference Series}}},\ \bibinfo
  {series} {American Institute of Physics Conference Series}, Vol.\ \bibinfo
  {volume} {735},\ \bibinfo {editor} {edited by\ \bibinfo {editor}
  {\bibfnamefont {R.}~\bibnamefont {{Fischer}}}, \bibinfo {editor}
  {\bibfnamefont {R.}~\bibnamefont {{Preuss}}}, \ and\ \bibinfo {editor}
  {\bibfnamefont {U.~V.}\ \bibnamefont {{Toussaint}}}}\ (\bibinfo {year}
  {2004})\ pp.\ \bibinfo {pages} {195--206}\BibitemShut {NoStop}%
\bibitem [{\citenamefont {{Streit}}(2010)}]{Streit:2010}%
  \BibitemOpen
  \bibfield  {author} {\bibinfo {author} {\bibfnamefont {R.~L.}\ \bibnamefont
  {{Streit}}},\ }\href@noop {} {\emph {\bibinfo {title} {Poisson Point
  Processes}}}\ (\bibinfo  {publisher} {Springer},\ \bibinfo {year}
  {2010})\BibitemShut {NoStop}%
\bibitem [{\citenamefont {Blanchet}(2014)}]{Blanchet:2013haa}%
  \BibitemOpen
  \bibfield  {author} {\bibinfo {author} {\bibfnamefont {L.}~\bibnamefont
  {Blanchet}},\ }\href {\doibase 10.12942/lrr-2014-2} {\bibfield  {journal}
  {\bibinfo  {journal} {Living Rev. Rel.}\ }\textbf {\bibinfo {volume} {17}},\
  \bibinfo {pages} {2} (\bibinfo {year} {2014})}\BibitemShut {NoStop}%
\bibitem [{\citenamefont {Maggiore}(2007)}]{Maggiore:2007}%
  \BibitemOpen
  \bibfield  {author} {\bibinfo {author} {\bibfnamefont {M.}~\bibnamefont
  {Maggiore}},\ }\href@noop {} {\emph {\bibinfo {title} {Gravitational Waves.
  Vol. 1: Theory and Experiments}}},\ Oxford Masters Series in Physics\
  (\bibinfo  {publisher} {Oxford University Press},\ \bibinfo {year}
  {2007})\BibitemShut {NoStop}%
\end{thebibliography}%


\appendix


\begin{widetext}

\section{Bayesian inference of cosmological parameters from BNS mergers}
\label{section:derivation_app}

The task of inferring the Hubble constant from BNS merger 
events can be thought of as a special case of the more general
problem of cosmological parameter inference from such data
(\eg, \cite{Markovic:1993,del_Pozzo:2012,delPozzo:2017,Belgacem:2018}).
Combined with the fact that
a Euclidean approximation is insufficient even for current low-redshift
BNS merger samples, 
taking a fully general approach and then making low-redshift
approximations ensures that any such simplifications are done rigorously.
The BNS merger population model (\app{model_app}),
data (\app{data_app}) and sample selection (\app{sample_app})
have the same structure as the more specific simulation 
described in \sect{model}.
These ingredients are then combined self-consistently
to obtain both the full posterior distribution
and the marginalized posterior in the 
cosmological parameters (\app{bayes_app}). 


\subsection{Physical model}
\label{section:model_app}

\begin{figure*}
\includegraphics[width=13cm]{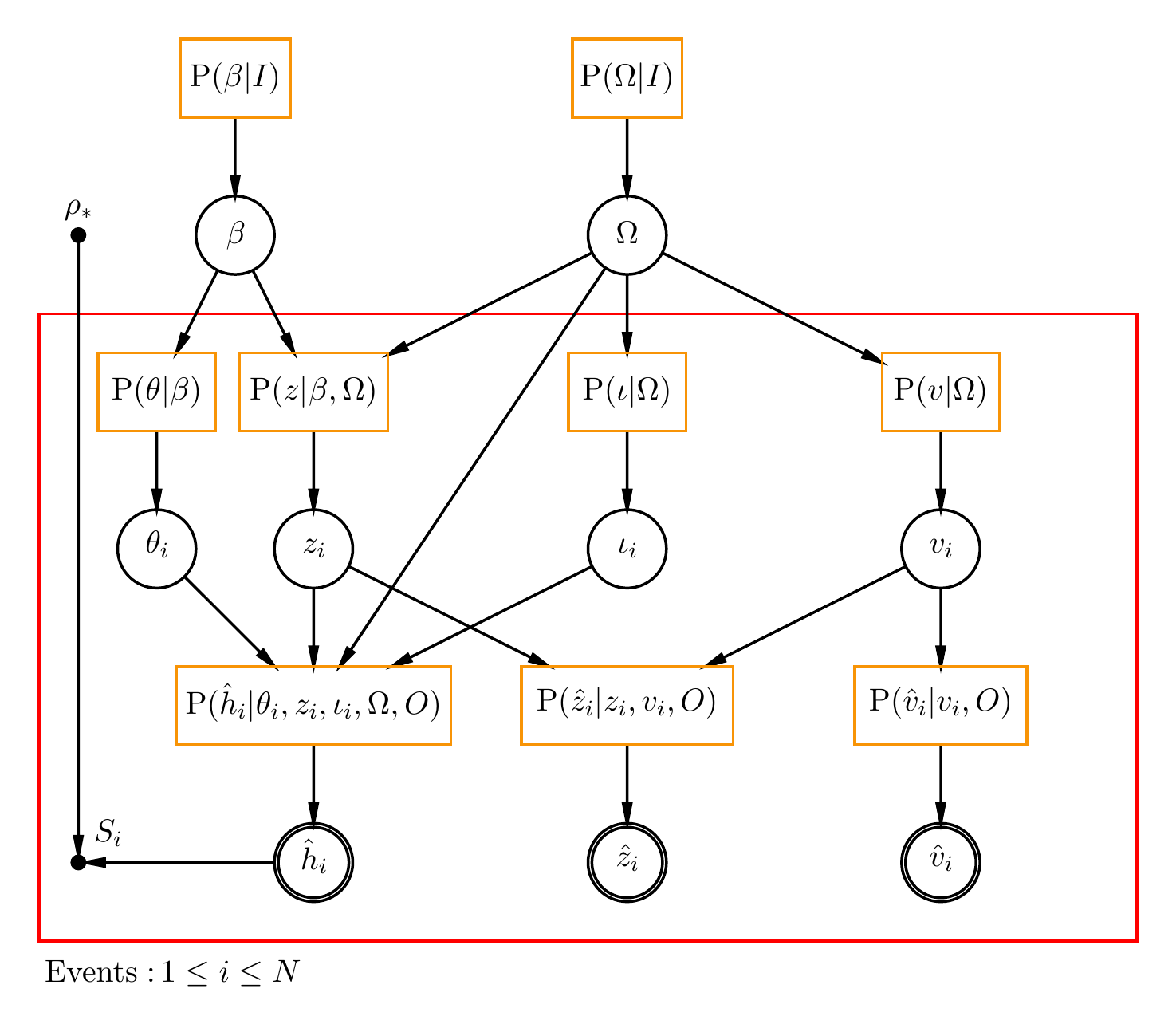}
\caption{Network diagram for the hierarchical model describing the BNS
  population and data.
  Quantities in single circles are model parameters;
  quantities in double circles are measured values.
  These quantities are linked by the probability distributions
  in orange rectangles: the top row are parameter priors;
  the middle row describe the population model;
  the bottom row define the (object-level) likelihood.
  The arrows linking the quantities and distributions define
  the forward/generative model that could be used to simulate
  samples and measurements.
  The quantities inside the red rectangular plate are specific
  to a single BNS merger event;
  the detected events are indexed by $i \in \{1, 2, \ldots N\}$.
}
\label{figure:network}
\end{figure*}

A necessary ingredient for any Bayesian inference formalism 
is a generative model
which could be used to simulate a mock dataset.
For the case of BNS mergers 
this includes both a cosmological model
(\app{cosmology_app})
and the BNS population (\app{bnspop_app}),
although it is primarily the
dependence
structure of the model, 
rather than specific functional forms, 
which are required at this stage.
This physical model includes neither 
measurements (\app{data_app})
or sample selection (\app{selection_app}), 
which are kept distinct as they play different roles in the 
simulation and inference.
The overall structure of the model is summarized in
\fig{network}.


\subsubsection{Cosmology}
\label{section:cosmology_app}

GW signals from BNS events can, in principle, be seen to cosmological 
distances and so must in general
be analyzed in the context of a
full
cosmological model.
This is described by a set of parameters 
$\csmpar = (\hubble, \Omega_{\rm m}, \Omega_\Lambda, \omega, \ldots)$\
which between them 
specify the contents of the universe and its expansion history.
None of the results
presented here depend strongly on the particular 
type of cosmological model
adopted, so $\csmpar$ is used in order to preserve generality,
although it is implicit that \hubble\ is always included in this list.

The radial coordinate used to specify
position along the line-of-sight is the 
(cosmological) redshift, $z$, 
defined strictly in terms of the ratio of the cosmological scale factor
to the current value.
This is distinct from the observable redshift,
$\zspec$,
which in general differs from $z$ due to the source's peculiar velocity
relative to the Hubble-Lema\^{i}tre flow.
Assuming this motion is non-relativistic,
the observable redshift is given in terms of 
the line-of-sight component of the source's peculiar velocity, 
$v$, as
\begin{equation}
\label{equation:z_app}
\zspec = (1 + z) \, \left(1 + \frac{v}{c}\right) - 1
  = z + (1 + z) \, \frac{v}{c},
\end{equation}
where positive $v$ hence corresponds to motion away from the observer.

The redshift and cosmological model combine to specify both the 
the luminosity distance,
$\dl(z, \csmpar)$
and 
the co-moving volume element 
$\diff V / \diff z\, (\csmpar)$.
Expressions for 
$\dl(z, \csmpar)$
and
$\diff V / \diff z\, (\csmpar)$
are available for standard cosmological models 
\cite{Carroll_etal:1992,Wright:2006},
although the model is kept general here.
The numerical marginalization scheme described in \sect{cosmol}
requires the function 
$z(\dl, \csmpar)$, defined such that
$z[\dl(z, \csmpar), \csmpar] = \dl[z(\dl, \csmpar), \csmpar] = 1$
for any cosmological model 
(and any non-negative value of $z$ or $\dl$).
These inverses exist
in standard cosmological models as 
the luminosity distance
increases monotonically with $z$ \cite{Weinberg:2008}.


\subsubsection{The BNS merger population}
\label{section:bnspop_app}

The BNS population is defined primarily
by the rate of mergers
per unit proper time 
per unit co-moving volume,
$\rate(z, \gwpoppar)$,
where $\gwpoppar$ are the parameters which 
describe
the population model.
The expected number of events 
in the redshift range $z$ to $z + \diff z$
that would be registered by a perfect all-sky detector 
in time interval $\diff t$ is hence
\begin{equation}
\diff \nexp = 
 \frac{\rate(z, \gwpoppar)}{1 + z}
 \, \frac{\diff \vol}{\diff z}(\csmpar)
 \, \diff z \, \diff t,
\end{equation}
where the reduction by
a factor of $(1 + z)$ comes about due to time dilation from
the source frame, and $\diff V / \diff z (\csmpar)$ is 
the co-moving volume element defined in \app{cosmology_app}.

The properties of a single BNS merger are separated into 
intrinsic system parameters, $\gwpar$
(the NS masses, spins, \etc),
and observer-dependent quantities,
taken here to be
the inclination of the system with respect to the line-of-sight, $\inc$,
and the line-of-sight peculiar velocity of the host galaxy, $v$
(and $z$, although this is already incorporated in the overall rate).
For a given redshift, 
the population model is 
a (normalized) probability
distribution of the form
\begin{equation}
\prob(\gwpar, v, \inc | z, \gwpoppar, \csmpar) 
  =
  \prob(\gwpar | z, \gwpoppar) \,
  \prob(v | z, \csmpar) \,
  \step(\inc) \, \step(\pi - \inc) \, \frac{\sin(\inc)}{2} 
.
\end{equation}
This encodes the assumption that 
BNS systems are oriented randomly, 
leading
to the sinusoidal distribution in $\inc$.

Some care is required in treating the 
peculiar velocity of the BNS merger,
as it has two distinct contributions: 
the motion of the host galaxy relative to the 
Hubble-Lema\^{i}tre flow;
and the orbital
motion of the merger system within the host galaxy.
(Only BNS mergers with counterparts are considered here,
\cf\ \cite{Nissanke_etal:2013b}, 
and all
such systems have, by definition, a host galaxy.)
Both these motions are expected to have typical
speeds of $\sim 10^2\ \kms$, 
although they enter the inference formalism in distinct ways.
In physical terms 
the peculiar velocity of the host galaxy,
$v$, 
is determined by the local distribution of matter,
but even in the absence of source-specific data (\app{vobs_app}),
knowledge of cosmological structure formation
implies a distribution of the form $\prob(v | z, \csmpar)$.
While there is a formal link between the peculiar velocity
and the cosmological parameters,
this is negligible compared to the link through the GW data,
and so the dependence of 
$\prob(v | z, \csmpar)$ on $\csmpar$ can be ignored.
This reflects the status of the host's peculiar velocity as a nuisance
parameter in this context.


\subsection{BNS merger data}
\label{section:data_app}

Three distinct types of measurement provide information
about the properties of a BNS merger event:
GW strain time-series data
(\app{gwobs_app});
a spectroscopic redshift for the host galaxy
(\app{zobs_app});
and, potentially, an estimated (line-of-sight) peculiar velocity
for the host galaxy
(\app{vobs_app}).
All of these measured quantities have associated uncertainties
(and other details associated with the measurement process);
for brevity, all such quantities are combined into
a single parameter, $O$, that characterizes the observations.


\subsubsection{GW data}
\label{section:gwobs_app}

The GW signal from a merger event comprises two orthogonal polarizations,
$\strain_\pol(t; \gwpar, \inc, v, z, \csmpar)$,
with $\pol \in \{\plus, \times\}$,
where the form of the $t$ dependence is determiend by
the intrinsic merger properties, 
$\gwpar$, 
the observer-dependent quantities
$\inc$, $v$ and $z$, 
and the cosmological parameters, $\csmpar$.
The full waveforms depend on the complicated non-linear physics of the merger,
but the dependence on $v$ and $z$ is determined 
purely by the physics of GW propagation in an expanding universe.
The GW signal in the far field regime
is subject to a time dilation by a factor of 
$(1 + z) \, (1 + v / c)$
and the amplitude scales as $1 / \dprop(z, \csmpar)$,
where $\dprop(z, \csmpar)$ is the proper distance to redshift $z$,
so the expression
$\dprop(z, \csmpar) \, \strain_\pol
[(t - t_0) / (1 + z)\,(1 + v /c); \gwpar, \inc, v, z, \csmpar]$
is independent of redshift and peculiar velocity.
The dependence of the strain on the peculiar velocity
is sufficiently small that it can be ignored (\app{linear}),
so 
$\strain_\pol(t; \gwpar, \inc, v, z, \csmpar)
\rightarrow 
\strain_\pol(t; \gwpar, \inc, z, \csmpar)$
is assumed.

The GW data for a merger comprise a discretized time-series
of measured strains, denoted $\dgw$ for simplicity,
along with associated uncertainties,
implicitly included in $O$.
The likelihood for the strain data hence has the form
$\prob(\dgw | \gwpar, \inc, z, \csmpar, O)$,
which is kept general in this derivation;
the specific form used in the simulations described 
in \sect{model} is detailed in \app{linear}.


\subsubsection{Redshift measurement}
\label{section:zobs_app}

The measured spectroscopic redshift
of a BNS host, $\zobs$,
is linked to both its cosmological redshift,
$z$, and its line-of-sight peculiar velocity, $v$,
leading to a likelihood of the form
$\prob(\zobs | z, v, O)$,
with the observable redshift given in \eq{z_app}.


\subsubsection{Peculiar velocity estimates}
\label{section:vobs_app}

It is possible that the (line-of-sight) peculiar velocity
of a BNS host can be estimated from the positions and/or motions
of nearby galaxies, 
yielding an estimate $\vobs$.
In the absence of further information, this contribution to the likelihood 
has the form
$\prob(\vobs | v, O)$,
where the information on the positions and/or velocities of nearby
galaxies is left implicit.
It is unrealistic to consider the case that the 
uncertainty is small or negligible,
as there is no credible way to get precise peculiar velocity information;
but it is useful to consider the limit that 
the uncertainty is infinite, which is equivalent to there
being no useful peculiar velocity data at all for a given system.
This situation could also be recovered
by removing 
$\vobs$ and the associated likelihood from the calculation altogether.

It is also possible in principle that the line-of-sight 
component of the 
orbital motion of the merger relative to its host galaxy 
could be estimated (\eg, through the combination of a precise location
and rotation curve),
although the impact of this on the GW data is minor anyway 
(\app{linear}) so this is not explored further here.


\subsection{Sample of BNS events}
\label{section:sample_app}

The full BNS merger dataset from a GW survey 
(and follow-up observations) 
is a catalog of 
$N$ selected merger events, 
each with associated GW data, 
$\dgws = (\dgw_1, \dgw_2, \ldots, \dgw_N)$
redshift measurements,
$\zobss = (\zobs_1, \zobs_2, \ldots, \zobs_N)$
and (possibly)
host peculiar velocity information,
$\vobss = (\vobs_1, \vobs_2, \ldots, \vobs_N)$,
as described in \app{data_app}.
The additional model ingredients needed to 
define the sample generation process is the
selection rule (\app{selection_app}),
which then determines the 
the number of selected events (\app{number_app}).


\subsubsection{Selection}
\label{section:selection_app}

The selection of a BNS merger event into a sample is
assumed to be determined by the GW data alone,
and to take the form of a hard cut on some statistic calculated
from the GW data, $\dgw$. 
This function is denoted
$\ston(\dgw, O)$ and can be thought of as the 
observed SNR (or a proxy for this).
The form of $\ston(\dgw, O)$ 
is not as important as the fact
that it is deterministic, meaning the selection probability
can be written as
\begin{equation}
\label{equation:psel_app}
\prob(S | \dgw, O) = \step[\ston(\dgw, O) - \stonmin].
\end{equation}
The fact that the selection is  
a deterministic function of the data 
means that for any event in a selected sample,
$\hat{\ston} = \ston(\dgw, O) \geq \stonmin$
and so $\prob(S | \dgw, O) = 1$,
a critical fact \cite{Loredo:2004} 
which simplifies the parameter inference calculation (\sect{bayes}).


\subsubsection{Number of events} 
\label{section:number_app}

Modeling the BNS merger population
as a realization of a Poisson point process
\cite{Streit:2010}, 
the number of detected events in a sample is drawn from
the Poisson distribution 
\begin{equation}
\label{equation:poisson}
\prob(N | \csmpar, \gwpoppar, O)
  = \step(N) \,
  \frac{[\nexp(\gwpoppar, \csmpar, O)]^N \,
  \exp[- \nexp(\gwpoppar, \csmpar, O)]}{N!},
\end{equation}
which is characterized purely by the expected number of events,
$\nexp(\csmpar, \gwpoppar, O)$.
This is obtained by integrating the product of the 
event rate and the detection probability 
(assumed to depend only on the GW data) over the BNS merger
properties, which gives
\begin{equation}
\label{equation:nexp_app}
\nexp(\gwpoppar, \csmpar, O)
  =
  T
  \int_0^\infty \diff z \,
  \frac{\rate(z, \gwpoppar)}{1 + z}
  \frac{\diff \vol}{\diff z}(\csmpar)
  \int \diff \gwpar \,
  \prob(\gwpar | z, \gwpoppar) \,
  \int_0^\pi \diff \inc \, \frac{\sin(\inc)}{2}
  \int \diff \dgw \,
  \prob(\dgw | \gwpar, \inc, z, \csmpar, O) \,
  \step[\ston(\dgw, O) - \stonmin],
\end{equation}
where the integral with respect to $t$ yields 
the observing time, $T$,
which is included in $O$ along with $\stonmin$.
In general, this integral must be evaluated numerically, 
\eg, using a Monte Carlo approach
such as that described in \sect{simulation},
although in some applications they can be avoided 
(\eg, \cite{Taylor_etal:2012}).


\subsection{Parameter inference}
\label{section:bayes_app}

The general inference task here is to obtain constraints on the
full set of model parameters given all 
the data on a sample of $N$ BNS merger events,
indexed by $i \in \{1, 2, \ldots N\}$.
The 
model, defined 
in \sect{model},
is characterized by
the cosmological parameters, $\csmpar$, 
the BNS population parameters, $\gwpoppar$, 
and the BNS mergers' physical properties,
$\gwpars = (\gwpar_1, \gwpar_2, \ldots, \gwpar_N)$,
host redshifts,
$\zs = (z_1, z_2, \ldots, z_N)$, 
and peculiar velocities,
$\vs = (v_1, v_2, \ldots, v_N)$;
the data for the $N$ events are,
as described in \sect{data},
the GW measurements
$\dgws = (\dgw_1, \dgw_2, \ldots, \dgw_N)$,
redshift measurements
$\zobss = (\zobs_1, \zobs_2, \ldots, \zobs_N)$
and (possibly) peculiar velocity information,
$\vobss = (\vobs_1, \vobs_2, \ldots, \vobs_N)$.
Combined with a description the sample selection (\app{sample_app}),
this is sufficient to calculate the joint posterior 
distribution in all the parameters (\app{joint_app}),
and the marginalized distribution in the 
parameters of interest (\app{marginal_app}).


\subsubsection{Joint posterior distribution}
\label{section:joint_app}

The constraints on the cosmological parameters, BNS population
and the individual events implied by 
a sample of $N$ detected BNS mergers 
are fully described by the 
joint posterior distribution in all the model parameters,
$\prob(
  \gwpars, \incs, \zs, \vs, \gwpoppar, \csmpar
  |
  N, \dgws, \zobss, \vobss, O, I)$,
where $I$ is the prior information assumed about $\csmpar$ and $\gwpoppar$.
The  knowledge about the cosmological model and the BNS
population are assumed to be independent, so that the prior 
distribution
factorizes as 
$\prob(\csmpar, \gwpoppar | I) = \prob(\csmpar | I) \, \prob(\gwpoppar | I)$;
the prior information on all the other parameters is specified
by $\csmpar$ and $\gwpoppar$, and so is conditionally-independent of $I$.
The posterior is conditioned not only on the obvious 
data for each of the merger events 
(\ie, $\dgws$, $\zobss$ and $\vobss$), 
but also on the 
size of the sample (\ie, the value of $N$)
and hence implicitly 
on the fact that the GW data from each 
detected
event 
must satisfy
the selection criterion outlined in \app{selection_app}.

The task now is to write the joint posterior distribution 
in terms of the functions/distributions defined in 
\app{model_app}, \app{data_app} and \app{sample_app}.
It is useful to introduce explicitly the fact that each of the 
merger events was selected, 
denoted 
$\sels = (S_1, S_2, \ldots, S_N)$ following \app{selection_app}.
This can be added to the list of quantities being conditioned on
because the selection process defined in \eq{psel} is 
deterministic -- whether $S$ is true can be
determined from $\dgw$ --
which means that including $\sels$ does not 
add any extra information:
$\prob(
  \gwpars, \incs, \zs, \vs, \gwpoppar, \csmpar
  |
  N, \dgws, \zobss, \vobss, O, I)
  =
  \prob(
  \gwpars, \incs, \zs, \vs, \gwpoppar, \csmpar
  |
  N, \sels, \dgws, \zobss, \vobss, O, I)$.
The reason for including $\sels$ is that it gives the freedom to condition 
on selection alone, 
which allows the 
the unnormalized posterior to be written as
\begin{equation}
\label{equation:post_intermediate}
  \prob(
  \gwpars, \incs, \zs, \vs, \gwpoppar, \csmpar
  |
  N, \dgws, \zobss, \vobss, O, I)
  \propto
  \prob(\csmpar | I) \, \prob(\gwpoppar | I) \,
  \prob(N | \csmpar, \gwpoppar, O) \,
  \prod_{i = 1}^N
  \prob(\gwpar_i, \inc_i, z_i, v_i, \dgw_i, \zobs_i, \vobs_i |
  S_i, \gwpoppar, \csmpar, O).
\end{equation}
The first two terms are the prior distributions on 
the cosmological and BNS population parameters;
the 
third
term encodes the constraints provided by the number 
of detected events (\sect{number});
but the terms in the product
still require some manipulation to be written in terms of the 
distributions which define the model.
Taking any one such term and successively
applying Bayes's theorem, the chain rule, and the law of total
probability then yields the joint distribution in the intrinsic
and observed properties of a selected merger event as
\begin{align}
& \prob(\gwpar, \inc, z, v, \dgw, \zobs, \vobs |
  S, \gwpoppar, \csmpar, O)
\\
&
  = \frac{1}{\nexp(\csmpar, \gwpoppar, O)} \,
  \frac{ \rate(z, \gwpoppar)}{1 + z} \, 
  \frac{\diff V }{\diff z} (\csmpar) \,
  \prob(\gwpar | z, \gwpoppar) \, 
  \step(\inc) \, \step(\pi - \inc) \, \frac{\sin(\inc)}{2} \,
  \prob(v | z, \csmpar) \,
  \prob(\dgw | \gwpar, \inc, z, \csmpar, O)
  \, \prob(\zobs | z, v, O) \,
  \prob(\vobs | v, O) ,
\nonumber
\end{align}
where the fact that $\prob(S | \dgw, O) = 1$ for selected events has been
used to omit this term, and
the normalizing constant is 
equal to the expected number of events in the sample, 
given in \eq{nexp_app}. 

Inserting this object-level distribution
and the Poisson distribution of the 
number of events in \eq{poisson}
into \eq{post_intermediate}
then gives the unnormalized joint posterior distribution 
in all the model parameters as 
\begin{align}
\label{equation:fullpost}
& \prob(
  \gwpars, \incs, \zs, \vs, \gwpoppar, \csmpar
  |
  N, \dgws, \zobss, \vobss, O, I)
  \propto
  \prob(\csmpar | I) \, \prob(\gwpoppar | I) \,
  \exp[- \nexp(\gwpoppar, \csmpar, O)] \,
  \\
  & \times
  \prod_{i = 1}^N
  \frac{\rate(z_i, \gwpoppar)}{1 + z_i} \,
  \frac{\diff V}{\diff z_i} (\csmpar) \,
  \step(\inc_i) \, \step(\pi - \inc_i) \, \sin(\inc_i) \, 
  \prob(\gwpar_i | z_i, \gwpoppar) \, 
  \prob(\dgw_i | \gwpar_i, \inc_i, z_i, \csmpar, O) \,
  \prob(v_i | z_i, \csmpar) \,
  \prob(\zobs_i | z_i, v_i, O) \,
  \prob(\vobs_i | v_i, O),
  \nonumber
\end{align}
where the selection indicators $\sels$ have been omitted
because (as argued above) they do not contain any extra information
beyond that already encoded in $\dgws$.
This has the standard structure for the full likelihood 
of a Poisson point process \cite{Streit:2010}, in 
particular with the expected number of detected sources appearing 
only in the exponential term,
although the number of different component distributions in the 
product somewhat obscures the link to this standard statistical model.
This posterior has the characteristic Poisson point
process structure that also appears in 
\eg, Refs.~\cite{Taylor_etal:2012,Mandel_etal:2018}.

Equation~\ref{equation:fullpost}
represents the main result of the above
model formulation as it
encodes all the links between the measured data and the
quantities of interest.
Other than the assumption of random orientations encoded
in the sinusoidal inclination distribution, 
it is deliberately kept completely general, 
with only the structure of the conditional dependencies from
\fig{network} enforced.
Importantly, 
this result is valid in a fully cosmological context, 
which means it also provides a rigorous route to obtaining 
low-redshift approximations without the need for any
heuristic arguments about the relationship
between distance, redshift and line-of-sight velocity.


\subsubsection{Marginalization over nuisance parameters}
\label{section:marginal_app}

The main scientific aim here is inference of the cosmological
parameters, and specifically \hubble, 
which casts all the individual and population-level
BNS quantities as nuisance parameters to be integrated out.
For the simple model described in \sect{model}
this is done by sampling the joint posterior distribution,
as described in \sect{bayes},
and so there is no need to 
explicitly calculate the marginalized posterior distribution.
The alternative approach 
\cite{Abbott_etal:2017a,Chen_etal:2018,Feeney_etal:2018},
which will likely be needed when analyzing real data,
is to separately marginalize over the redshift
parameters constrained by the EM data and 
then use samples from the individual merger posteriors 
to marginalize over those parameters.
Equation~\ref{equation:fullpost}
can be used as the basis for this approach, 
with the only potential difference coming from how the 
pre-factor of 
$\exp[- \nexp(\gwpoppar, \csmpar, O)]$ is handled.
Fortunately, this term has only a weak dependence on 
\hubble\ for low-redshift samples 
(\cite{Abbott_etal:2017a,Chen_etal:2018} and \sect{number}),
so such choices should not strongly affect the final 
marginal posterior distribution in $\csmpar$.


\section{Likelihood for the merger inspiral}
\label{section:linear}

A BNS merger is 
controlled
by non-linear physics in dynamical
and
strongly curved spacetime, and 
is described by a large number of parameters.
Within the frequency band of the current GW detectors,
however,
the emission relevant for distance constraints arise primarily from
the comparatively simple inspiral phase preceding the merger,
which can be modeled using the post-Newtonian appoximation 
to
general relativity \eg, \cite{Blanchet:2013haa}.
For the type of system under consideration here
several additional simplifications can be made.
Most importantly,
it is assumed that an EM counterpart
has been identified,
meaning that the sky position is known.
As the known Galactic BNSs that would merge within a Hubble
time have low dimensionless spin parameters of $\la 0.04$
\cite{Abbott_etal:2017d},
the spins of the merging NSs are ignored and set to zero.
Adopting these simplifications,
the GW signal from a merger event takes the form of two
orthogonal strain waveforms that, to
leading order in the post-Newtonian expansion parameter,
can be written as 
(\eg, \cite{Maggiore:2007})
\begin{align}
\label{equation:strains}
\strain_\plus(t)
& =
\frac{G \, \mchirp / c^2}{\dprop(z, \csmpar)} \,
\frac{1 + \cos^2(\inc)}{2} \,\,
\phi_\plus\!\!\left[ 
  t; \frac{5 \, G \, \mchirp}{c^3} \, (1 + z) 
  \left(1 + \frac{u}{c} + \frac{v}{c} \right) ,
  \tc, 
  \phic
\right]
\nonumber \\
\\
\strain_\times(t)
& =
\frac{G \, \mchirp / c^2}{\dprop(z, \csmpar)} \,
[- \cos(\inc)] \,\,
\phi_\times\!\!\left[ 
  t; \frac{5 \, G \, \mchirp}{c^3} \, (1 + z) 
  \left(1 + \frac{u}{c} + \frac{v}{c} \right) , 
  \tc,
  \phic
\right]
,
\nonumber
\end{align}
with the detector-frame time dependence encoded in the functions
\begin{align}
\label{equation:straint}
\phi_\plus(t; \tau, \tc, \phic) & = 
\left(\frac{\tc - t}{\tau}\right)^{-1/4}
\cos\left\{2 \left[
\phic - \left(\frac{\tc - t}{\tau}\right)^{5/8}
\right]
\right\}
\nonumber \\
\\
\phi_\times(t; \tc, \tau, \phic) & = 
\left(\frac{\tc - t}{\tau}\right)^{-1/4}
\sin\left\{2 \left[
\phic - \left(\frac{\tc - t}{\tau}\right)^{5/8}
\right]
\right\},
\nonumber 
\end{align}
where,
with $M_1$ and $M_2$ as the masses of the two NSs,
$\mchirp = (M_1 \, M_2)^{3/5} / (M_1 + M_2)^{1/5}$
is the chirp mass, 
$\tc$ is the detector-frame time of coalescence,
$\phic$ is the orbital phase at this time,
$\inc$ is the inclination of the system relative to the line-of-sight, 
$z$ is the (cosmological) redshift of the system, 
$v$ is the (line-of-sight) peculiar velocity of the host galaxy,
$u$ is the (line-of-sight) velocity of the merger relative to the host,
and 
$\dprop(z, \csmpar)$ the proper distance to redshift $z$
given the cosmological parameters $\csmpar$.
The detector-frame coalescence time-scale, $\tau$, 
is equal to the source-frame value of $5 \, G \, \mchirp / c^3$
dilated by a factor of $(1 + z) \, (1 + u / c + v / c)$.

The GW data from a single detector is a 
time-series of measurements
given by a linear combination of the 
orthogonal polarization waveforms given in \eq{strains},
where the weightings depend on the merger's sky position
and the detector's orientation and geometry,
to which detector noise is added.
If a single merger is observed by multiple detectors 
then the joint dataset can be used 
to extract information about both polarizations
(\eg, \cite{Maggiore:2007}).
Taking the full GW data on a merger to be $\dataset$, 
the likelihood has the form
$\prob(\dataset | \mchirp, \tc, \phic, \inc, z, u, v, \csmpar, O)$,
where 
the merger and cosmological parameters are defined above,
and $O$ encodes all the relevant information about the 
observations, such as detector geometries and noise properties.
The sky position of the merger is not included in the list of
parameters here as they are taken to be determined precisely
by the identification of a host galaxy from EM data.
The general expression for the GW likelihood is complicated,
but
for a well-measured event,
such as GW~170807,
a comparatively simple form can be adopted.
The oscillatory nature of the several thousand cycles of the inspiral signal
allows precise constraints to be placed on the parameters,
$\tau$, $\tc$ and $\phic$, that determine the 
time-dependence 
of the waveform
according to \eq{straint}.
The measured amplitude(s) then place
more uncertain constraints on the
pre-factors of the two strain waveforms given in \eq{strains}.
The parameter-dependence
of the likelihood can hence be well captured by
calculating a few (nearly) sufficient statistics
(\eg, Refs.~\cite{Cutler_Flanagan:1994,Nissanke_etal:2010}):
the precisely measured detector-frame coalescence time-scale, 
$\hat{\tau}(\dataset)$;
the precisely measured detector-frame coalescence time, $\tcobs(\dataset)$;
the precisely measured coalescence phase, $\phicobs(\dataset)$;
and estimates of the two orthogonal amplitudes, 
$\hat{\amp}_\plus(\dataset)$ and $\hat{\amp}_\times(\dataset)$.
The uncertainties on these amplitudes,
$\sigma_{\!A_\plus}\!(O)$ and $\sigma_{\!A_\times}\!(O)$,
are determined by a combination of the 
sky position of the source (assumed to be known) and the 
detector's geometry and noise properties (encoded in $O$).
These two uncertainties cannot be ignored,
as they are the dominant contributions to whether a merger
is detected in the first place and, 
for those which pass the selection criteria,
the distance uncertainty.
Assuming no uncertainty on the temporal parameters,
and
that the noise on the amplitudes is Gaussian and uncorrelated,
the GW likelihood can be written 
in the form (\cf\ \eq{strains})
\begin{align}
\label{equation:dnorm_app_t}
&
\prob(\dataset
  | \mchirp, \tc, \phic, \inc, z, u, v, \csmpar, O)
  =
  \delta\!\left[\tcobs(\dataset) - \tc\right]
  \,
  \delta\!\left[\phicobs(\dataset) - \phic\right]
\\
&
\times 
  \,
  \delta\!\!\left[\hat{\tau}(\dataset) \! - \frac{5\,G}{c^3} (1 + z) \!
  \left(1 \! + \! \frac{u}{c} \! + \! \frac{v}{c} \right) \! \mchirp \right]
  \normal\!\!\left[ \hat{\amp}_\plus(\dataset);
\frac{G \, \mchirp / c^2}{\dprop(z, \csmpar)}
\frac{1 + \cos^2(\inc)}{2},
  \sigma_{\!A_\plus}^2\!(O) \!
  \right]
  \normal\!\!\left[ \hat{\amp}_\times(\dataset);
-\frac{G \, \mchirp / c^2}{\dprop(z, \csmpar)}
\cos(\inc),
  \sigma_{\!A_\times}^2\!(O) \!
  \right]
\!.
\nonumber
\end{align}
Calculating these statistics from real data is a challenging
numerical task, although a number of approaches have been
demonstrated that can achieve this
(\eg, \cite{Cutler_Flanagan:1994,Nissanke_etal:2010}).

Fortunately,
for the purposes of the bias analysis presented here,
there is no need to actually simulate full time streams and
go through this complicated procedure to calculate these statistics;
it is sufficient instead to 
include 
$\hat{\tau}$,
$\tcobs$,
$\phicobs$, 
$\hat{\amp}_\plus$ and $\hat{\amp}_\times$
in the formalism as if these quantities were measured directly.
It is hence possible to 
replace $\dataset$ 
with the above statistics and similarly give
the noise amplitudes in place of $O$,
meaning that the likelihood is given entirely in terms of 
explicitly defined quantities.
Also defining the (accurately) measured redshifted chirp mass as
$\mzobs = \hat{\tau} / (5 \, G / c^2)$,
\eq{dnorm_app_t} can be rewritten as 
\begin{align}
\label{equation:dnorm_app}
&
\prob(\tcobs, \phicobs, \mzobs, \hat{\amp}_\plus, \hat{\amp}_\times 
  | \mchirp, \tc, \phic, \inc, z, u, v, \csmpar, \sigma_{\!A_\plus}, 
  \sigma_{\!A_\times})
\\
& 
  \! =
  \delta(\tcobs - \tc)
  \,
  \delta(\phicobs - \phic)
  \, 
  \delta\!\!\left[\mzobs \! - \! (1 + z) \!
  \left(1 \! + \! \frac{u}{c} \! + \! \frac{v}{c} \right) \! \mchirp \right]
  \normal\!\!\left[ \hat{\amp}_\plus;
\frac{G \, \mchirp / c^2}{\dprop(z, \csmpar)} 
\frac{1 + \cos^2(\inc)}{2},
  \sigma_{\!A_\plus}^2 \!
  \right]
  \normal\!\!\left[ \hat{\amp}_\times;
-\frac{G \, \mchirp / c^2}{\dprop(z, \csmpar)} 
\cos(\inc),
  \sigma_{\!A_\times}^2 \!
  \right]
\!.
\nonumber
\end{align}
This could be used either as the sampling distribution 
for a merger with specified properties to
generate the effective data
(\ie, $\tcobs$, $\phicobs$, $\mzobs$, $\hat{\amp}_\plus$ and 
$\hat{\amp}_\times$)
or as the basis for simulating parameter constraints,
in either case without any reference to a full timestream.

In the context of cosmological parameter estimation several
further simplifications can usefully be made:

\begin{itemize}

\item
Given that
both
the galactic motions
relative to the Hubble flow 
and the orbital speeds within galaxies 
are typically a few hundred km/s,
the peculiar velocity terms $u/c$ and $v/c$
produce negligible relative offsets of $\la 0.1 \%$.
As such, 
in the analysis of the GW data
it is reasonable to ignore 
the peculiar velocities
completely
(\ie, 
setting $u = v = 0$).
(The peculiar velocity of the host galaxy cannot be ignored
as its relative contribution to the spectroscopic redshift
is given by the ratio $|v| / (c \, z)$,
which can be $\!\ga 10\%$ at $z \la 0.01$.
It is, however, 
the host galaxy's peculiar velocity that is relevant
here, as it is assumed that no redshift is obtained for the 
merger/burst itself. 
Hence it is only 
$v$ which affects the measured 
redshift;
$u$ can be ignored in the analysis of the EM data as well as
the GW data.)

\item
The coalescence time, $\tc$, and coalescence phase, $\phic$, 
are
both nuisance parameters that do not provide any useful 
information about the properties of the merger.
They would hence be marginalized over in any cosmological
or population analysis, potentially increasing the uncertainty
on other parameters.
However, under the approximation that they are constrained perfectly
by the time evolution of the GW signal,
$\tc$ and $\phic$ can simply be 
omitted from the likelihood completely.

\end{itemize}

Applying these simplifications leaves the 
the significant part of \eq{dnorm_app} as 
\begin{align}
\label{equation:final}
& \prob(\mzobs, \hat{\amp}_\plus, \hat{\amp}_\times
  | \mchirp, \inc, z, \csmpar, \sigma_{\!A_\times}, \sigma_{\!A_\plus})
\nonumber \\
& \simeq
  \delta\!\left[\mzobs - (1 + z) \, \mchirp \right]
  \,
  \normal\!\left[ \hat{\amp}_\plus;
\frac{G \, (1 + z) \, \mchirp / c^2}{\dl(z, \csmpar)} \,
\frac{1 + \cos^2(\inc)}{2},
  \sigma_{\!A_\plus}^2
  \right]
  \,
  \normal\!\left[ \hat{\amp}_\times;
-\frac{G \, (1 + z) \, \mchirp / c^2}{\dl(z, \csmpar)} \,
\cos(\inc),
  \sigma_{\!A_\times}^2
  \right],
\end{align}
where
$\dl(z, \csmpar) = (1 + z) \, \dprop(z, \csmpar)$ is the
luminosity distance to redshift $z$.
This change is made so that 
it is the tightly constrained redshifted chirp mass 
$(1 + z) \, \mchirp$ that appears in all three terms;
in an inference context the substitution 
$(1 + z) \, \mchirp \rightarrow \mzobs$ could be made,
leaving only $\dl(z, \csmpar)$ and $\inc$ to be constrained
by the measured amplitudes $\hat{\amp}_\plus$ and $\hat{\amp}_\times$.
Equation~\ref{equation:final} is 
the form of the likelihood used in \sect{data}
and hence is the basis for all the simulations in this paper.

\end{widetext}


\end{document}